\documentclass[fleqn,usenatbib]{mnras}

 \usepackage{mathptmx}

\usepackage[T1]{fontenc}

\usepackage{graphicx}   
\usepackage{amsmath}   
\usepackage{amssymb}   

\usepackage{threeparttable}   

\title[Revising the gravity and temperature dependence of the Wilson-Bappu effect]{On the physical nature of the Wilson-Bappu effect: revising the gravity and temperature dependence}

\author[F. Rosas-Portilla, K. -P. Schr\"oder, D. Jack]{
  F. Rosas-Portilla$^{1}$\thanks{email: fd.rosasportilla@ugto.mx},
  K. -P. Schr\"oder$^{1,2}$, 
  D. Jack$^{1}$
  \\
  $^{1}$Departamento de Astronomía, Universidad de Guanajuato,
  Callejón de Jalisco S/N Col. Valenciana, Guanajuato 36023, México
  \\
  $^{2}$Sterrewacht Leiden, Universiteit Leiden, Nils Bohrweg 2,
  Leiden 2333CA, Netherlands
}

\date{Accepted XXX. Received YYY; in original form ZZZ}

\pubyear{2021}

\begin{document}
\label{firstpage}
\pagerange{\pageref{firstpage}--\pageref{lastpage}}
\maketitle

\begin{abstract}
  We present a sample of 32 stars of spectral types G and K, and luminosity classes I to V, with moderate activity levels, covering four orders of magnitude of surface gravity and a representative range of effective temperature. For each star we obtained high S/N TIGRE-HEROS spectra with a spectral resolving power of $R\approx20,000$ and have measured the \ion{Ca}{ii} K line-widths of interest, $W_0$ and $W_1$. The main physical parameters are determined by means of \textsc{iSpec} synthesis and \textit{Gaia} EDR3 parallaxes. Mass estimates are based on matching to evolution models. Using this stellar sample, that is highly uniform in terms of spectral quality and assessment, we derive the best-fit relation between emission line width and gravity $g$, including a notable dependence on effective temperature $T_{\rm eff}$, of the form $W_1 \propto g^{-0.229} T_{\rm eff}^{+2.41}$. This result confirms the physical interpretation of the Wilson-Bappu effect as a line saturation and photon redistribution effect in the chromospheric \ion{Ca}{ii} column density, under the assumption of hydrostatic equilibrium at the bottom of the chromosphere. While the column density (and so $W_1$) increases towards lower gravities, the observed temperature dependence is then understood as a simple ionization effect -- in cooler stars, \ion{Ca}{ii} densities decrease in favor of \ion{Ca}{i}.
\end{abstract}

\begin{keywords}
  stars: chromospheres -- stars: late-type -- stars: activity -- stars: fundamental parameters -- techniques: spectroscopic
\end{keywords}

\section{Introduction}

Since \citet{Eberhard1913}, chromospheric emission in the \ion{Ca}{ii} H\&K doublet lines ($\lambda_{\text{H}}$ = 3968.47 \AA{} and $\lambda_{\text{K}}$ = 3933.68 \AA{}) is understood to be a universal phenomenon, connected with stellar magnetic activity, and shared by very different kind of late-type stars, which reaches over a wide range of gravities, from Sun to red giants and supergiant stars.

Much of what we know today about chromospheric emission is based on the pioneering work of Olin C. Wilson and his group at the Mount Wilson Observatory, spanning 6 decades of his life. Obviously, strength and width of that emission contains vital information of the chromospheric heating processes and fundamental physical properties of this otherwise elusive atmospheric layer.

One of the most recognized publications of O. C. Wilson's scientific legacy is his work with M. K. Vainu Bappu of 1957 \citep{Wilson1957}, in which an empirical and apparently universal relation was found between the line width $\log W_0$ of the \ion{Ca}{ii} K chromospheric emission, taken at half-peak intensity, and the stellar absolute visual magnitudes for a sample of 185 late-type stars. Hampered by the lack of precise distances, \citet{Wilson1967} tried to improve this empirical relation. The same lack of precision in giant luminosities, whose parallaxes were too small to be measured directly in those days, made it impossible to test for the dependence on effective temperature and other physical parameters, and to this day it also remains debated, how precise and universal that relation is.  

A major factor in the formation of the observed chromospheric emission line profile is the numerous photon absorption and re-emission processes occurring between the bottom of chromosphere, where the emission is produced, and the top, where the observed line profile emerges. In the absence of considerable collision rates for the \ion{Ca}{ii} H\&K transition because of the low chromospheric densities, almost always an absorbed photon is re-emitted. As laid out by \citet{Ayres1975}, during this process the photons migrate towards the wings of the line profile function, where their escape probability is greater, since the optical depth there is less.

Consequently, there is a line broadening effect, caused by the large optical depth and saturation at the centers of the \ion{Ca}{ii} H\&K lines, which depends on the column density present in each particular chromosphere. This widening effect is larger for stars with lower gravity, notably giant stars. More luminous giants have a lower gravity, for which hydrostatic equilibrium at the bottom of the chromosphere demands larger scale heights and consequently, a broader \ion{Ca}{ii} H\&K emission. 

In particular, the many absorption and re-emission processes that each emerging photon undergoes, is favoring a migration into the wings of the line profile function, which results (as briefly explained above) in the density broadening of the emission line. The simple explanation of this process by \citet{Ayres1975} and \citet{Ayres1979} suggests a clear relation between column density $N$ and gravity $g$ ($N \propto 1/\sqrt{g}$) and yields a dependence of the foot-to-foot emission-line width $W_1$ with gravity of the order of $W_0 \propto g^{-0.25}$. Comparing giants with known physical parameters, this is essentially consistent with observations (see, e.g. \citet{Ayres1975}, \citet{Lutz1982}, or \citet{Park2013}, to name a small selection of work done on this topic). However, in such work, hitherto a 1:1 relation between the two respective line widths $W_1$ and $W_0$ has always been implied, simply relying on the idea of homologous emission line profiles. 

The other important factor, apart from gravity, in the dependence of emission-line width on physical parameters is the effective temperature $T_\text{eff}$, since the ionization balance \ion{Ca}{ii}:\ion{Ca}{i} matters for the \ion{Ca}{ii} column density as well. Underestimating this term can severely reduce the apparent dependence on gravity, since the giants higher up in the RGB and AGB are not only characterized by a lower gravity but also a lower effective temperature. This establishes cross-talk and the necessity to solve very well for the dependence on \textit{both} these physical stellar parameters. However, the notoriously difficult task to assess $T_\text{eff}$ with cool giants has hindered exactly all efforts in this regard.

Another complication arises from the effects of strong activity on the chromospheric emission line width, as found with very active binaries (see \citet{Montes1994}). Apparently, the emission line width can be broaden by local plasma motions along magnetic loops. This adds to the above density broadening in the non-active chromosphere and, when not taken into consideration, can offset efforts to find the correct relation of the pure density broadening with the fundamental physical parameters. In this context, it is important to recognize that chromospheric emission increases, by virtue of rising magnetic activity, on the upper AGB \citep{Schroder2018}. On the other hand, to obtain a precise width measurement, it is desirable to have an emission that exceeds the basal flux. Hence, moderate activity seems acceptable in a suitable stellar sample, but strong activity must be avoided. 

This study attempts to be conscious of the above described problems and is an attempt to improve the empirical basis for the physical relation behind the Wilson-Bappu effect. Our motivation is to verify the interpretation of \cite{Ayres1975} as a density broadening effect to a higher level of confidence than previous studies. The alternative suggestion of a Doppler broadening by turbulence (e.g. \citet{Reimers1973}, and see extensive discussion in \citet{Lutz1982}), seems unlikely already because the inferred turbulent velocities would have to become clearly supersonic in giants and supergiants. 

In order to deliver a most precise assessment of the physical relation underlying the Wilson-Bappu effect, we here present a sample of cool stars, covering four orders of magnitude of gravity and a representative range of effective temperatures, for which we obtained high S/N (typically 50-70 at \ion{Ca}{ii} K), $R\approx20,000$ TIGRE spectra of a uniform quality across the sample. In addition, we assess the physical parameters uniformly with a consistent method, as described below, using the precise \textit{Gaia} EDR3 parallaxes to minimize spectroscopic analysis ambiguities and cross talk between poorly derived physical parameters. We developed a method, dealing well with the residual noise on the observed emission line profiles, to measure both line widths of interest ($W_0$ and $W_1$). Verifying their relation, which is not exactly 1:1, we find this to be an additional, hitherto neglected factor in the comparison of theoretical prediction (using $W_1$ with observation based on O.C. Wilson's $W_0$, which can be measured to higher precision).

\section{Observations}

\subsection{The stellar sample}

In order to revise, how the widths of the K emission-line of \ion{Ca}{ii} behave as a function of (a) the surface gravity $\log g$ and (b) the effective temperature $T_\text{eff}$, we selected a sample of 32 well know stars with spectral types G and K, and luminosity classes I to V. This representative selection covers the large range of $\log g$ from 0.9 to 4.5 dex, while $T_\text{eff}$ values reach from 3844 to 6013 K (see Table \ref{table:stellar_parameters}). The latter is restricted by the tested reliability of spectroscopic assessment consistency (details are described below).

The state of activity is mostly low to moderate, see the plots of the chromospheric \ion{Ca}{ii} K emission line profiles in Appendix \ref{sec:CaII_widths}. Apart from the aforementioned complications with the emission from very active regions in the chromosphere, we also avoid another problem: the rotational velocities and their variation across the sample remain within the spectral resolution of TIGRE-HEROS (see below), which is equivalent to 7.5 km/s. Only some of the most active giants reach rotation velocities of over 8 km/s, see e.g. \citet{Auriere2015} (Table 3 therein). The same can be said about the turbulent velocities (of the order of several km/s in both, the photosphere and lower chromosphere), which should -- by comparison to the widening by the Wilson Bappu effect -- not be of much significance.

Included in our sample are the four Hyades K giants, which mark -- according to their X-ray and other activity properties, see \citet{Schroder2020} -- low to strong solar activity. The other sample stars fall into this range as well. Please note, that (1) the scales in Appendix \ref{sec:CaII_widths} are not the same, (2) the photospheric line depths vary a lot, giving large contrast in a cool giant like Arcturus to even the pure basal chromospheric flux, and (3) even the strongest solar activity, in the wider stellar context, is very moderate by comparison.

Thanks to the recently published parallaxes of the \textit{Gaia} EDR3 catalogue \citep{GaiaCollaboration2020} of the ESA \textit{Gaia} mission \citep{GaiaCollaboration2016} and using reasonable estimates of the stellar mass $M_{\star}$ from evolution tracks that match the luminosity $L$ and $T_\text{eff}$ of each respective star, derived provisionally by a spectral fitting process, we obtained the parallax-based surface gravities, which in turn help to maximize the non-spectroscopic information on each star. In an iterative approach, we then improved the effective temperature derived for each star from spectral synthesizing by selecting for those least residuals solutions, which reproduce the respective calculated gravity.

Only in four cases, with the three very bright stars $\alpha$ Tau (HD 29139), $\alpha$ Hya (HD 81797) and $\alpha$ Boo (HD 124897), as well as the bright eclipsing binary $\zeta$ Aur (HD 32068), \textit{Gaia}'s EDR3 parallaxes are not reliable. In these cases, we used the \textit{Hipparcos} parallaxes \citep{ESA1997}, instead, since their accuracy is not limited so much for the brightest stars, and for $\zeta$ Aur it is in good agreement with the distance obtained from the observed binary system information (see \citet{Schroder1997}).

\subsection{Spectroscopic observations with TIGRE}

To start with a homogeneous set of spectra of comparable quality, all stars of this sample were observed with the 1.2 m robotic telescope TIGRE \citep{Schmitt2014} located near Guanajuato, central Mexico. It is equipped with the HEROS echelle spectrograph, which has a moderately-high uniform resolution of $R \approx 20,000$ over the broad wavelength range from about 3800 to 8800 \AA{}, and produces spectra of good quality -- typically S/N $\sim$ 100 or more, averaged over the spectrum, which corresponds to 50-70 in the \ion{Ca}{ii} H\&K line region of a cool star.

The broad wavelength coverage of HEROS is achieved by two separate channels and cameras (for \textit{red} and \textit{blue} light, named channel R and channel B). In this way, the cameras and the cross-dispersers are individually optimized for each wavelength region in sensitivity, the echelle orders well separated, and resolution kept uniform. There is only a small gap of about 120 \AA{} around 5800 \AA{} (located on the blue side of the sodium D line), caused by the characteristics of the dichromatic beam splitter.

The advantage of TIGRE-HEROS spectra is the versatility of their use. While we obtain information on the chromospheric \ion{Ca}{ii} H\&K emission from the B channel spectra, the R channel spectral range is more suitable to spectral synthesizing and the R camera has the best S/N. Flexible scheduling of the robotic telescope provided us with the required observations, including several takes of each star, over the course of about a year.

To improve our measurements of \ion{Ca}{ii} K emission-line widths, we typically added two or three spectra of the B channel, following a quality control \textit{by eye} to discard anomalous differences between different takes. With this approach, we avoided the effects of electronic noise or under-exposure in nights of bad transparency, optimize S/N, as well as our measurements of the emission line widths.

\section{Deriving the physical parameters}

\subsection{Calculation of parallax-based gravities}
\label{subsec:parallax-based gravities}

In a first step to calculate the parallax-based gravities, we derived absolute visual magnitudes via the \textit{Gaia} EDR3 parallaxes and available photometry, and estimated stellar masses $M_\star$ using an approximate preliminary analysis of its position in the Hertzsprung-Russell diagram to find the matching evolution track. We used the photometric apparent magnitudes from the SIMBAD database in the $V$ band ($m_V$), as well as the color index $B-V$. The parallaxes in milliarcseconds (and distances in parsecs) were taken from the \textit{Gaia} EDR3 archive. We applied the parallax zero-point correction of \citet{Lindegren2020}, using their published Python code. No interstellar extinction correction was applied, because the average distance of our sample stars is below $\sim 100$ pc.

To determine the bolometric magnitudes and, finally, each luminosity, we applied the bolometric corrections (BC) from \citet{Flower1996} for which the effective temperature $T_\text{eff}$, or the color index $B-V$, is required. As a start value (to be refined later) for the $T_\text{eff}$ we used the average of the PASTEL catalogue entries \citep{Soubiran2016} for each sample star.

\citet{Flower1996} expresses the bolometric correction in the $V$-band ($\text{BC}_V$) as a function of the effective temperature of the form: $\text{BC}_V = a + b(\log T_\text{eff}) + c(\log T_\text{eff})^{2} + ...$, for three different temperature ranges. The coefficients $a, b, c, ...$ are shown in his Table 6 for each interval, but because of a typographical error, the powers of those coefficients were omitted. A fact that was clarified and corrected by \citet{Torres2010}. In addition, \citet{Flower1996} calculates $\log T_\text{eff}$ as a function of color index of the form $\log T_\text{eff} = a + b(B-V) + c(B-V)^{2} + ...$ where the coefficients $a, b, c, ... $ are taken separately for supergiants or for main sequence stars, subgiants and giants. However, the values of $\text{BC}_V$ differ a bit, depending on whether the color index or the effective temperature is used. To have a better approximation and eliminate possible differences, we used the average of the resulting two values of $\text{BC}_V$, derived from $B-V$ and $T_\text{eff}$, respectively. Finally, the luminosity in solar units was calculated using the relation $\log L = (4.74 - M_V - \text{BC}_V)/2.5$, where we applied a solar visual absolute magnitude of 4.74.

Using the stellar luminosities and effective temperatures, the stars were located in a Hertzsprung-Russell diagram (HRD). The $\text{M}_\star$ were estimated using a grid of stellar evolution tracks that we computed with the Cambridge (UK) Eggleton code in its updated version (\citet{Pols1997}, \citet{Pols1998}), with the quasi-solar metallicity of $Z=0.02$ (see Fig. \ref{fig:HRD_masses}).

According to the PASTEL catalogue, our sample has an average metallicity of $-0.05$ dex, i.e. close to solar metallicity. However, taking into account the residual differences, we need to consider that any smaller metallicity makes a star of given mass and age more luminous and less cool, caused by its lower opacities, and the stellar position in the HRD is shifted to the lower left, see e.g. \citet{Schroder2013}. Consequently, because of such miss-matches based on the residual differences of the stellar metallicites from solar, we considered an uncertainty for each stellar mass estimate of about 10 percent.

Finally, the surface gravity of each sample star was calculated from the respective stellar luminosity, mass, and (initial) effective temperature as follows. If $R_\star$ is the stellar radius, $\log L \propto  \log(R_\star^{2} T_\text{eff}^{4}) \propto \log(\text{M}_\star g^{-1} T_\text{eff}^{4})$. Finally, if $\log g_\odot = 4.437$ (in units of cgs) is the solar surface gravity, the surface gravity is given by

\begin{equation}
  \log g = 4.437 + \log \left( \dfrac{\text{M}_{\star}}{\text{M}_{\sun}} \right) 
  - \log \left( \dfrac{L}{L_{\sun}} \right) 
  + 4 \log \left( \dfrac{T_\text{eff}}{T_{\sun}}\right).
  \label{eq:gravity}
\end{equation}

Using equation \ref{eq:gravity}, this resulted in preliminary gravities for the sample. We used this value to derive the effective temperatures by means of spectral synthesizing (see below), avoiding solutions with wrong gravity and the resulting cross-talk into $T_\text{eff}$, see \citep{Schroder2021}. Then the calculated gravities were updated with these self-determined $T_\text{eff}$ values. We repeated this procedure to finally refine the effective temperatures as well, where needed. Table~\ref{table:parallax-based_gravities} summarizes our final results for all 32 stars used in this study.

\subsection{Deriving effective temperatures by spectral synthesizing}

To derive stellar parameters, $T_\text{eff}$ in particular from the R channel of TIGRE-HEROS, we used the spectral analysis toolkit \textsc{iSpec} \citep{Blanco-Cuaresma2014} in its latest Python 3 version (v2020.10.01) \citep{Blanco-Cuaresma2019}. It allows a comparison of an observed spectrum with synthetic spectra, interpolated from atmospheric models of different physical parameters. The best match, defined by the least differences ($\chi^2$ method) then defines, at least in principle, the corresponding physical parameters. However, in practice there are a number of complications, mostly of physical nature, as pointed out by \citet{Schroder2021}. Therefore, information external to the spectroscopic synthesizing is vital to allow testing and to filter seemingly equivalent best-fit solutions. 

These problems are mostly caused by a large minimal $\chi^2$ level, produced by the residual mismatches of line strengths (atomic $f$-values), as well as by subtle, unnoticed blends with other lines -- even in a well-selected, representative subset of seemingly reliable lines. Furthermore, any such subset is only representative and usable for a limited range of effective temperatures. In addition, towards the low end of stellar effective temperatures, NLTE effects may be mismatched by the model libraries that \textsc{iSpec} can use. These complications set limits and require a careful handling of the parameter assessment, applied consistently to all sample stars. 

We applied the spectral synthesis process to determine the stellar parameters only in the R-channel of the spectrum, given its better S/N in comparison with the B-channel of HEROS, and its lesser line density, better avoiding line blends interfering with the analysis. We then developed an \textsc{iSpec} script (\textsc{iSPar} v4.4.3), written in Python 3, to perform the spectral synthesizing by an algorithm, which uses the very useful \textsc{iSpec} tools in a consistent manner.

We used the grid of \textsc{marcs} atmosphere models \citep{Gustafsson2008} included in \textsc{iSpec}. For the reference solar abundances, we used the values published in \citet{Grevesse2007}. The synthetic spectra were calculated using the radiative transfer code \textsc{turbospectrum} version 15.1 (\citet{Alvarez1998}, \citet{Plez2012}), which uses already spherical models instead of the plane-parallel approximation (a better approach for giant stars).

When spectral synthesis methods are applied to spectroscopic analysis, parameter determination is complicated by having to fit several of them at the same time, as this approach causes a multitude of ambiguous best-fit solutions. These are caused by the fact that one mismatched parameter can be compensated in its effect by the mismatch of others, named as \textit{cross-talk}, still resulting in a seemingly good representation of the observed spectrum. \textsc{iSpec} therefore finds different solutions in parameter space, depending on the choice of the initial values, which all need to be explored and tested. We based our algorithm therefore on the considerations of \citet{Schroder2021}, who showed a fast method for determining the physical parameters of main sequence stars with moderate resolution spectra, where that was possible. For example, there are no reliable $ubvy$-photometry-derived metallicity values available for giant stars. Still, our script aims on minimizing cross-talk by maximizing the use of non-spectroscopic information to filter out those best-fit solutions with parameters contradicting that information external to the synthesizing process.

A very delicate procedure preceding the synthesizing step is the fit of the pseudo-continuum of the observed stellar spectrum in question and the line selection for the synthesizing procedure with \textsc{iSpec}. After several comparative tests, we found that the impact of this step on the final results of the stellar parameters is not negligible, since it affects the $\chi^2$ sums and their minima in parameter space. We explain in detail the pseudo-continuum adjustment and the line selection in the appendix \ref{sec:pseudo-continuum_and_line_selection}. The latter is important for the successful spectroscopic synthesizing and is a compromise between completeness and quantity versus line data quality.

\subsubsection{Final stellar parameter determination}

For the final analysis of the now readjusted, normalized observed spectrum by parameter synthesizing, we set the start values to the averages of $T_\text{eff}$ and $\text{[M/H]}$ in the PASTEL catalogue, while the initial gravities were calculated using equation \ref{eq:gravity}. Initial $v_\text{mic}$ values -- as detailed in the appendix \ref{sec:pseudo-continuum_and_line_selection} for the synthetic spectra used to guide the continuum readjustment -- are set to the velocities used for the \textsc{phoenix} library \citep{Husser2013}, while $v_\text{mac}$ is set again by equation \ref{eq:vmac}.

The best-fit results of \textsc{iSpec} synthesizing are sensitive to the choice of the initial parameters. The further exploration of the parameter space therefore follows the strategy of \citet{Blanco-Cuaresma2014}, varying the initial values in steps. Our \textsc{iSpec} script implements an analysis that varies the initial parameters by as much as follows: $\pm 100$ K for $T_\text{eff}$, $\pm 0.15$ dex for $\log g$, and $\pm 0.15$ dex for $\text{[M/H]}$, each time repeating the process for a total of 27 calculations around the initial values. Then stellar parameters with the best $\chi^2$ result of all solutions is chosen. These step-sizes and the following sequence performed best in our tests described in subsection \ref{subsec:GBS_Test}.

Based on the strategy of \cite{Schroder2021}, the best way to determine the stellar parameters is to leave only one of them free at a time, while the others remain fixed, going through the following sequence in a partially iterative procedure: (1) refine $v_\text{mic}$; (2) having re-set $v_\text{mac}$ through equation \ref{eq:vmac}, based on the now refined $v_\text{mic}$, refine gravity ($\log{g}$); (3) now refine $v\sin i$; (4) refine $T_\text{eff}$, $\text{[M/H]}$ and $\text{[}\alpha\text{/Fe]}$; (5) refine $v_\text{mac}$; (6) refine again $v_\text{mic}$; (7) refine again $T_\text{eff}$, $\text{[M/H]}$ and $\text{[}\alpha\text{/Fe]}$; (8) refine again $v\sin i$ and finally, (9) refine again $\text{[Fe/H]}$. However, with the resolution of TIGRE–HEROS spectra in mind, we are aware that different turbulence and rotation velocity values under 3 km/s bear no physical meaning, considering our modest spectral resolution. Still, there is a need to adjust them in a tested way to minimize cross talk (see below). We should also note, for the complications with cross-talk in combination with the limited spectral resolution, that we do not consider the rotation and turbulence velocity values derived here of any credibility because they are typically half the instrumental line profile width. Our tests only demonstrate the reliability of the main physical parameters, namely effective temperature, gravity, and metallicity.

\subsection{Consistency check with parallax-based gravities}

\begin{figure}
  \includegraphics[width=\columnwidth]{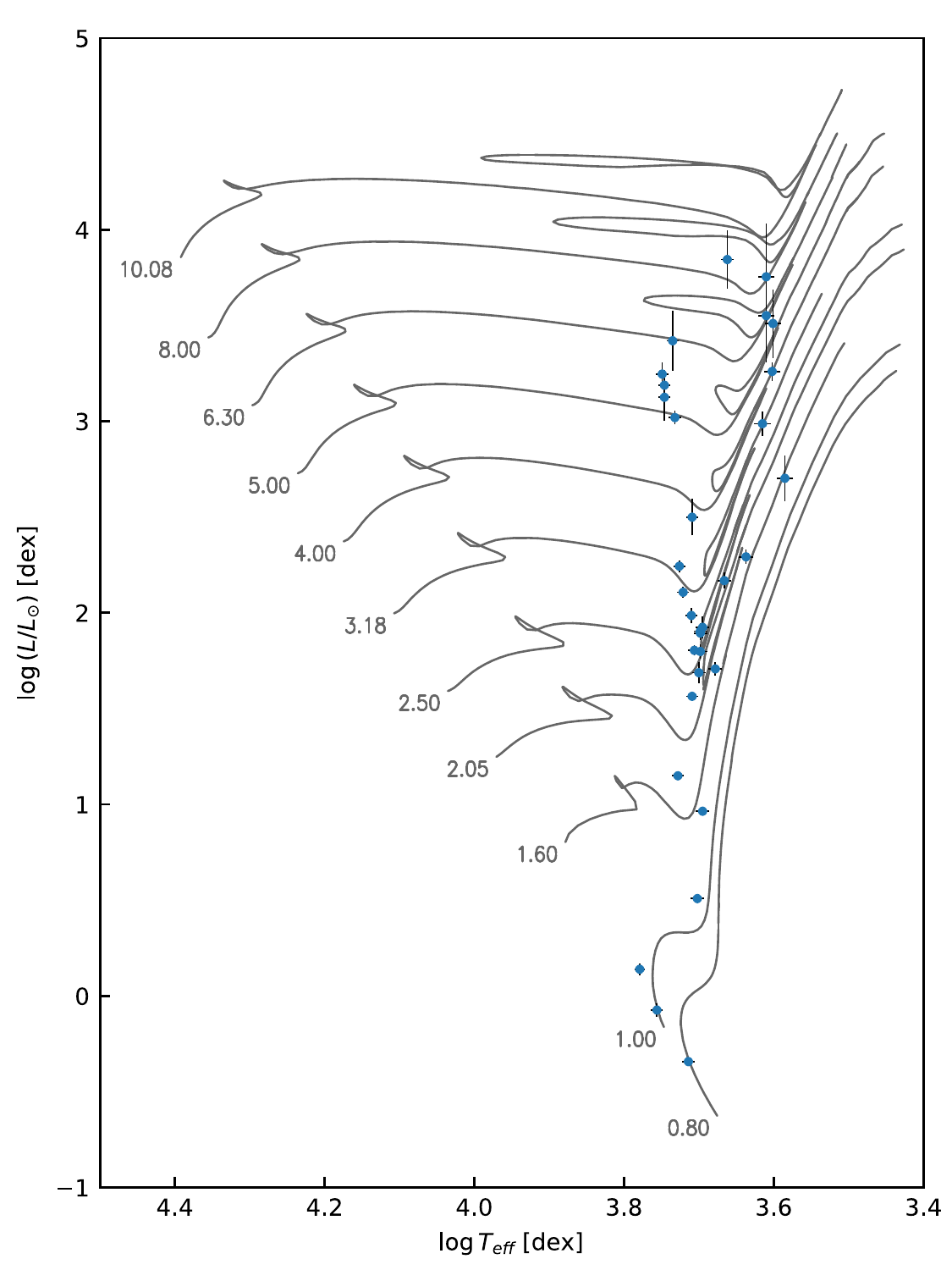}
  \caption{Grid of evolution tracks with solar metallicity for different stellar masses in solar mass units. The grid was calculated using the Cambridge (UK) Eggleton code in its version described and tested by \citet{Pols1997} and \citet{Pols1998}.}
  \label{fig:HRD_masses}
\end{figure}

The improved values of $T_\text{eff}$ obtained from spectral synthesis with our \textsc{iSpec} script give slightly different positions in the HR diagram. We therefore repeated, at this point, the mass estimates and, consequently, recalculated the parallax-based gravities (in the same way as in subsection \ref{subsec:parallax-based gravities}), and finally also repeated all synthesizing steps of above. Table \ref{table:stellar_parameters} summarizes the final stellar parameters obtained for the sample stars with this method, while Table \ref{table:parallax-based_gravities} details the distances, magnitudes, physical parameters and parallax-based gravities. Fig. \ref{fig:HRD_masses} pictures the final estimation of the masses for the stellar sample on a grid of stellar evolution tracks, computed with the Cambridge (UK) Eggleton code for solar metallicity.

\begin{figure}
  \includegraphics[width=\columnwidth]{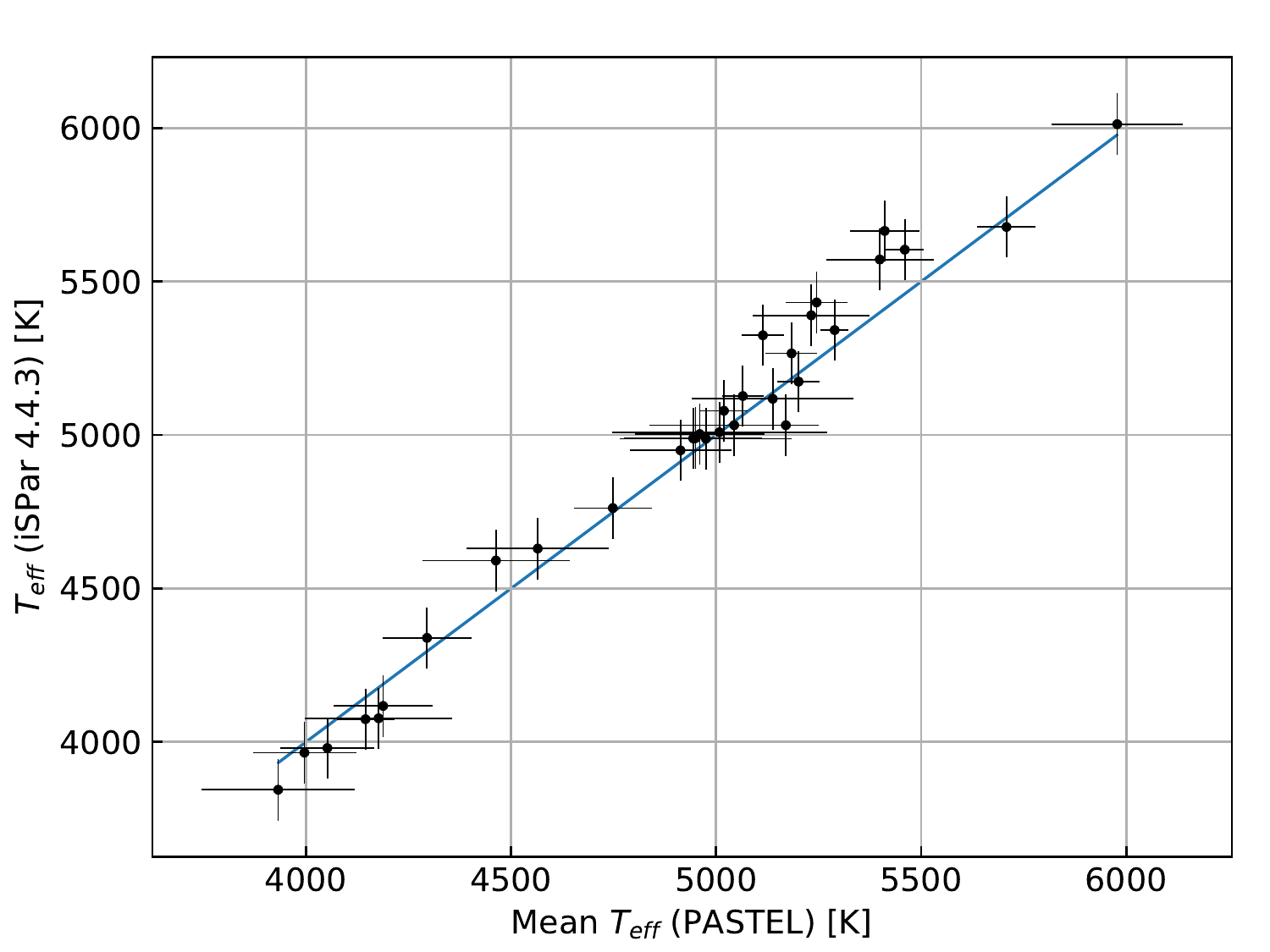}
  \caption{Temperature comparison between the values calculated by spectral synthesis using our \textsc{iSpec} script and the average values in the PASTEL catalogue.}
  \label{fig:Teff_comparison}
\end{figure}

\begin{figure}
  \includegraphics[width=\columnwidth]{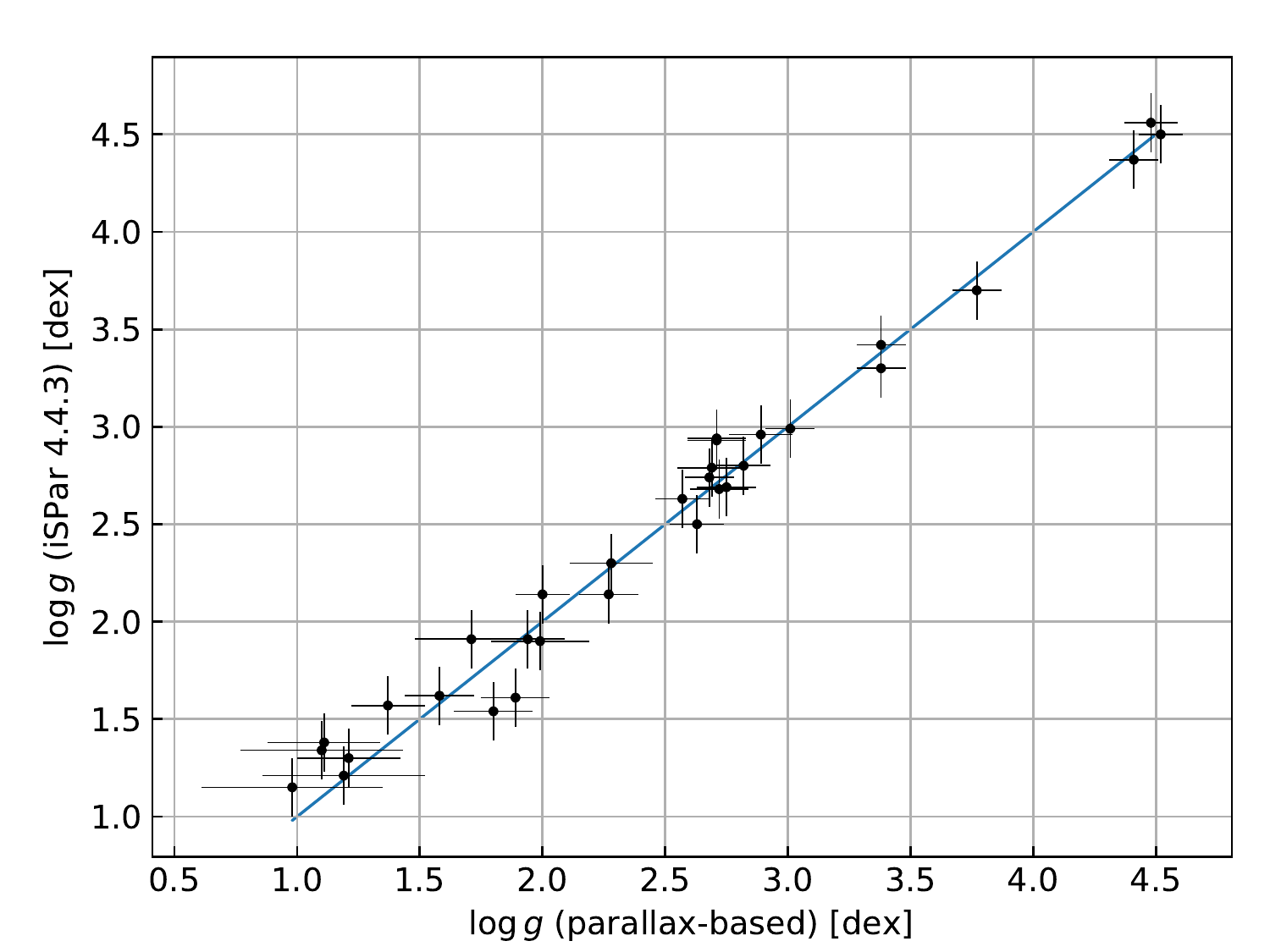}
  \caption{Surface gravity comparison between the values calculated by spectral synthesis using our \textsc{iSpec} script and the parallax-based gravities using the equation \ref{eq:gravity}.}
  \label{fig:logg_comparison}
\end{figure}

\begin{figure}
  \includegraphics[width=\columnwidth]{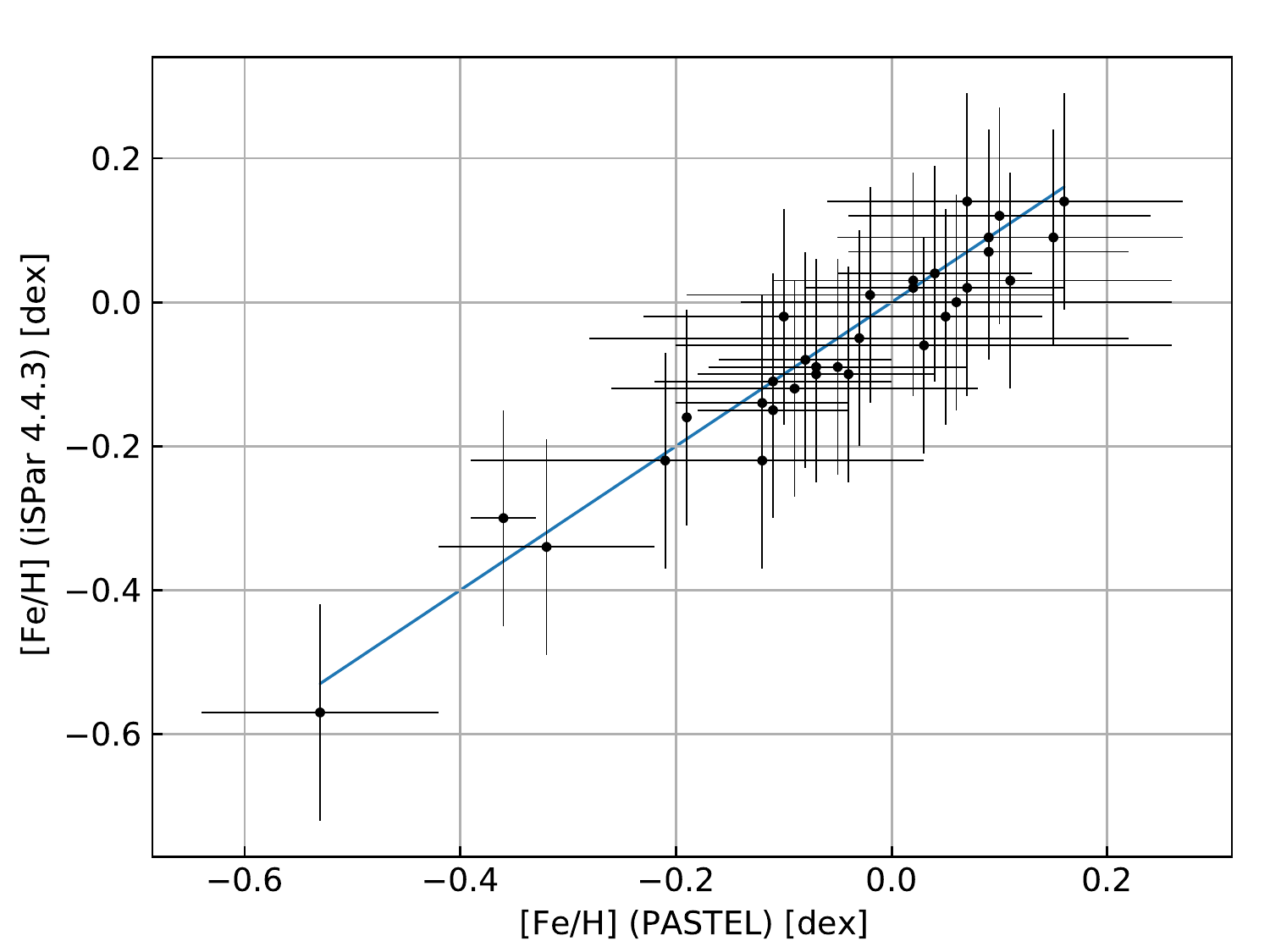}
  \caption{Metallicity comparison between the values calculated by spectral synthesis using our \textsc{iSpec} script and the average values in the PASTEL catalogue.}
  \label{fig:FeH_comparison}
\end{figure}

To ensure that our assessment of the stellar parameters of the sample is sufficiently reliable, we tested our results as follows: Fig. \ref{fig:Teff_comparison} and Fig. \ref{fig:FeH_comparison} show a comparison between our results for $T_\text{eff}$ and $\text{[Fe/H]}$ by spectral synthesis (as described above) and the average values in the PASTEL catalogue. Fig. \ref{fig:logg_comparison} shows a comparison between our results for $\log g$ by spectral synthesis and the parallax-based gravities of Table \ref{table:parallax-based_gravities}. The blue line in these graphs represents the ideal of a 1:1 relation; the uncertainties of our estimations were calculated as explained in subsection \ref{subsec:GBS_Test}.

\begin{table*}
  \centering
  \caption{Stellar parameters derived with our \textsc{iSpec} script.}
  \label{table:stellar_parameters}
  \begin{threeparttable}
    \begin{tabular}{llcccccccc}
      \hline
      Star & Type$^a$ & $T_\text{eff}$$^b$ & $\log g$$^c$ & $\text{[M/H]}$ & $\text{[}\alpha\text{/Fe]}$ & $v_\text{mic}$$^d$ & $v_\text{mac}$$^d$ & $v\sin{i}$$^d$ & $\text{[Fe/H]}$  \\
      & & [K] & [dex] & [dex] & [dex] & [km/s] & [km/s] & [km/s] & [dex] \\ \hline
      HD 8512 & K0IIIb & 4762 (17) & 2.50 (0.04) & -0.15 (0.01) & 0.04 (0.02) & 1.66 & 3.44 & 0.00 & -0.15 (0.01) \\
      
      HD 10476 & K1V & 5174 (24) & 4.50 (0.02) & -0.10 (0.02) & 0.01 (0.02) & 0.96 & 0.00 & 0.00 & -0.10 (0.02) \\
      
      HD 18925 & G9III & 5032 (26) & 2.30 (0.00) & -0.17 (0.02) & -0.04 (0.03) & 1.39 & 3.67 & 1.20 & -0.16 (0.01) \\
      
      HD 20630 & G5V & 5678 (43) & 4.56 (0.05) & -0.03 (0.04) & 0.01 (0.04) & 1.31 & 3.70 & 0.60 & -0.02 (0.02) \\
      
      HD 23249 & K0+IV & 5032 (28) & 3.70 (0.06) & 0.03 (0.02) & 0.02 (0.03) & 1.18 & 0.48 & 0.00 & 0.00 (0.02) \\
      
      HD 26630 & G0Ib & 5572 (31) & 1.90 (0.11) & 0.01 (0.02) & 0.02 (0.03) & 3.25 & 5.79 & 9.98 & 0.01 (0.02) \\
      
      HD 27371 & G9.5IIIab & 4988 (31) & 2.93 (0.06) & 0.10 (0.03) & -0.10 (0.03) & 1.58 & 3.22 & 0.00 & 0.07 (0.02) \\
      
      HD 27697 & G9.5III & 4989 (32) & 2.68 (0.06) & 0.12 (0.03) & -0.03 (0.03) & 1.58 & 3.78 & 0.00 & 0.09 (0.02) \\
      
      HD 28305 & G9.5III & 4950 (25) & 2.79 (0.07) & 0.18 (0.02) & -0.12 (0.03) & 1.65 & 2.51 & 0.00 & 0.14 (0.02) \\
      
      HD 28307 & G9III & 5009 (28) & 2.96 (0.04) & 0.14 (0.02) & -0.09 (0.03) & 1.67 & 2.46 & 0.00 & 0.12 (0.02) \\
      
      HD 29139 & K5+III & 3844 (18) & 1.30 (0.04) & -0.19 (0.01) & -0.14 (0.02) & 1.65 & 3.07 & 3.94 & -0.22 (0.02) \\
      
      HD 31398 & K3II-III & 4077 (13) & 1.34 (0.03) & -0.07 (0.01) & -0.17 (0.01) & 2.11 & 4.02 & 3.93 & -0.09 (0.01) \\
      
      HD 31910 & G1Ib & 5665 (26) & 1.91 (0.11) & -0.04 (0.02) & 0.05 (0.03) & 3.79 & 7.78 & 10.66 & -0.02 (0.01) \\
      
      HD 32068 & K5II & 3980 (18) & 1.21 (0.03) & -0.05 (0.01) & -0.17 (0.02) & 1.86 & 2.84 & 6.35 & -0.06 (0.01) \\
      
      HD 48329 & G8Ib & 4591 (11) & 1.38 (0.03) & 0.05 (0.01) & -0.13 (0.01) & 2.93 & 5.37 & 9.08 & 0.03 (0.01) \\
      
      HD 71369 & G5III & 5266 (27) & 2.74 (0.07) & -0.11 (0.02) & 0.05 (0.03) & 1.76 & 4.32 & 0.00 & -0.10 (0.02) \\
      
      HD 81797 & K3IIIa & 4117 (18) & 1.57 (0.03) & -0.08 (0.01) & -0.15 (0.02) & 1.89 & 3.31 & 1.67 & -0.09 (0.01) \\
      
      HD 82210 & G5III-IV & 5342 (58) & 3.42 (0.15) & -0.30 (0.05) & 0.03 (0.05) & 1.69 & 5.40 & 0.00 & -0.30 (0.03) \\
      
      HD 96833 & K1III & 4630 (23) & 2.14 (0.06) & -0.10 (0.02) & 0.03 (0.03) & 1.65 & 3.41 & 0.00 & -0.11 (0.02) \\
      
      HD 104979 & G8III & 4990 (30) & 2.69 (0.08) & -0.33 (0.03) & 0.05 (0.03) & 1.43 & 0.98 & 0.00 & -0.34 (0.02) \\
      
      HD 109379 & G5IIB & 5325 (28) & 2.63 (0.06) & 0.14 (0.02) & 0.01 (0.03) & 1.51 & 2.78 & 5.90 & 0.14 (0.02) \\
      
      HD 114710 & F9.5V  & 6013 (52) & 4.37 (0.06) & 0.01 (0.04) & 0.01 (0.03) & 1.12 & 1.94 & 3.37 & 0.02 (0.02) \\
      
      HD 115659 & G8IIIa & 5127 (30) & 2.94 (0.05) & 0.03 (0.03) & -0.04 (0.03) & 1.57 & 3.96 & 0.46 & 0.02 (0.02) \\
      
      HD 124897 & K1,5III & 4339 (18) & 1.54 (0.04) & -0.55 (0.02) & 0.20 (0.03) & 1.75 & 2.86 & 0.00 & -0.57 (0.02) \\
      
      HD 148387 & G8-IIIab & 5079 (36) & 2.80 (0.08) & -0.07 (0.03) & 0.04 (0.04) & 1.53 & 1.70 & 0.00 & -0.08 (0.02) \\
      
      HD 159181 & G2Ib-IIa & 5390 (33) & 2.14 (0.10) & 0.02 (0.03) & -0.01 (0.03) & 2.72 & 8.22 & 7.26 & 0.03 (0.02) \\
      
      HD 164058 & K5III & 3965 (14) & 1.62 (0.02) & -0.01 (0.01) & -0.16 (0.01) & 1.70 & 3.29 & 2.62 & -0.05 (0.01) \\
      
      HD 186791 & K3II & 4074 (15) & 1.15 (0.03) & -0.11 (0.01) & -0.15 (0.02) & 2.03 & 3.61 & 3.52 & -0.12 (0.01) \\
      
      HD 198149 & K0IV & 5003 (29) & 3.30 (0.06) & -0.21 (0.02) & -0.01 (0.03) & 1.21 & 3.05 & 1.47 & -0.22 (0.02) \\
      
      HD 204867 & G0Ib & 5604 (27) & 1.61 (0.07) & 0.02 (0.02) & 0.04 (0.03) & 2.96 & 3.34 & 10.87 & 0.04 (0.01) \\
      
      HD 205435 & G8III & 5118 (33) & 2.99 (0.01) & -0.14 (0.03) & 0.05 (0.04) & 1.77 & 2.18 & 0.00 & -0.14 (0.02) \\
      
      HD 209750 & G2Ib & 5432 (19) & 1.91 (0.08) & 0.08 (0.02) & 0.01 (0.02) & 3.14 & 6.70 & 9.79 & 0.09 (0.01) \\
      \hline
    \end{tabular}
    \begin{tablenotes}
    \item[a] Spectral type from SIMBAD database.
    \item[b] The nominal and very small error on $T_{\rm eff}$ reported by \textsc{iSpec} does not enter our uncertainty of parallax-based gravities. We rather apply a more realistic uncertainty of $\pm 100$ K according to our tests in subsection \ref{subsec:GBS_Test}.
    \item[c] The surface gravity values derived by \textsc{iSpec} won't be used in our analysis, we show them only as a reference to the consistency achieved.
    \item[d] Because of our modest spectral resolution ($ R \approx 20,000$), turbulences and rotation velocities derived of 3 km/s are not physically meaningful. Nevertheless, the average nominal error reported by \textsc{iSpec} is $\sim 0.5$ km/s.
    \end{tablenotes}
  \end{threeparttable}
\end{table*}

\begin{table*}
  \centering
  \caption{Distances, magnitudes, physical parameters and parallax-based gravities.}
  \label{table:parallax-based_gravities}
  \begin{threeparttable}
    \begin{tabular}{llcccccc}
      \hline
      Star & $D$ & $B-V$ & $V$ & \text{BC} & $\log (L/L_{\sun})$ & $M_{\star}$ & $\log g$$^b$ \\
      & [pc] & [mag] & [mag] & [mag] & [dex] & [M$_{\sun}$] & [dex] \\ \hline
      HD 8512 & 34.48 (0.26) & 1.06 (0.05) & 0.902 (0.026) & -0.439 (0.059) & 1.707 (0.034) & 1.70 (0.17) & 2.63 (0.11) \\
      
      HD 10476 & 7.65 (0.01) & 0.84 (0.03) & 5.823 (0.013) & -0.235 (0.031) & -0.343 (0.017) & 0.85 (0.09) & 4.52 (0.09) \\
      
      HD 18925 & 70.98 (3.85) & 0.70 (0.26) & -1.326 (0.128) & -0.208 (0.112) & 2.505 (0.096) & 3.90 (0.39) & 2.28 (0.17) \\
      
      HD 20630 & 9.27 (0.02) & 0.67 (0.07) & 5.014 (0.054) & -0.101 (0.033) & -0.073 (0.035) & 1.00 (0.10) & 4.48 (0.11) \\
      
      HD 23249 & 9.09 (0.02) & 0.92 (0.03) & 3.747 (0.014) & -0.287 (0.041) & 0.508 (0.022) & 1.20 (0.12) & 3.77 (0.10) \\
      
      HD 26630 & 256.12 (17.25) & 0.96 (0.31) & -2.882 (0.156) & -0.201 (0.153) & 3.125 (0.124) & 5.50 (0.55) & 1.99 (0.20) \\
      
      HD 27371 & 46.23 (0.39) & 0.99 (0.06) & 0.325 (0.028) & -0.331 (0.065) & 1.894 (0.037) & 2.62 (0.26) & 2.71 (0.12) \\
      
      HD 27697 & 49.32 (0.55) & 0.98 (0.07) & 0.295 (0.034) & -0.324 (0.072) & 1.904 (0.043) & 2.75 (0.28) & 2.72 (0.12) \\
      
      HD 28305 & 44.79 (0.34) & 1.01 (0.09) & 0.274 (0.067) & -0.354 (0.082) & 1.924 (0.059) & 2.75 (0.28) & 2.69 (0.14) \\
      
      HD 28307 & 40.42 (0.71) & 0.94 (0.09) & 0.807 (0.047) & -0.294 (0.087) & 1.687 (0.054) & 2.42 (0.24) & 2.89 (0.13) \\
      
      HD 29139$^a$ & 20.43 (0.32) & 1.54 (0.09) & -0.692 (0.044) & -1.333 (0.257) & 2.702 (0.120) & 1.50 (0.15) & 1.21 (0.21) \\
      
      HD 31398 & 138.62 (7.27) & 1.53 (0.25) & -3.019 (0.124) & -1.128 (0.481) & 3.551 (0.242) & 6.50 (0.65) & 1.10 (0.33) \\
      
      HD 31910 & 259.88 (10.86) & 0.93 (0.20) & -3.054 (0.101) & -0.177 (0.096) & 3.184 (0.079) & 5.20 (0.52) & 1.94 (0.15) \\
      
      HD 32068$^a$ & 240.96 (16.84) & 1.22 (0.32) & -3.160 (0.162) & -0.844 (0.285) & 3.494 (0.179) & 5.10 (0.51) & 1.04 (0.26) \\
      
      HD 48329 & 269.47 (13.10) & 1.41 (0.23) & -4.173 (0.116) & -0.708 (0.267) & 3.844 (0.153) & 8.20 (0.82) & 1.11 (0.23) \\
      
      HD 71369 & 55.85 (0.45) & 0.85 (0.06) & -0.315 (0.028) & -0.223 (0.042) & 2.107 (0.028) & 3.20 (0.32) & 2.68 (0.10) \\
      
      HD 81797$^a$ & 55.28 (0.55) & 1.45 (0.06) & -1.743 (0.032) & -0.994 (0.129) & 2.987 (0.064) & 3.20 (0.32) & 1.37 (0.15) \\
      
      HD 82210 & 32.13 (0.12) & 0.77 (0.04) & 2.036 (0.018) & -0.179 (0.030) & 1.149 (0.019) & 1.70 (0.17) & 3.38 (0.10) \\
      
      HD 96833 & 43.11 (0.47) & 1.14 (0.07) & -0.163 (0.033) & -0.526 (0.074) & 2.168 (0.043) & 2.40 (0.24) & 2.27 (0.12) \\
      
      HD 104979 & 51.40 (0.44) & 0.99 (0.06) & 0.565 (0.028) & -0.330 (0.066) & 1.798 (0.038) & 2.30 (0.23) & 2.75 (0.12) \\
      
      HD 109379 & 45.51 (0.48) & 0.88 (0.07) & -0.651 (0.033) & -0.227 (0.046) & 2.243 (0.032) & 3.30 (0.33) & 2.57 (0.11) \\
      
      HD 114710 & 9.20 (0.01) & 0.59 (0.07) & 4.431 (0.053) & -0.049 (0.026) & 0.139 (0.032) & 1.10 (0.11) & 4.41 (0.10) \\
      
      HD 115659 & 39.22 (0.42) & 0.92 (0.07) & 0.033 (0.033) & -0.268 (0.064) & 1.986 (0.039) & 2.90 (0.29) & 2.71 (0.12) \\
      
      HD 124897$^a$ & 11.26 (0.06) & 1.23 (0.04) & -0.308 (0.022) & -0.692 (0.071) & 2.292 (0.037) & 1.40 (0.28) & 1.80 (0.16) \\
      
      HD 148387 & 28.01 (0.22) & 0.91 (0.05) & 0.503 (0.027) & -0.284 (0.043) & 1.804 (0.028) & 2.55 (0.26) & 2.82 (0.11) \\
      
      HD 159181 & 119.87 (1.55) & 0.98 (0.08) & -2.584 (0.038) & -0.233 (0.052) & 3.019 (0.036) & 5.00 (0.50) & 2.00 (0.11) \\
      
      HD 164058 & 46.31 (0.03) & 1.53 (0.02) & -1.099 (0.008) & -1.210 (0.113) & 2.815 (0.049) & 4.10 (0.41) & 1.58 (0.14) \\
      
      HD 186791 & 179.93 (12.43) & 1.51 (0.32) & -3.556 (0.160) & -1.099 (0.541) & 3.754 (0.280) & 8.00 (0.80) & 0.98 (0.37) \\
      
      HD 198149 & 14.37 (0.03) & 0.91 (0.02) & 2.622 (0.011) & -0.289 (0.040) & 0.959 (0.020) & 1.40 (0.14) & 3.38 (0.10) \\
      
      HD 204867 & 168.47 (6.02) & 0.82 (0.18) & -3.243 (0.088) & -0.143 (0.069) & 3.246 (0.063) & 5.60 (0.56) & 1.89 (0.14) \\
      
      HD 205435 & 38.59 (0.15) & 0.89 (0.04) & 1.087 (0.019) & -0.267 (0.037) & 1.564 (0.022) & 2.20 (0.22) & 3.01 (0.10) \\
      
      HD 209750 & 203.54 (17.57) & 0.96 (0.39) & -3.603 (0.197) & -0.218 (0.191) & 3.420 (0.155) & 6.30 (0.63) & 1.71 (0.23) \\
      \hline
    \end{tabular}
    \begin{tablenotes}
    \item[a] Distance calculated using the \textit{Hipparcos} parallaxes \citep{ESA1997}
    \item[b] The surface gravity was calculated using Equation \ref{eq:gravity}.
    \end{tablenotes}
  \end{threeparttable}
\end{table*}

\subsection{Test with \textit{Gaia} FGK Benchmark Stars}
\label{subsec:GBS_Test}

The \textit{Gaia} FGK Benchmark Stars (GBS onwards, \citet{Jofre2014}, \citet{Blanco-Cuaresma2014a} \citet{Heiter2015}, \citet{Jofre2015a}, \citet{Hawkins2016}, \citet{Jofre2017}) is a high resolution and high S/N library of a set of calibration star spectra. This library constitutes a set of very well-known stars especially convenient for spectral analysis assessments. It covers a wide range in effective temperature (3500 to 6600 K), surface gravity ($0.50$ to $4.60$ dex) and metallicity ($-2.70$ to $0.30$ dex). In addition, the stellar parameters of the set was mainly obtained from methods independent of spectroscopy.

We used four well-know stars from the GBS, $\delta$ Eri (HD23249), $\alpha$ Tau (HD 29139), \textit{Arcturus} (HD 124897) and the \textit{Sun}, to test the reliability of our method, as well as to see, if the use of TIGRE-HEROS spectra might somehow affect our results for reasons based on any instrumental issues. Therefore, spectra of three different resolutions and S/N (see \citet{Blanco-Cuaresma2014a} and references therein) are tested with our script and line list, after downgrading their resolution to the one of TIGRE-HEROS, $R=20,000$. The spectra in the GBS cover a wavelength range from 4800 \AA{} to 6800 \AA{}. For a meaningful comparison, we cut both GBS and R-channel TIGRE-HEROS spectra down to the same range of 5800 \AA{} to 6800 \AA{}.

The tests allow us to prove the reliability of our \textsc{iSpec} script on spectra other than TIGRE-HEROS, as well as the use of our line list in Table \ref{table:iSpec-lines}. We performed tests for many scenarios, including different resolutions, added spectra or using just one, automatic determination of parameters, varying the initial parameters, different wavelength ranges, etc. However, for the purpose of this publication, we only present the results of the first two tests, since the other test results do not differ much.

Table \ref{table:GBS_test} summarizes the test results for $T_\text{eff}$, $\log\,g$, and $\text{[Fe/H]}$, comparing the outcome of our \textsc{iSpec} script for different resolutions. To test for any changes in the stellar parameter determination, whether we add or not several TIGRE-HEROS spectra, we used three spectra for each star with an average S/N $\geq 100$. 

These tests prove the reliability and robustness of the physical parameters obtained from the spectral synthesis approach taken with our \textsc{iSpec} script, employed on different resolutions. Therefore, we consider the total uncertainty of stellar parameters with this method as: $T_\text{eff} = \pm 100$ K, $\log g = \pm 0.15$ dex and $\text{[Fe/H]} = \pm 0.15$.

\begin{table*}
  \centering
  \caption{Comparison tests with \textit{Gaia} FGK Benchmark Stars. The original resolution of spectra has been downgraded to $R=20,000$.}
  \label{table:GBS_test}
  \begin{threeparttable}
    \begin{tabular}{llllcccccc}
      \hline
      Star & Instrument & Resolution & Add$^a$ & $T_\text{eff}$ & $\Delta T_\text{eff}$$^b$ & $\log g$ & $\Delta \log g$$^b$ & $\text{[Fe/H]}$ & $\Delta \text{[Fe/H]}$$^b$ \\
      & & & & [K] & [K] & [dex] & [dex] & [dex] & dex \\ \hline
      $\delta$ Eri (HD 23249) & TIGRE-HEROS & 20524 & False & 4969 (29) & 15 & 3.60 (0.04) & -0.16 & -0.01 (0.02) & -0.07 \\ 
      
      $\delta$ Eri (HD 23249) & TIGRE-HEROS & 20524 & True & 4974 (25) & 20 & 3.59 (0.03) & -0.17 & 0.00 (0.01) & -0.06 \\ 
      
      $\delta$ Eri (HD 23249) & UVES.POP & 85000 & False & 4972 (7) & 18 & 3.64 (0.01) & -0.12 & 0.05 (0.01) & -0.01 \\ 
      
      $\delta$ Eri (HD 23249) & HARPS & 115000 & False & 4971 (7) & 17 & 3.62 (0.01) & -0.14 & 0.04 (0.00) & -0.02 \\ \hline
      
      $\alpha$ Tau (HD 29139) & TIGRE-HEROS & 20366 & False & 3805 (20) & -122 & 1.13 (0.03) & 0.02 & -0.36 (0.02) & 0.01 \\ 
      
      $\alpha$ Tau (HD 29139) & TIGRE-HEROS & 20366 & True & 3798 (16) & -129 & 1.16 (0.03) & 0.05 & -0.37 (0.01) & 0.00 \\ 
      
      $\alpha$ Tau (HD 29140) & NARVAL & 80000 & False & 3821 (8) & -106 & 1.25 (0.02) & 0.14 & -0.27 (0.01) & 0.10 \\ 
      
      $\alpha$ Tau (HD 29141) & HARPS & 115000 & False & 3838 (17) & -89 & 1.24 (0.03) & 0.13 & -0.30 (0.01) & 0.07 \\ \hline
      
      Arcturus (HD 124897) & TIGRE-HEROS & 20435 & False & 4341 (19) & 55 & 1.60 (0.05) & -0.04 & -0.57 (0.02) & -0.04 \\ 
      
      Arcturus (HD 124897) & TIGRE-HEROS & 20435 & True & 4342 (15) & 56 & 1.61 (0.04) & -0.03 & -0.57 (0.02) & -0.04 \\ 
      
      Arcturus (HD 124897) & UVES.POP & 80000 & False & 4338 (3) & 52 & 1.67 (0.01) & 0.03 & -0.52 (0.00) & 0.01 \\ 
      
      Arcturus (HD 124897) & HARPS & 115000 & False & 4334 (5) & 48 & 1.64 (0.02) & 0.00 & -0.50 (0.01) & 0.03 \\ \hline
      
      Sun & TIGRE-HEROS & 20501 & False & 5750 (41) & -22 & 4.43 (0.06) & -0.01 & -0.05 (0.02) & -0.05 \\ 
      
      Sun & TIGRE-HEROS & 20501 & True & 5743 (34) & -29 & 4.43 (0.04) & -0.01 & -0.05 (0.01) & -0.05 \\ 
      
      Sun & UVES & 78000 & False & 5758 (14) & -14 & 4.45 (0.02) & 0.01 & -0.03 (0.01) & -0.03 \\ 
      
      Sun & HARPS & 115000 & False & 5756 (12) & -16 & 4.44 (0.01) & 0.00 & -0.02 (0.01) & -0.02 \\
      \hline
    \end{tabular}
    \begin{tablenotes}
    \item[a] Adding three TIGRE-HEROS spectra.	
    \item[b] Differences with values in \textit{Gaia} FGK Benchmark Stars library.
    \end{tablenotes}
  \end{threeparttable}
\end{table*}




\section[The Ca II emission line widths of interest and their measurements]{The \ion{Ca}{ii} emission line widths of interest and their measurements}

\subsection[Formation of the Ca II K emission line and where to take its width]{Formation of the \ion{Ca}{ii} K emission line and where to take its width}

The very cores of the \ion{Ca}{ii} H\&K lines are of such huge optical depth, that they provide one of very few windows into the chromospheres of cool stars. While most line cores in the optical spectrum reflect the physical conditions of the upper photosphere, the \ion{Ca}{ii} doublet allows us to trace the temperature structure beyond its minimum at the base of the chromosphere, well into the rise, which is manifested by the \ion{Ca}{ii} H\&K line core reversals into emission. An early, extensive review of the chromospheric physics learned from this doublet was given by \citet{Linsky1970}, and recent reviews are by \cite{Linsky2017} and \cite{Ayres2019}. The focus is usually on the K line, often omitting its twin, in order to avoid any confusion which may arise from the overlap of the \ion{Ca}{ii} H line with the hydrogen H$_{\epsilon}$ line.

Since \ion{Ca}{ii} is more easily ionized than \ion{Mg}{ii} and hydrogen, its formation does not reach into the increasingly warmer, uppermost chromosphere. There, magnetically 
dominated phenomena like prominences make the interpretation of, for example, H$_{\alpha}$ much more difficult. The K-lines reversal, its K2 emission, further shows a self-reversal (K3), which however must not be interpreted in terms of a temperature profile. Rather, this reversal is the product of extreme NLTE photon line scattering and geometrical distribution in the higher chromospheric layers -- therefore much more prominent in luminous giants with their larger chromospheric extent. Hence, for a physical understanding of the lower chromosphere, our interest must be aimed on the outer proportions of the K2 chromospheric emission profile.

Consequently, \citet{Wilson1957} based their empirical relation between the chromospheric \ion{Ca}{ii} K emission line width and visual absolute magnitude $M_\text{V}$ on their measurements of $W_0$ -- chosen simply for the benefit of the best measurement precision, when taken as the positions of the outer edges of the calcium emission features. This is particularly true for measurements on photographic plates, as used in those days. Using the same spectrograms of \citet{Wilson1957}, \citet{Lutz1970} finally shows that the width at half intensity of the K emission line peak of \ion{Ca}{ii} is better correlated with $M_\text{V}$ unlike the original method used by \citet{Wilson1957}. Those peaks are also known as K2. Dealing with noisy spectra, the half-peak points are relatively well defined on the steep flanks of even a noisy emission line. 

By contrast, the foot-points of the broad photospheric absorption line (K1) to either side of the chromospheric emission are not so well defined, especially in a noisy spectrum, because the deepest point in a shallow depression is ambiguous. But these K1 points are of theoretical significance: They are formed right at the base of the chromosphere, in the temperature-minimum. Assuming the source function there is still dominated by the Planck function, the minimal spectral flux observed in these deepest points of the K line on either side of the chromospheric emission reveals the physical conditions in temperature minimum.

However, as pointed out and illustrated by \citet{Shine1975} (see their Fig. 1), the exact distance and depths of these K1 points depends a lot on partial redistribution (PRD) effects on the K2 wings -- becoming considerable, given the large number of photon line scattering events in this strong line prior to each escaping photon. PRD effects further increase with the larger chromospheric column densities in the more luminous giants. In that respect, \citet{Shine1975} already made it clear, that not considering PRD can lead to significant underestimates of the temperature in the temperature-minimum. Likewise, we conclude from their Fig. 1 that PRD and stellar gravity impact on the relation between $W_0$, O.C. Wilson's half-maximum width, and the distance between the K1 points of the chromospheric emission line profile, hereafter referred to as $W_1$. 

In their layout of a density-broadening explanation of the Wilson-Bappu effect -- by the very numerous chromospheric absorption and re-emission events preceding each photon reaching the observer -- \cite{Ayres1975} refer to the emission line width as taken between these two K1 foot points (i.e., $W_1$), simply for its significance by characterizing the base of the chromosphere. In assuming (without saying), as a first-order approximation, homologous emission line profiles across the giant branch, their explanation of the Wilson-Bappu effect, based on the theoretical arguments involving $W_1$, is implied to be true for O.C. Wilson's $W_0$ as well.

However this simplification is not entirely true, as the theoretical work of \citet{Shine1975} already predicted, and as our measurements confirm below. In particular, when looking at lower gravities, $W_1$ starts to grow a bit faster that $W_0$. Because we are seeking a more precise comparison of theory with observation, this detail needs to be considered as well -- which to our knowledge was never done before. The reason for that may simply stem from the difficulty to measure $W_1$. Hence, we here present our method to determine well both these line widths, handling well the presence of noise on the observed emission line profile, and used them here consistently for the whole sample.

\subsection[Measuring the widths of the chromospheric Ca II K emission line]{Measuring the widths of the chromospheric \ion{Ca}{ii} K emission line}
\label{subsec:line_width_measurements} 

The best way to use all information in an observed line profile with noise, is to empirically represent the whole profile by a best-fit analytical function, and then derive key quantities from that function. This approach avoids dependence on local noise effects, a particular advantage for measuring the shallow K1 minima by the foot points of the chromospheric emission, defining $W_1$ -- provided, the representative function covers them well, too.

The physical features of the K emission line of \ion{Ca}{ii}, however, are difficult to match by the types of profiles mostly used elsewhere in the analysis of spectral lines: Gaussian, Lorentzian, or a convolution of both, called \textit{Voigt} profile. \citet{Lutz1970} shows for the simplest case of a \ion{Ca}{ii} K emission line, which does not have central self-absorption, that even there fitting a Lorentzian profile does not match correctly neither the \textit{wings} of the line, nor the region of the \textit{peak}. On the other hand, while a Gaussian fit matches better the line wings, it cannot represent the peak area either, because it is wider than the wing-matching Gaussian function. To fully consider the additional complexity of chromospheric \ion{Ca}{ii} emission line profiles with central self-absorption, as required by this work, none of those three profiles are remotely adequate, and their use would only result in totally incorrect line width measurements.

To measure the width of the K emission line of \ion{Ca}{ii} in the spectra of our stellar sample, we wrote a dedicated code, \textsc{Half-Intensity Emission Width} (\textsc{HIEW} v4.0), developed in C++. Our script \textsc{HIEW} works along a routine, which involves five main stages: (1) reading the configuration file and spectrum; (2) K-line representation by cubic splines to reduce variations caused by electronic noise in the spectrum; (3) determination of the minima and maxima in each of \textit{blue} B and \textit{red} R peaks of the K-line profile; (4) determination of the half peak intensity level, based on the average of the B and R peak intensity; (5) calculation of $W_0$ as the half-intensity emission width, and $W_1$ as the foot point separation of the representative emission line profile spline function.

In the observed spectrum, the real emission line profile is folded with the instrumental profile, and in this way the observed line is broadened by the limited spectral resolution. The instrumental profile width of TIGRE-HEROS is a function of wavelength and the instrumental FWHM in the \ion{Ca}{ii} K line region is $\sim$ 0.19 \AA{} (\cite{Schmitt2014}, \citet{Czesla2014}). However, the way of correction of the measured K-line widths by the instrumental profile width is a delicate matter, and has been discussed in detail already by \citet{Lutz1970}.  

To consider in more detail the effects of the TIGRE-HEROS resolution in our measurements, we simulated an emission K line-profile using, as a first approximation, a sum of analytical functions (gaussian, voigt and both) even when, as \citet{Lutz1970} clearly showed, the complexity of chromospheric \ion{Ca}{ii} emission line profiles does not adequately fit any of these analytic functions. However, our tests showed that the use of two gaussian functions - with positive amplitude to fit the emission (K2) and negative amplitude to fit the autoabsorption (K3) - appears to match sufficiently well a number of representative cases of observed emission K line-profiles of our sample.
		
\begin{figure}
	\includegraphics[width=\columnwidth]{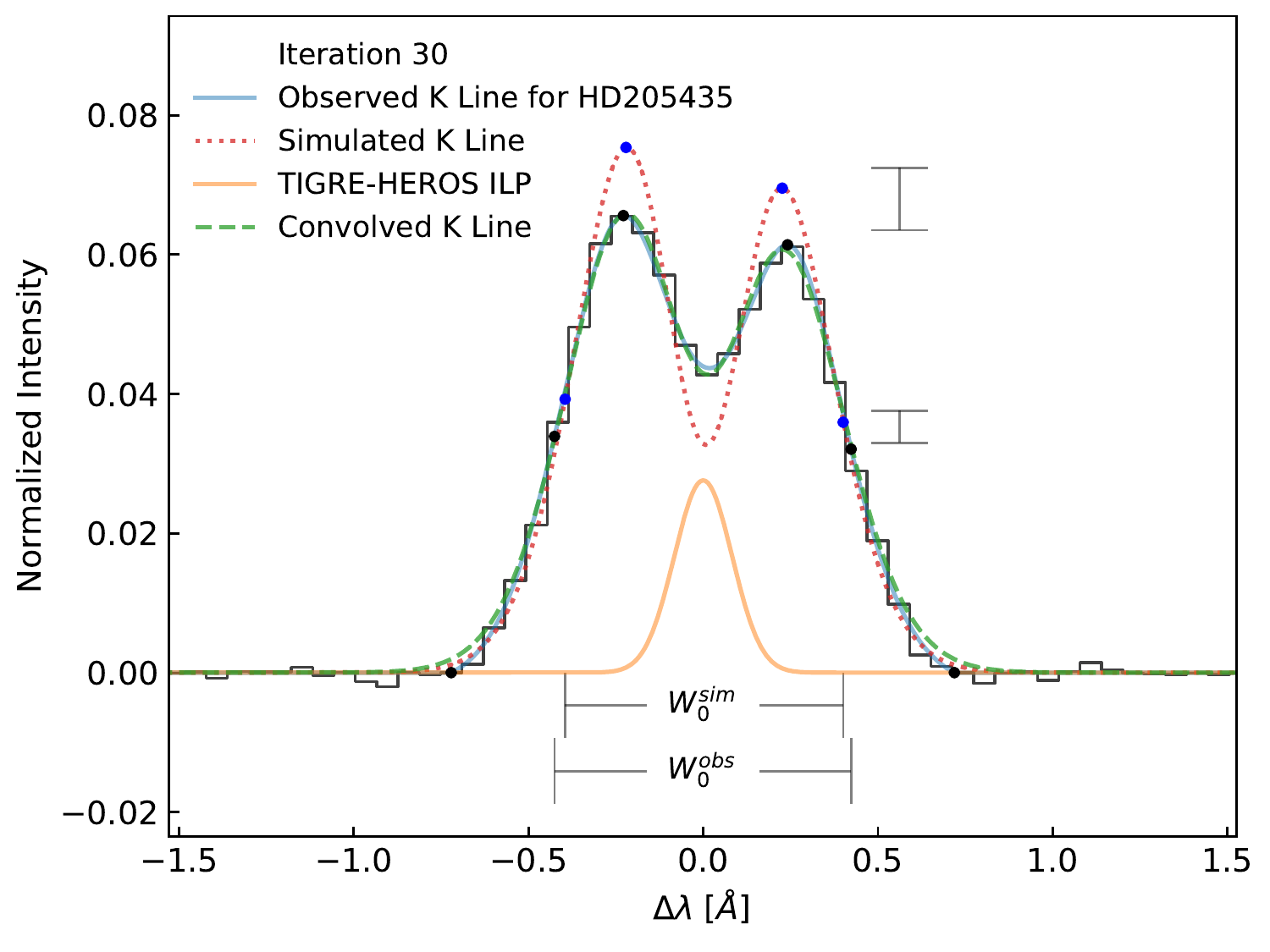}
	\caption{Simulation of the normalized \ion{Ca}{ii} K emission line for HD 205435. We used a sum of two gaussian functions to simulated a K line profile (dotted red line) without the effect of the instrumental line profile of TIGRE-HEROS (solid orange line). The convolution between the simulated K emission line and instrumental line profile (dashed green line) fits well enough to the observed K emission line for HD 205435. Points mark the minimum, maximum, and half-maximum of the simulated (blue dots) and observed (black dots) line profiles. In the right part, we show the differences of the maximum averages and half-maximum averages. In the bottom part, we show the $W_0$ for simulated and observed emission line.}
	\label{fig:HIEW_Convolution}
\end{figure}

In a first step, we normalized the observed emission line by a cubic spline fitting of the \ion{Ca}{ii} absorption. Then, we convoluted a simulated profile with the instrumental line profile (ILP) of TIGRE-HEROS and, by an RMS criterion, we improved recursively the analytical parameters of the functions considered. After 30 iterations the convolution of the simulated profile converges with the observed line profile. Fig. \ref{fig:HIEW_Convolution} shows the effect of the TIGRE-HEROS resolution in a simulated emission line for HD 205435 and the comparison with a observed \ion{Ca}{ii} K emission line. As we could have expected, the maximum, and half maximum points of the simulated and observed line profiles differ. Nevertheless, the difference of $W_0$ line widths for both K line-profiles are within the instrumental error ($\sim$ 0.11 \AA{}, equivalent to the standard deviation of TIGRE-HEROS' instrumental line-profile close to the K-line of \ion{Ca}{ii}).

The two extremes, to subtract the instrumental resolution linearly, or in quadrature, are correct only if either both the instrumental profile and the line profile are lorentzian functions, or in the latter case, gaussians. But the K2 emission line shape is clearly neither, as mentioned above, and in the case of the HEROS spectrograph, the line profile is not gaussian (see \citet{Czesla2014}). If anything, then the economic coverage of the nominal spectrograph resolution by only three pixels and the scattering characteristics in the fiber-feed may enhance the wings of the effective instrumental profile and bring it closer to a lorentzian function.

As already studied by \citet{Lutz1970}, the choice of this correction does matter a little bit, as it systematically affects the narrower lines more. We tested both approaches and find a maximum difference of 0.02 in the gravity exponent. In addition, the difference of widths of simulated, linear and quadrature subtraction are within the instrumental error of TIGRE-HEROS. However, given the analysis above, neither case is \textit{a priory} a choice more justifiable than the other. Interestingly, O. C. Wilson always used the simple linear subtraction, and the detailed revision of his measurements by \citet{Lutz1970} showed that it seemed to be more consistent than a trial with a subtraction in quadrature. Apparently, and by coincidence, another detail of the measuring process comes in to compensate any theoretical offset: washed-out (by limited resolution) narrower lines tend to be measured, effectively, further down than broad lines, when aiming at the 50 per cent level of the observed (and so broadened) profile. Hence, we settled for the simple linear correction method to arrive at our \textit{true} widths of the \ion{Ca}{ii} K emission line. The appendix \ref{sec:CaII_widths} contains the line width measurements, and Table \ref{table:K-line_widths} summarizes these results for the whole stellar sample.

\begin{table*}
  \centering
  \caption{Width measurements of the K emission line of \ion{Ca}{ii}.}
  \label{table:K-line_widths}
  \begin{threeparttable}
    \begin{tabular}{lcccccc}
      \hline
      Star & $W_{0}$$^a$ & $W_{0}$$^b$ & $\log W_0$$^c$ & $W_{1}$$^a$ & $W_{1}$$^b$ & $\log W_1$$^c$\\
      & [\AA] & [km/s] & [dex] & [\AA] & [km/s] & [dex] \\ \hline
      HD 8512 & 0.72 & 55 & 1.74 (0.07) & 1.14 & 87.18 & 1.94 (0.04) \\
      
      HD 10476 & 0.31 & 24 & 1.37 (0.16) & 0.72 & 55.12 & 1.74 (0.07) \\
      
      HD 18925 & 0.80 & 61 & 1.78 (0.06) & 1.16 & 88.55 & 1.95 (0.04) \\
      
      HD 20630 & 0.36 & 27 & 1.43 (0.14) & 0.95 & 72.07 & 1.86 (0.05) \\
      
      HD 23249 & 0.40 & 30 & 1.48 (0.12) & 0.62 & 47.34 & 1.68 (0.08) \\
      
      HD 26630 & 1.81 & 138 & 2.14 (0.03) & 2.50 & 190.67 & 2.28 (0.02) \\
      
      HD 27371 & 0.79 & 60 & 1.78 (0.06) & 1.32 & 100.46 & 2.00 (0.04) \\
      
      HD 27697 & 0.71 & 54 & 1.73 (0.07) & 1.16 & 88.55 & 1.95 (0.04) \\
      
      HD 28305 & 0.73 & 56 & 1.74 (0.07) & 1.17 & 89.47 & 1.95 (0.04) \\
      
      HD 28307 & 0.73 & 56 & 1.74 (0.07) & 1.15 & 87.64 & 1.94 (0.04) \\
      
      HD 29139 & 0.93 & 71 & 1.85 (0.05) & 1.67 & 127.48 & 2.11 (0.03) \\
      
      HD 31398 & 1.17 & 89 & 1.95 (0.04) & 2.11 & 160.45 & 2.21 (0.02) \\
      
      HD 31910 & 1.94 & 148 & 2.17 (0.03) & 2.60 & 198.46 & 2.30 (0.02) \\
      
      HD 32068 & 1.29 & 98 & 1.99 (0.04) & 2.32 & 176.48 & 2.25 (0.02) \\
      
      HD 48329 & 1.73 & 132 & 2.12 (0.03) & 2.59 & 197.70 & 2.30 (0.02) \\
      
      HD 71369 & 0.76 & 58 & 1.77 (0.06) & 1.13 & 86.30 & 1.94 (0.04) \\
      
      HD 81797 & 1.01 & 77 & 1.88 (0.05) & 1.85 & 141.22 & 2.15 (0.03) \\
      
      HD 82210 & 0.53 & 40 & 1.61 (0.09) & 1.34 & 102.34 & 2.01 (0.04) \\
      
      HD 96833 & 0.76 & 58 & 1.76 (0.07) & 1.37 & 104.12 & 2.02 (0.04) \\
      
      HD 104979 & 0.68 & 52 & 1.72 (0.07) & 0.93 & 70.83 & 1.85 (0.05) \\
      
      HD 109379 & 0.83 & 63 & 1.80 (0.06) & 1.13 & 86.26 & 1.94 (0.04) \\
      
      HD 114710 & 0.48 & 37 & 1.57 (0.10) & 0.75 & 56.95 & 1.76 (0.07) \\
      
      HD 115659 & 0.72 & 55 & 1.74 (0.07) & 1.13 & 86.26 & 1.94 (0.04) \\
      
      HD 124897 & 0.84 & 64 & 1.80 (0.06) & 1.35 & 102.75 & 2.01 (0.04) \\
      
      HD 148387 & 0.70 & 53 & 1.73 (0.07) & 0.97 & 73.90 & 1.87 (0.05) \\
      
      HD 159181 & 1.93 & 147 & 2.17 (0.03) & 2.82 & 215.00 & 2.33 (0.02) \\
      
      HD 164058 & 0.96 & 73 & 1.86 (0.05) & 1.86 & 142.13 & 2.15 (0.03) \\
      
      HD 186791 & 1.20 & 91 & 1.96 (0.04) & 2.28 & 173.73 & 2.24 (0.02) \\
      
      HD 198149 & 0.50 & 38 & 1.58 (0.10) & 0.71 & 54.21 & 1.73 (0.07) \\
      
      HD 204867 & 1.89 & 144 & 2.16 (0.03) & 2.46 & 187.47 & 2.27 (0.02) \\
      
      HD 205435 & 0.66 & 50 & 1.70 (0.08) & 1.27 & 96.79 & 1.99 (0.04) \\
      
      HD 209750 & 1.86 & 142 & 2.15 (0.03) & 2.62 & 199.38 & 2.30 (0.02) \\
      \hline
    \end{tabular}
    \begin{tablenotes}
    \item[a] The error of the line width measurements is equivalent to the standard deviation of TIGRE-HEROS' instrumental line-profile close to K-line of \ion{Ca}{ii} $\sim$ 0.11 \AA.
    \item[b] The error in km/s is equivalent to 8 km/s.
    \item[c] Logarithmic values correspond to widths in km/s.
    \end{tablenotes}
  \end{threeparttable}
\end{table*}

\subsection[Are W0 and W1 analogous measurements of the Ca II K emission line width?]{Are $W_0$ and $W_1$ analogous measurements of the \ion{Ca}{ii} K emission line width?}
\label{subsec:W1_and_W0_comparison}

As pointed out before, \citet{Ayres1975} and also \citet{Cram1978} compared and proposed empirical width-luminosity correlations, referring to the \ion{Ca}{ii} K$_1$ minimum features, which means the $W_1$ widths, not O.C. Wilson's $W_0$. However, in his work \citet{Ayres1979} argues that $W_0$ should be, much like $W_1$, proportional to $g^{1/4}$, too. That would be the case, e.g., if the emission line profiles were homologous. However, we want to revise this element of the comparison of theory with observation as well. Consequently, we compared the $W_0$ and $W_1$ widths to see, whether they are indeed analogous measurements of the chromospheric \ion{Ca}{ii} K line emission width, using our above described measurement method (see section \ref{subsec:line_width_measurements})

\begin{figure}
  \includegraphics[width=\columnwidth]{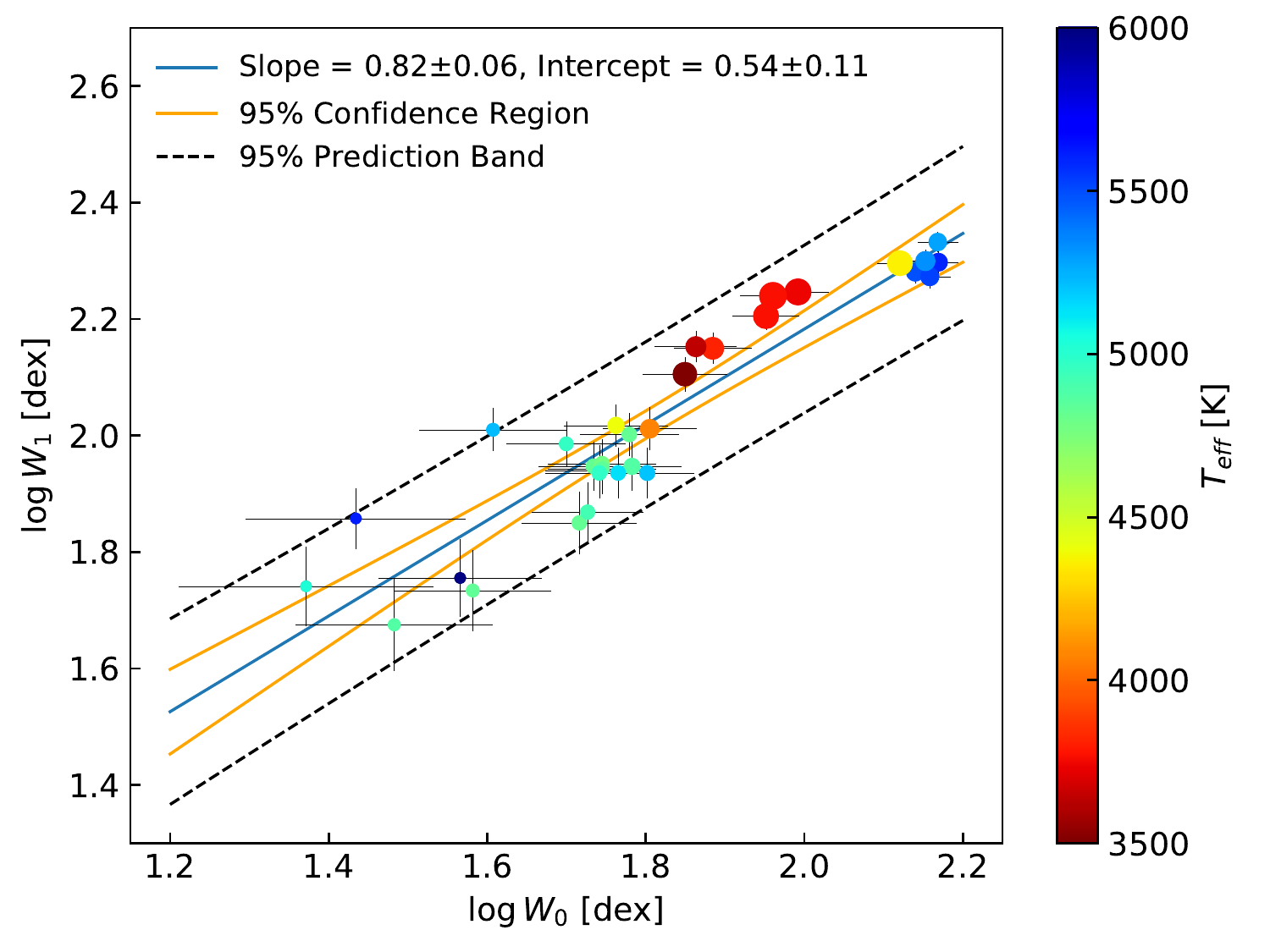}
  \caption{Comparison of line width measurements $\log W_1$ vs $\log W_0$ of the \ion{Ca}{ii} K emission line for the stellar sample. We used the logarithmic values of widths in km/s. The blue line corresponds to a linear fit. The color code represents the effective temperature, and the point-size represents the surface gravity with larger points for lower gravities.}
  \label{fig:log_W1_vs_log_W0}
\end{figure}

Using the width measurements of the \ion{Ca}{ii} K emission line listed in Table \ref{table:K-line_widths}, we compare and correlate $W_1$ with $W_0$ for our stellar sample. Fig. \ref{fig:log_W1_vs_log_W0} shows this comparison, resulting in a linear least squares fit with a high Pearson's correlation coefficient of 0.934. The best relation for $W_1$ versus $W_0$ is

\begin{equation}
  \log W_1 = (0.82 \pm 0.06) \log W_0 + (0.54\pm0.11).
  \label{eq:log_W1_vs_log_W0}
\end{equation}

This relation is not far from what one may expect of homologous emission line profiles (implying an exponent of 1.0), however it clearly proves that $W_1$ and $W_0$ scale differently over our sample. The effect is significant and must be taken into consideration by the comparison of theory with observation, see section below, in particular for the relation of line width with gravity.

\section{Gravity and temperature dependence: observation versus theory}

The relation between the width of the \ion{Ca}{ii} K emission line and the stellar parameters, i.e. effective temperature and surface gravity, has been studied by a number of authors during the last decades (\citet{Reimers1973}, \citet{Neckel1974}, \citet{Ayres1975}, \citet{Ayres1979}, \citet{Lutz1982}, \citet{Park2013}). The work of \citet{Ayres1975} demonstrated, how the Wilson-Bappu effect can be derived from of the relation of the chromospheric \ion{Ca}{ii} column density $N$ with gravity $g$, assuming hydrostatic equilibrium at the bottom of the chromosphere. $N$ in turn reflects on how the line width grows with lower gravity, by simple density-broadening -- the migration of the photons into the line wings after multiple absorption and re-emission processes. Assuming also that the mean continuum optical depth in the stellar temperature minimum is relatively independent of surface gravity, \citet{Ayres1975} arrived at the simple relation $W_1 \propto {A_\text{met}}^{1/4} g^{-1/4}$, where $A_\text{met}$ is the abundance of easily ionized metals relative to hydrogen. The simplification of the opacity problem, required to derive the above relation, was later supported by more detailed arguments, given in \citet{Ayres1979} and \citet{Ayres2019}. A key point is, irrespective of, e.g., the relative proportions of H$^-$ and line opacities in each chromosphere, that all relevant contributions to the opacity are controlled by the electron density in the same way.

We deliberately avoided studying the possible metallicity dependence, by the choice of a sample of mostly solar abundance, in order to have fewer ambiguity on the gravity dependence and to test the theoretical approach as such. A dependence of $W_1$ on $T_\text{eff}$ was still ignored by \citet{Ayres1975} because their initial stellar sample did not spread much over that parameter, only in gravity, but was then taken into consideration soon after by \citet{Ayres1979}. In consequence, the line broadening appears to depend on both, gravity and effective temperature. The latter effect is based on the ionization ratio of \ion{Ca}{i}:\ion{Ca}{ii}, consequently reducing the \ion{Ca}{ii} column density in the coolest chromospheres (comparing stars of same surface gravity). Still this is of secondary nature because the range of effective temperatures of cool stars with chromospheres is quite limited, while gravity covers more than four orders of magnitude.

There is now a problem with any empirical assessment of the line width relation, simultaneously with both these parameters: cross-talk from a mismatch of the effective temperature affecting the gravity term. That interrelation is caused by the orientation of the giant branches in the HR diagram: the coolest giants are also mostly those with the highest luminosity and, consequently, lowest gravity. Hence, to empirically evaluate the theoretical reasoning of \citet{Ayres1975} by means of the gravity term of the emission line width, one must get the temperature term right as well.

To test the interpretation of the Wilson-Bappu effect by \citet{Ayres1975} and \citet{Ayres1979}, we derived the width correlation of the \ion{Ca}{ii} K emission line with surface gravity and effective temperature in our sample, for both measurements, $W_0$ and $W_1$, using simple power-law relations in their logarithmic form:

\begin{equation}
  \log W = \alpha \log\,g + \beta \log T_\text{eff} + C.
  \label{eq:log_W_vs_logg_and_log_teff}
\end{equation}

\noindent where $\alpha$, $\beta$ and $C$ are fitting constants. Table \ref{table:gravity_and_temperature_dependence_of_WBE} shows the results for these values when we apply equation \ref{eq:log_W_vs_logg_and_log_teff} to $W_0$ and $W_1$.

\begin{table}
  \centering
  \caption{Results for the gravity and temperature dependence of the WBE considering different measurements of the K emission line of \ion{Ca}{ii}.}
  \label{table:gravity_and_temperature_dependence_of_WBE}
  \begin{threeparttable}
    \begin{tabular}{ccccc}
      \hline
      Width & $\alpha$ & $\beta$ & $C$ & $\chi^2$ \\ \hline
      $W_0$ & -0.279 (0.011) & 2.93 (0.22) & -8.3 (0.8) & 0.0360 \\
      $W_1$ & -0.220 (0.020) & 1.8 (0.4) & -4.1 (1.4) & 0.1040 \\
      $W_1^\prime$ & -0.229 (0.009) & 2.41 (0.18) & -6.3 (0.7) & 0.0216 \\
      \hline
    \end{tabular}
  \end{threeparttable}
\end{table}

Based on the best $\chi^2$, we observe that $W_0$ has a better correlation with gravity and temperature than $W_1$ for our stellar sample. However, this can be explained by the better definition of the former on the flanks of line profile. And as we showed in subsection \ref{subsec:W1_and_W0_comparison}, there is a clear relation between both widths, where $W_0$ rises a bit faster than $W_1$ (see equation \ref{eq:log_W1_vs_log_W0}). Therefore, the relation of $\log\,W_1$ has a smaller power with $\log\,g$ than $\log\,W_0$. 

\begin{figure}
  \includegraphics[width=\columnwidth]{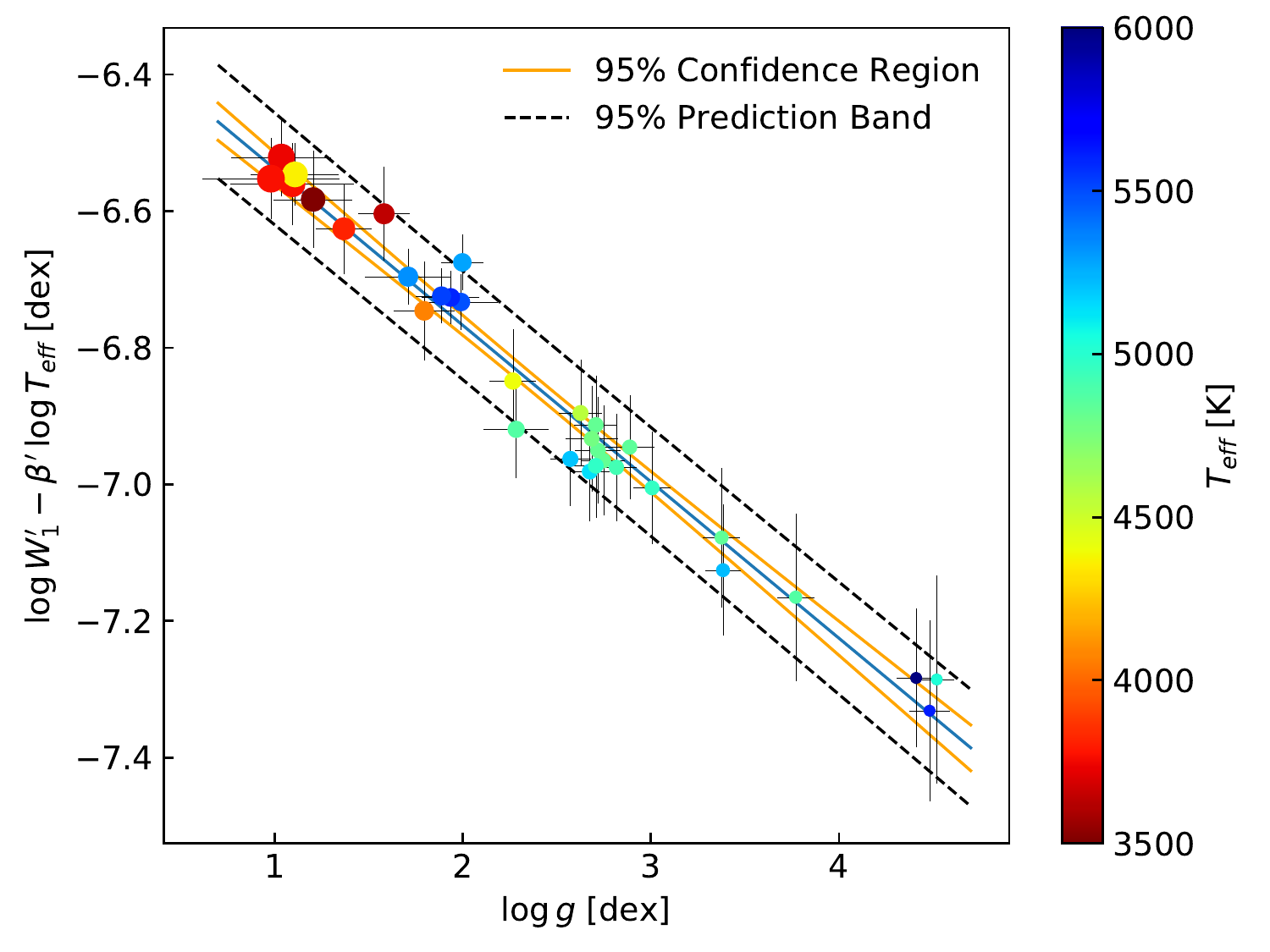}
  \caption{Dependence of $W_1^\prime$ on $\log\,g$ by subtracting the temperature contribution. The blue line corresponds to a linear fit. The color code represents the effective temperature, and the point-size represents the surface gravity, using larger points for lower gravities.}
  \label{fig:log_W1p_vs_log_g}
\end{figure}

\begin{figure}
  \includegraphics[width=\columnwidth]{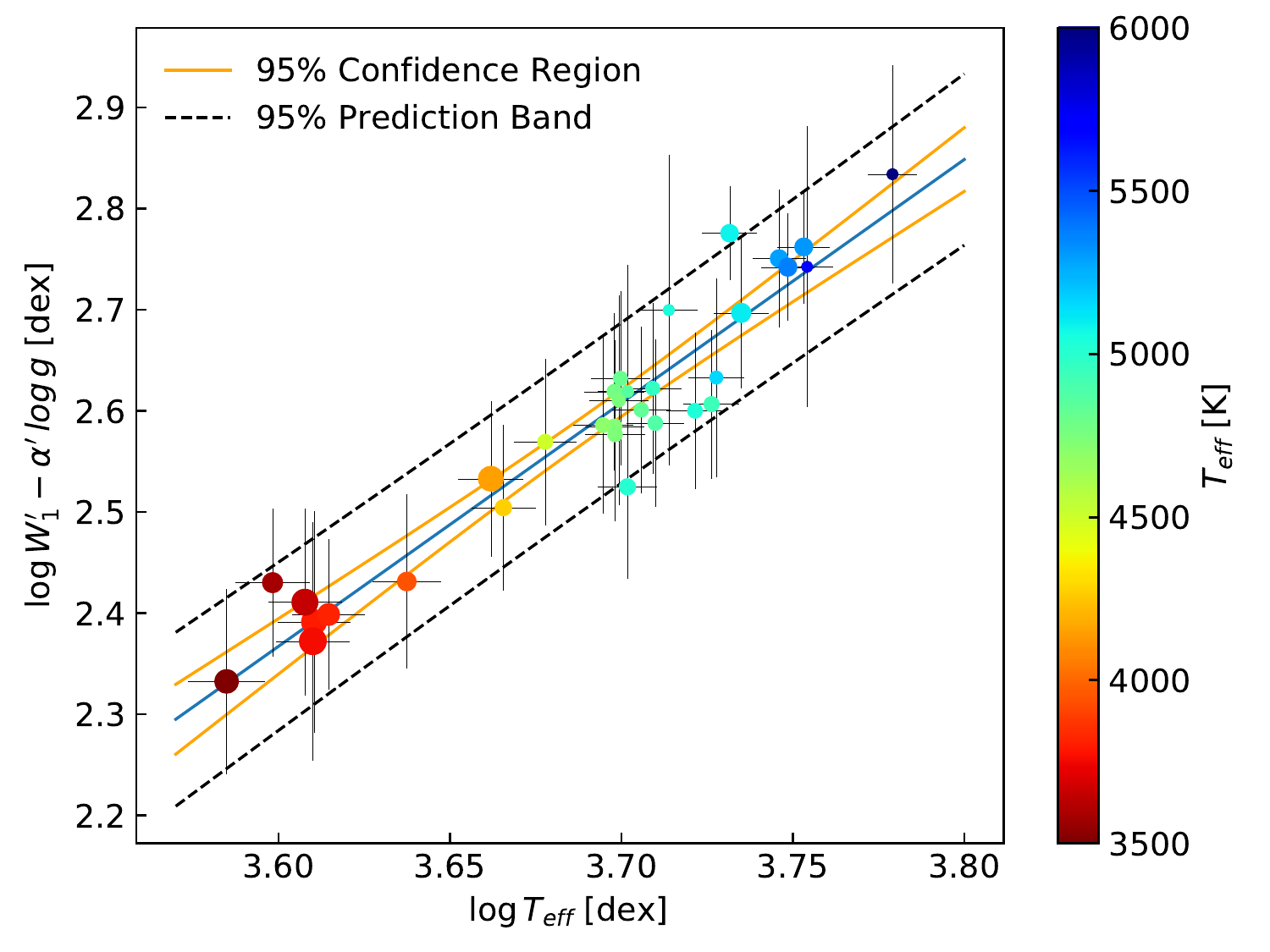}
  \caption{Dependence of $W_1^\prime$ on $\log\,T_\text{eff}$ by subtracting the gravity contribution. The blue line corresponds to a linear fit. The color code represents the effective temperature, and the point-size represents the surface gravity with larger points for lower gravities.}
  \label{fig:log_W1p_vs_log_Teff}
\end{figure}

We may also convert $W_0$ into $W_1$ values, using equation \ref{eq:log_W1_vs_log_W0}, in order to circumnavigate the larger uncertainties of individual $W_1$ values. We tested the dependence of these converted values $W_1^\prime$ with gravity and temperature in the same way as before and found that their correlation got two times better than $W_0$, based on the respective minimal $\chi^2$ sums. In addition, the gravity exponent calculated with $W_1^\prime$ is the closest to the theoretical prediction of \cite{Ayres1975}. This is pictured in Fig. \ref{fig:log_W1p_vs_log_g} and Fig. \ref{fig:log_W1p_vs_log_Teff}, showing the direct dependence of $\log\,W_1^\prime$ with $\log\,g$, as well as $\log\,T_\text{eff}$, subtracting the respective other contribution.

\section{Conclusions}

We developed and demonstrated a consistent analysis, applied uniformly on a representative stellar sample, for which we obtained good S/N TIGRE spectra of the moderately high spectral resolving power of $R \sim 20,000$. The new \text{Gaia} EDR3 parallaxes allowed us to calculate precise, parallax-based gravities, in combination with stellar mass estimates based on HRD-position matching evolution tracks. That approach facilitated the assessment of effective temperatures by means of spectral synthesizing, as we were able to exclude false solutions with erroneous gravities.

Furthermore, with the help of our here presented measurement technique for the chromospheric \ion{Ca}{ii} K emission line width, we found that the width used by O.C. Wilson, $W_0$ (the width at half-peak intensity), and the width of theoretical significance, $W_1$ (the K$_1$ minima separation, originating in the temperature minimum at the very bottom of the chromosphere), do not scale 1:1 across the HRD, but are related to each other by $\log W_1 = 0.82 \log W_0 + 0.53$. Hence, chromospheric emission lines are not exactly homologous over a wider range of gravity. Therefore, the relation between $W_0$ and $W_1$ is required to extend the empirical Wilson-Bappu effect, based on $W_0$, to a relation of the line width with physical parameters such as gravity and effective temperature, as predicted by theory for $W_1$. 

This distinction and the above described consistent approach to the physical parameter assessment for our sample stars, enables us to test the theoretical explanation of \citet{Ayres1975} and \citet{Ayres1979}, in particular by comparing the empirical findings to their predicted gravity dependence. In fact, ignoring the difference between the two emission line widths leads to a much worse agreement of theory with observation.

In conclusion, we confirm the dependence on gravity proposed by \cite{Ayres1975} and \citet{Ayres1979}, suggesting that the Wilson-Bappu effect is indeed a line saturation and photon redistribution effect, driven by the growing column densities in giants with lower gravities. That same interpretation provides a straight-forward explanation for the temperature term as well, namely the effect of the \ion{Ca}{ii}:\ion{Ca}{i} ionization balance. Even though it is difficult to quantify, because (i) the relevant temperature of the line-forming layers of each chromosphere cannot so easily be related to the effective temperature of the respective stellar photosphere underneath, and (ii) because of NLTE effects in any chromospheric ionization balance, the intuitive understanding is that cooler stars have a larger fraction of neutral Ca in their chromospheres, resulting in a reduced \ion{Ca}{ii} column density. This explains the positive sign of the temperature exponent, while the substantial effect over a relatively small temperature range calls for a large value -- as observed.

This nice agreement of our empirical study with the theoretical predictions of \cite{Ayres1975} also suggests the validity, at least to a good degree, of their assumption of hydrostatic equilibrium at the base of the chromosphere -- at least for stars, that are not overly active. In those cases, the chromospheric structure is not dominated by magnetic energy density and its dynamic fine structure. This is good news: chromospheric physics may not be entirely hopeless -- after all, some approximations do appear to be reasonable. At least, this is our interpretation of the meaning of the Wilson-Bappu effect and the agreement obtained with simple chromospheric physics.

\section*{Acknowledgements}

This study used the services of the Strasbourg astronomical data centre, and data from the European Space Agency (ESA) mission \textit{Gaia} (\url{https://www.cosmos.esa.int/gaia}), processed by the \textit{Gaia} Data Processing and Analysis Consortium (DPAC, \url{https://www.cosmos.esa.int/web/gaia/dpac/consortium}). Furthermore, this study made use of the \textit{Gaia} FGK Benchmark Stars Library \url{https://www.blancocuaresma.com/s/benchmarkstars}.

We wish to thank Thomas Ayres for his very helpful comments on the subject of this publication, prior to submission. Furthermore, we are grateful to Marco Mittag from the University of Hamburg for his permission to use the TIGRE spectra of HD 198149 and HD 20630 of his observing program. The authors also wish to acknowledge the valuable work of Sergi Blanco-Cuaresma, CfA, Harvard-Smithsonian, on \textsc{iSpec}, a very useful tool for spectral analysis and parameter synthesis. We are grateful to the referee, Jeffrey L.
Linsky, for his valuable recommendations of additional literature and more. We also appreciate the suggestions of the editor Dr. Helen Klus that improved the quality of our work. 

This work benefited from financial support of the bilateral project CONACyT-DFG No. 278156 for the TIGRE collaboration with the University of Hamburg, and from general support of our home institutions. In particular, we thank the University of Guanajuato for the grants for the projects 036/2021 and 105/2021 of the {\it Convocatoria Institucional de Investigaci\'on Cient\'ifica 2021 and 2022}. 

\section*{Data Availability}

The data underlying this article are available in the article.



\bibliographystyle{mnras}
\bibliography{bibliography}

\begin{thebibliography}{}
\makeatletter
\relax
\def\mn@urlcharsother{\let\do\@makeother \do\$\do\&\do\#\do\^\do\_\do\%\do\~}
\def\mn@doi{\begingroup\mn@urlcharsother \@ifnextchar [ {\mn@doi@}
  {\mn@doi@[]}}
\def\mn@doi@[#1]#2{\def\@tempa{#1}\ifx\@tempa\@empty \href
  {http://dx.doi.org/#2} {doi:#2}\else \href {http://dx.doi.org/#2} {#1}\fi
  \endgroup}
\def\mn@eprint#1#2{\mn@eprint@#1:#2::\@nil}
\def\mn@eprint@arXiv#1{\href {http://arxiv.org/abs/#1} {{\tt arXiv:#1}}}
\def\mn@eprint@dblp#1{\href {http://dblp.uni-trier.de/rec/bibtex/#1.xml}
  {dblp:#1}}
\def\mn@eprint@#1:#2:#3:#4\@nil{\def\@tempa {#1}\def\@tempb {#2}\def\@tempc
  {#3}\ifx \@tempc \@empty \let \@tempc \@tempb \let \@tempb \@tempa \fi \ifx
  \@tempb \@empty \def\@tempb {arXiv}\fi \@ifundefined
  {mn@eprint@\@tempb}{\@tempb:\@tempc}{\expandafter \expandafter \csname
  mn@eprint@\@tempb\endcsname \expandafter{\@tempc}}}

\bibitem[\protect\citeauthoryear{Alvarez \& Plez}{Alvarez \&
  Plez}{1997}]{Alvarez1998}
Alvarez R.,  Plez B.,  1997, A\&A, 330, 1109

\bibitem[\protect\citeauthoryear{Auri{\`{e}}re et~al.,}{Auri{\`{e}}re
  et~al.}{2015}]{Auriere2015}
Auri{\`{e}}re M.,  et~al., 2015, \mn@doi [A\&A] {10.1051/0004-6361/201424579},
  574, A90

\bibitem[\protect\citeauthoryear{Ayres}{Ayres}{1979}]{Ayres1979}
Ayres T.~R.,  1979, \mn@doi [ApJ] {10.1086/156873}, 228, 509

\bibitem[\protect\citeauthoryear{Ayres}{Ayres}{2019}]{Ayres2019}
Ayres T.~R.,  2019, in , The Sun as a Guide to Stellar Physics.
Elsevier, pp 27--57, \mn@doi{10.1016/B978-0-12-814334-6.00002-9}

\bibitem[\protect\citeauthoryear{Ayres, Shine  \& Linsky}{Ayres
  et~al.}{1975}]{Ayres1975}
Ayres T.~R.,  Shine R.~A.,   Linsky J.~L.,  1975, \mn@doi [ApJ]
  {10.1086/181724}, 195, L121

\bibitem[\protect\citeauthoryear{Blanco-Cuaresma}{Blanco-Cuaresma}{2019}]{Blanco-Cuaresma2019}
Blanco-Cuaresma S.,  2019, \mn@doi [MNRAS] {10.1093/mnras/stz549}, 486, 2075

\bibitem[\protect\citeauthoryear{Blanco-Cuaresma, Soubiran, Jofr{\'{e}}  \&
  Heiter}{Blanco-Cuaresma et~al.}{2014a}]{Blanco-Cuaresma2014a}
Blanco-Cuaresma S.,  Soubiran C.,  Jofr{\'{e}} P.,   Heiter U.,  2014a, \mn@doi
  [A\&A] {10.1051/0004-6361/201323153}, 566, 1

\bibitem[\protect\citeauthoryear{Blanco-Cuaresma, Soubiran, Heiter  \&
  Jofr{\'{e}}}{Blanco-Cuaresma et~al.}{2014b}]{Blanco-Cuaresma2014}
Blanco-Cuaresma S.,  Soubiran C.,  Heiter U.,   Jofr{\'{e}} P.,  2014b, \mn@doi
  [A\&A] {10.1051/0004-6361/201423945}, 569, A111

\bibitem[\protect\citeauthoryear{Catanzaro}{Catanzaro}{1997}]{Catanzaro1997}
Catanzaro G.,  1997, \mn@doi [Ap\&SS] {10.1023/A:1001197016750}, 257, 161

\bibitem[\protect\citeauthoryear{Cram \& Ulmschneider}{Cram \&
  Ulmschneider}{1978}]{Cram1978}
Cram L.~E.,  Ulmschneider P.,  1978, A\&A, 62, 239

\bibitem[\protect\citeauthoryear{Czesla}{Czesla}{2014}]{Czesla2014}
Czesla S.,  2014, Technical Report~1, {The instrumental line-profile of
  TIGRE/HEROS}.
Hamburger Sternwarte, Hamburg, Germany

\bibitem[\protect\citeauthoryear{ESA}{ESA}{1997}]{ESA1997}
ESA 1997, {The Hipparcos and Tycho Catalogues}.
ESA SP-1200

\bibitem[\protect\citeauthoryear{Eberhard \& Schwarzschild}{Eberhard \&
  Schwarzschild}{1913}]{Eberhard1913}
Eberhard G.,  Schwarzschild K.,  1913, \mn@doi [ApJ] {10.1086/142037}, 38, 292

\bibitem[\protect\citeauthoryear{Flower}{Flower}{1996}]{Flower1996}
Flower P.~J.,  1996, \mn@doi [ApJ] {10.1086/177785}, 469, 355

\bibitem[\protect\citeauthoryear{{Gaia Collaboration} et~al.,}{{Gaia
  Collaboration} et~al.}{2016}]{GaiaCollaboration2016}
{Gaia Collaboration} et~al., 2016, \mn@doi [A\&A]
  {10.1051/0004-6361/201629272}, 595, 1

\bibitem[\protect\citeauthoryear{{Gaia Collaboration} et~al.,}{{Gaia
  Collaboration} et~al.}{2021}]{GaiaCollaboration2020}
{Gaia Collaboration} et~al., 2021, \mn@doi [A\&A]
  {10.1051/0004-6361/202039657}, 649, A1

\bibitem[\protect\citeauthoryear{Grevesse, Asplund  \& Sauval}{Grevesse
  et~al.}{2007}]{Grevesse2007}
Grevesse N.,  Asplund M.,   Sauval A.~J.,  2007, \mn@doi [Space Sci. Rev.]
  {10.1007/s11214-007-9173-7}, 130, 105

\bibitem[\protect\citeauthoryear{Gustafsson, Edvardsson, Eriksson,
  J{\o}rgensen, Nordlund  \& Plez}{Gustafsson et~al.}{2008}]{Gustafsson2008}
Gustafsson B.,  Edvardsson B.,  Eriksson K.,  J{\o}rgensen U.~G.,  Nordlund
  {\AA}.,   Plez B.,  2008, \mn@doi [A\&A] {10.1051/0004-6361:200809724}, 486,
  951

\bibitem[\protect\citeauthoryear{Hawkins et~al.,}{Hawkins
  et~al.}{2016}]{Hawkins2016}
Hawkins K.,  et~al., 2016, \mn@doi [A\&A] {10.1051/0004-6361/201628268}, 592, 1

\bibitem[\protect\citeauthoryear{Heiter, Jofr{\'{e}}, Gustafsson, Korn,
  Soubiran  \& Th{\'{e}}venin}{Heiter et~al.}{2015}]{Heiter2015}
Heiter U.,  Jofr{\'{e}} P.,  Gustafsson B.,  Korn A.~J.,  Soubiran C.,
  Th{\'{e}}venin F.,  2015, \mn@doi [A\&A] {10.1051/0004-6361/201526319}, 582,
  1

\bibitem[\protect\citeauthoryear{Husser, {Wende-von Berg}, Dreizler, Homeier,
  Reiners, Barman  \& Hauschildt}{Husser et~al.}{2013}]{Husser2013}
Husser T.~O.,  {Wende-von Berg} S.,  Dreizler S.,  Homeier D.,  Reiners A.,
  Barman T.,   Hauschildt P.,  2013, \mn@doi [A\&A]
  {10.1051/0004-6361/201219058}, 553, 1

\bibitem[\protect\citeauthoryear{Jofr{\'{e}} et~al.,}{Jofr{\'{e}}
  et~al.}{2014}]{Jofre2014}
Jofr{\'{e}} P.,  et~al., 2014, \mn@doi [A\&A] {10.1051/0004-6361/201322440},
  564, 1

\bibitem[\protect\citeauthoryear{Jofr{\'{e}} et~al.,}{Jofr{\'{e}}
  et~al.}{2015}]{Jofre2015a}
Jofr{\'{e}} P.,  et~al., 2015, \mn@doi [A\&A] {10.1051/0004-6361/201526604},
  582, 1

\bibitem[\protect\citeauthoryear{Jofr{\'{e}} et~al.,}{Jofr{\'{e}}
  et~al.}{2017}]{Jofre2017}
Jofr{\'{e}} P.,  et~al., 2017, \mn@doi [A\&A] {10.1051/0004-6361/201629833},
  601, 1

\bibitem[\protect\citeauthoryear{Lindegren et~al.,}{Lindegren
  et~al.}{2021}]{Lindegren2020}
Lindegren L.,  et~al., 2021, \mn@doi [A\&A] {10.1051/0004-6361/202039709}, 649,
  A2

\bibitem[\protect\citeauthoryear{Linsky}{Linsky}{2017}]{Linsky2017}
Linsky J.~L.,  2017, \mn@doi [ARA\&A] {10.1146/annurev-astro-091916-055327},
  55, 159

\bibitem[\protect\citeauthoryear{Linsky \& Avrett}{Linsky \&
  Avrett}{1970}]{Linsky1970}
Linsky J.~L.,  Avrett E.~H.,  1970, \mn@doi [PASP] {10.1086/128904}, 82, 169

\bibitem[\protect\citeauthoryear{Lutz}{Lutz}{1970}]{Lutz1970}
Lutz T.~E.,  1970, AJ, 75, 1007

\bibitem[\protect\citeauthoryear{Lutz \& Pagel}{Lutz \& Pagel}{1982}]{Lutz1982}
Lutz T.~E.,  Pagel B. E.~J.,  1982, MNRAS, 199, 1101

\bibitem[\protect\citeauthoryear{Montes, Fernandez-Figueroa, de Castro  \&
  Cornide}{Montes et~al.}{1994}]{Montes1994}
Montes D.,  Fernandez-Figueroa M.~J.,  de Castro E.,   Cornide M.,  1994, A\&A,
  285, 609

\bibitem[\protect\citeauthoryear{Neckel}{Neckel}{1974}]{Neckel1974}
Neckel H.,  1974, A\&A, 35, 99

\bibitem[\protect\citeauthoryear{Park, Kang, Lee  \& Lee}{Park
  et~al.}{2013}]{Park2013}
Park S.,  Kang W.,  Lee J.~E.,   Lee S.~G.,  2013, \mn@doi [ApJ]
  {10.1088/0004-6256/146/4/73}, 146, 73

\bibitem[\protect\citeauthoryear{Piskunov, Kupka, Ryabchikova, Weiss  \&
  Jeffery}{Piskunov et~al.}{1995}]{Piskunov1995}
Piskunov N.~E.,  Kupka F.,  Ryabchikova T.,  Weiss W.~W.,   Jeffery C.~S.,
  1995, A\&AS, 112, 525

\bibitem[\protect\citeauthoryear{Plez}{Plez}{2012}]{Plez2012}
Plez B.,  2012, {Turbospectrum: Code for spectral synthesis}

\bibitem[\protect\citeauthoryear{Pols, Tout, Schr{\"{o}}der, Eggleton  \&
  Manners}{Pols et~al.}{1997}]{Pols1997}
Pols O.~R.,  Tout C.~A.,  Schr{\"{o}}der K.~P.,  Eggleton P.~P.,   Manners J.,
  1997, MNRAS, 289, 869

\bibitem[\protect\citeauthoryear{Pols, Schr{\"{o}}der, Hurley, Tout  \&
  Eggleton}{Pols et~al.}{1998}]{Pols1998}
Pols O.~R.,  Schr{\"{o}}der K.~P.,  Hurley J.~R.,  Tout C.~A.,   Eggleton
  P.~P.,  1998, MNRAS, 298, 525

\bibitem[\protect\citeauthoryear{Reimers}{Reimers}{1973}]{Reimers1973}
Reimers D.,  1973, A\&A, 24, 79

\bibitem[\protect\citeauthoryear{Ryabchikova et~al.,}{Ryabchikova
  et~al.}{2016}]{Ryabchikova2016}
Ryabchikova T.,  et~al., 2016, \mn@doi [MNRAS] {10.1093/mnras/stv2725}, 456,
  1221

\bibitem[\protect\citeauthoryear{Schmitt et~al.,}{Schmitt
  et~al.}{2014}]{Schmitt2014}
Schmitt J. H. M.~M.,  et~al., 2014, \mn@doi [Astron. Nachr.]
  {10.1002/asna.201412116}, 335, 787

\bibitem[\protect\citeauthoryear{Schr{\"{o}}der, Pols  \&
  Eggleton}{Schr{\"{o}}der et~al.}{1997}]{Schroder1997}
Schr{\"{o}}der K.~P.,  Pols O.~R.,   Eggleton P.~P.,  1997, \mn@doi [MNRAS]
  {10.1093/mnras/285.4.696}, 285, 696

\bibitem[\protect\citeauthoryear{Schr{\"{o}}der, Mittag, Hempelmann  \&
  Schmitt}{Schr{\"{o}}der et~al.}{2013}]{Schroder2013}
Schr{\"{o}}der K.~P.,  Mittag M.,  Hempelmann A.,   Schmitt J. H. M.~M.,  2013,
  A\&A, 554, A50

\bibitem[\protect\citeauthoryear{Schr{\"{o}}der, Schmitt, Mittag,
  G{\'{o}}mez-Trejo  \& Jack}{Schr{\"{o}}der et~al.}{2018}]{Schroder2018}
Schr{\"{o}}der K.~P.,  Schmitt J. H. M.~M.,  Mittag M.,  G{\'{o}}mez-Trejo V.,
   Jack D.,  2018, \mn@doi [MNRAS] {10.1093/mnras/sty1942}, 480, 2137

\bibitem[\protect\citeauthoryear{Schr{\"{o}}der, Mittag, Jack, Jim{\'{e}}nez
  \& Schmitt}{Schr{\"{o}}der et~al.}{2020}]{Schroder2020}
Schr{\"{o}}der K.~P.,  Mittag M.,  Jack D.,  Jim{\'{e}}nez A.~R.,   Schmitt J.
  H. M.~M.,  2020, \mn@doi [MNRAS] {10.1093/mnras/stz3476}, 492, 1110

\bibitem[\protect\citeauthoryear{Schr{\"{o}}der, Mittag, {Flor Torres}, Jack
  \& Snellen}{Schr{\"{o}}der et~al.}{2021}]{Schroder2021}
Schr{\"{o}}der K.~P.,  Mittag M.,  {Flor Torres} L.~M.,  Jack D.,   Snellen I.,
   2021, \mn@doi [MNRAS] {10.1093/mnras/staa2261}, 501, 5042

\bibitem[\protect\citeauthoryear{Shine, Milkey  \& Mihalas}{Shine
  et~al.}{1975}]{Shine1975}
Shine R.~A.,  Milkey R.~W.,   Mihalas D.,  1975, ApJ, 201, 222

\bibitem[\protect\citeauthoryear{Soubiran, {Le Campion}, Brouillet  \&
  Chemin}{Soubiran et~al.}{2016}]{Soubiran2016}
Soubiran C.,  {Le Campion} J.~F.,  Brouillet N.,   Chemin L.,  2016, \mn@doi
  [A\&A] {10.1051/0004-6361/201628497}, 591, 1

\bibitem[\protect\citeauthoryear{Torres}{Torres}{2010}]{Torres2010}
Torres G.,  2010, \mn@doi [AJ] {10.1088/0004-6256/140/5/1158}, 140, 1158

\bibitem[\protect\citeauthoryear{Wilson}{Wilson}{1967}]{Wilson1967}
Wilson O.~C.,  1967, \mn@doi [PASP] {10.1086/128434}, 79, 46

\bibitem[\protect\citeauthoryear{Wilson \& Bappu}{Wilson \&
  Bappu}{1957}]{Wilson1957}
Wilson O.~C.,  Bappu M. K.~V.,  1957, \mn@doi [ApJ] {10.1086/146339}, 125, 661

\makeatother
\end{thebibliography}



\appendix

\section{Pseudo-continuum readjustment and line selection for synthezising procedure with \textsc{iSpec}}
\label{sec:pseudo-continuum_and_line_selection}

\subsection{Guided readjustment of the pseudo-continuum} 

To find suitable continuum sections, which are needed to guide the readjustment of the pseudo-continuum of the observed spectrum, ideally at the value of 1.0, we used a synthetic spectrum that matches the observed spectrum reasonably well. To produce it, the initial values of $T_\text{eff}$, $\log g$ and $\text{[Fe/H]}$ were set to the averages of the respective entries in the PASTEL catalogue. Furthermore, the resolution of the synthetic spectrum was set to the one of the TIGRE-HEROS target spectrum ($R\sim20,000$), and the projected rotation velocity $v \sin i$ was fixed at the small value of 1.6 km/s. This is a reasonable choice for giants, as well as for aged and relatively inactive main sequence stars, which form our sample. Individual differences of a few km/s in this rotation rate cannot be resolved by TIGRE-HEROS spectra, anyway. 

At this point, another choice was required, namely for the initial values of micro and macro turbulent velocities. Despite these being below the resolution limit of TIGRE-HEROS spectra, their choices make subtle but systematic differences in the $\chi^2$ sums of later best-fit solutions. Therefore, in the final synthesizing of the physical parameters (see below), the turbulence values are allowed to be refined, see Table \ref{table:stellar_parameters}. But for the initial model spectrum required here only to define suitable continuum sections, we needed reasonable fixed values. 

Initially, we tested the macro and micro-turbulence velocities, $v_\text{mac}$ and $v_\text{mic}$, used for the models of the \textsc{phoenix} library \citep{Husser2013}, which give reasonably realistic spectra for stars with solar abundances, over a range of effective temperatures and surface gravities. Therefore, like \citep{Husser2013}, we first fixed the macro-turbulence parameters at twice the micro-turbulence values. However, after several tests with well-studied stars (see subsection \ref{subsec:GBS_Test}), we refined the initial choices of macro-turbulence velocity $v_\text{mac}$ by the following empirical representation, according to the findings in main sequence stars by \cite{Ryabchikova2016}, while still using the micro-turbulence values of \cite{Husser2013} in this step:

\begin{equation}
  v_\text{mac} = \left\lbrace
  \begin{array}{ll}
    2.00 v_\text{mic} & \textup{if } \log g \leq 2.0\\
    3.00 v_\text{mic} & \textup{if } 2.0 < \log g < 3.75\\
    3.75 v_\text{mic} & \textup{if } \log g \geq 3.75.
  \end{array}
  \right.
  \label{eq:vmac}
\end{equation}

In this so-defined, representative synthetic spectrum, we then selected continuum sections according to the following criteria: (1) Maximum standard deviation from 1.0 of 0.1 per cent, and (2) minimum size of the region equivalent to FWHM of the instrumental profile of TIGRE-HEROS ($\sim$ 0.35 \AA{} in the center of R-channel, $\sim$ 0.23 \AA{} in the centre of B-channel). (3) To avoid confusion, spectral regions with abundant telluric lines were also discarded \citep{Catanzaro1997}. 

These continuum sections were finally used to readjust the observed stellar pseudo-continuum, employing the \textsc{iSpec} algorithm designed for this. According to \citet{Blanco-Cuaresma2014}, this subroutine applies a median and maximum filter with different window sizes. The former smooths out the noise in the continuum, while the latter ignores slightly smaller fluxes, which rather belong to shallow absorption lines -- i.e., the continuum may effectively be placed slightly above or below the simple flux average in the respective continuum section, depending on the values of those parameters. Then, this algorithm applies a third-degree spline, assigning one node every 20 \AA{} only within these selected continuum regions. This procedure is repeated recursively to different median and maximum filters, ten steps for each one from 0.1 \AA{} to 1.0 \AA{} for the first, and 1.5 \AA{} to 15 \AA{} for the last one. A total of up to 100 possible normalizations are calculated. 

The script finds the residuals between the synthetic and observed spectrum, and calculates the RMS only in the continuum regions. The readjusted spectrum with the best RMS is finally chosen. Finally, a second fine-tuning normalization is applied to the observed spectrum. Using again the continuum sections, this time as a template in the observed spectrum, the \textsc{iSpec} script further improves the normalization by the use of a median filter equal to the instrumental profile width of TIGRE-HEROS. As a whole, this guided continuum readjustment procedure improves the original normalization of the observed spectrum significantly, because the former can be mislead by agglomerations of absorption lines.

\subsection{Selection of representative lines for synthesizing procedure}

We selected a representative, reliable subset of lines for the synthesizing procedure, aiming on small $\chi^2$ minima, less impacted by line blends or mismatching atomic line strengths ($f$-values). For our sample stars, we based that selection on the following criteria: (1) the S/N in each line must be greater than 5 times; (2) a cross correlation is undertaken between the information of the Vienna Atomic Line Data Base (VALD) \citep{Piskunov1995} and the observed lines, avoiding differences in wavelength larger than the width of the instrumental profile of TIGRE-HEROS ($\sim 0.35$ \AA{}); (3) lines within spectral regions known to host many telluric lines are discarded; (4) were the observed line strength disagrees with the atomic data of the VALD list, or would even be zero according to the synthetic spectrum, such lines are discarded (see below).

We applied these criteria to a good solar spectrum, taken by TIGRE-HEROS, using a strategy similar to the line-by-line differential approach of \citet{Blanco-Cuaresma2019}, that is: a line can be considered good, when synthesizing arrives at a close-to-solar abundance (i.e. reasonably close to zero). After several tests, if a spectral element has more than 20 lines, we considered an abundance mismatch-margin of $\pm 0.15$ dex, otherwise accepted a margin of $\pm 0.50$ dex. Hence, lines producing an error larger than 0.50 dex are discarded. This tolerated metallicity mismatch, reflecting atomic data mismatches of the line list, is a compromise between quantity and quality of the selection. By this approach, only lines from iron peak elements (iron, chromium, nickel) and alpha elements (silicon, calcium, titanium) would qualify. For the here required solar reference model, we used the solar parameters\footnote{Values from NASA Sun Fact Sheet \url{https://nssdc.gsfc.nasa.gov/planetary/factsheet/sunfact.html}} to $T_\text{eff} = 5772$ K, $\log g = 4.44$ dex and $\text{[M/H]} = 0.00$ dex. Again the initial $v_\text{mic}$ was determined using the \textsc{phoenix} library, $v_\text{mac}$ with equation \ref{eq:vmac}, and $v \sin i$ was fixed to 1.6 km/s. 

From this process, a total of 202 lines were selected for the R-channel spectra of HEROS (covering 5800 to 8700 \AA{}). The final line list is shown in Table \ref{table:iSpec-lines}. In the same way of \citet{Blanco-Cuaresma2019}, these lines are not blindly used on the observed spectra. Before starting the parameter synthesizing process, the script checks if each of these lines exists in the target spectrum, and using a gaussian fit, then applies the equivalent width criterion as stated above, before including the line in the process. According to our tests, these lines worked well enough for our sample from main sequence stars to supergiants (see subsection \ref{subsec:GBS_Test}).

\begin{table}
  \centering
  \caption{List of lines selected for use with the \textsc{iSpec} script.}
  \label{table:iSpec-lines}
  \begin{threeparttable}
    \begin{tabular}{cccc}
      \hline
      $\lambda_0$ & $\lambda_\text{blue}$ & $\lambda_\text{red}$ & Note \\ %
	  {[nm]} & {[nm]} & {[nm]} & \\ \hline
	  580.52548	&	580.48018	&	580.55266	&	Ni 1	\\
	  580.57078	&	580.55266	&	580.61608	&	Fe 1	\\
	  580.67044	&	580.61608	&	580.73386	&	Fe 1	\\
	  580.92412	&	580.84258	&	581.01472	&	Fe 1	\\
	  581.47677	&	581.43147	&	581.55831	&	Fe 1	\\
	  581.63079	&	581.57643	&	581.67609	&	Fe 1	\\
	  583.16193	&	583.10757	&	583.21629	&	Ni 1	\\
	  583.83236	&	583.78706	&	583.89578	&	Fe 1	\\
	  584.70212	&	584.64776	&	584.75648	&	Ni 1	\\
	  584.81084	&	584.75648	&	584.85614	&	Fe 1	\\
	  585.74402	&	585.64436	&	585.81650	&	Ca 1	\\
	  585.96145	&	585.90709	&	586.04299	&	Fe 1	\\
	  586.64095	&	586.58659	&	586.68625	&	Ti 1	\\
	  586.75873	&	586.68625	&	586.79497	&	Ca 1	\\
	  587.32045	&	587.26609	&	587.35669	&	Fe 1	\\
	  592.77455	&	592.72925	&	592.81079	&	Fe 1	\\
	  593.01916	&	592.98292	&	593.07352	&	Fe 1	\\
	  593.46310	&	593.41780	&	593.51746	&	Fe 1	\\
	  595.27510	&	595.21168	&	595.37475	&	Fe 1	\\
	  596.58879	&	596.51631	&	596.63409	&	Ti 1	\\
	  597.85718	&	597.82094	&	597.92966	&	Ti 1	\\
	  598.48232	&	598.43702	&	598.58198	&	Fe 1	\\
	  599.13464	&	599.09840	&	599.16182	&	Fe 2	\\
	  599.67824	&	599.62388	&	599.70542	&	Ni 1	\\
	  600.72919	&	600.68389	&	600.75637	&	Ni 1	\\
	  600.79261	&	600.75637	&	600.81979	&	Fe 1	\\
	  604.20822	&	604.11762	&	604.28070	&	Fe 1	\\
	  605.36789	&	605.31353	&	605.47661	&	Ni 1	\\
	  605.60345	&	605.55815	&	605.67593	&	Fe 1	\\
	  606.54569	&	606.50039	&	606.62723	&	Fe 1	\\
	  608.41204	&	608.33956	&	608.45734	&	Fe 2	\\
	  608.52076	&	608.45734	&	608.57512	&	Fe 1	\\
	  608.62948	&	608.57512	&	608.72008	&	Ni 1	\\
	  609.11872	&	609.07342	&	609.14590	&	Ti 1	\\
	  609.36334	&	609.32710	&	609.39958	&	Fe 1	\\
	  609.82539	&	609.78915	&	609.94317	&	Fe 1	\\
	  610.26933	&	610.24215	&	610.30557	&	Ca 1	\\
	  610.32369	&	610.30557	&	610.41429	&	Fe 1	\\
	  612.21722	&	612.12662	&	612.30782	&	Ca 1	\\
	  612.49808	&	612.42560	&	612.56150	&	Si 1	\\
	  612.62492	&	612.56150	&	612.67928	&	Ti 1	\\
	  612.78800	&	612.71552	&	612.85142	&	Fe 1	\\
	  612.89672	&	612.85142	&	612.94202	&	Ni 1	\\
	  613.17758	&	613.08698	&	613.26818	&	Si 1	\\
	  613.66682	&	613.61246	&	613.73024	&	Fe 1	\\
	  614.24665	&	614.21947	&	614.29195	&	Si 1	\\
	  614.50033	&	614.44597	&	614.58187	&	Si 1	\\
	  614.78119	&	614.69965	&	614.82649	&	Fe 1	\\
	  614.92615	&	614.88085	&	614.97145	&	Fe 2	\\
	  615.16171	&	615.10735	&	615.20701	&	Fe 1	\\
	  615.76873	&	615.70531	&	615.85933	&	Fe 1	\\
	  616.13113	&	616.10395	&	616.16737	&	Ca 1	\\
	  616.22173	&	616.16737	&	616.29421	&	Ca 1	\\
	  616.35763	&	616.29421	&	616.42105	&	Ca 1	\\
	  616.53882	&	616.49352	&	616.59318	&	Fe 1	\\
	  616.64754	&	616.59318	&	616.73814	&	Ca 1	\\
	  616.90122	&	616.81968	&	616.92840	&	Ca 1	\\
	  616.95558	&	616.92840	&	617.00994	&	Ca 1	\\
	  617.05524	&	617.00994	&	617.16396	&	Fe 1	\\
	  617.33610	&	617.27268	&	617.38140	&	Fe 1	\\		
	  \hline
    \end{tabular}
  \end{threeparttable}
\end{table}

\begin{table}
  \centering
  \contcaption{List of lines selected for use with the \textsc{iSpec} script.}
  \begin{threeparttable}
    \begin{tabular}{cccc}
      \hline
      $\lambda_0$ & $\lambda_\text{blue}$ & $\lambda_\text{red}$ & Note \\ %
	  {[nm]} & {[nm]} & {[nm]} & \\ \hline
          617.53542	&	617.45388	&	617.58978	&	Ni 1	\\
          617.68038	&	617.61696	&	617.78004	&	Ni 1	\\	  
          618.02466	&	617.97030	&	618.07902	&	Fe 1	\\
          618.35988	&	618.30552	&	618.42330	&	Fe 1	\\
          619.15715	&	619.04843	&	619.27493	&	Fe 1	\\
          619.54673	&	619.50143	&	619.60109	&	Si 1	\\
          620.03597	&	619.98161	&	620.08127	&	Fe 1	\\
          621.34060	&	621.28624	&	621.42214	&	Fe 1	\\
          622.02916	&	621.97480	&	622.04728	&	Ti 1	\\
          622.40062	&	622.33720	&	622.43686	&	Ni 1	\\
          622.67242	&	622.59088	&	622.71772	&	Fe 1	\\
          622.92610	&	622.85362	&	622.97140	&	Fe 1	\\
          623.00764	&	622.97140	&	623.02576	&	Ni 1	\\
          623.07106	&	623.02576	&	623.16166	&	Fe 1	\\
          623.73243	&	623.57841	&	623.78679	&	Si 1	\\
          623.84115	&	623.78679	&	623.89551	&	Fe 2	\\
          624.06765	&	623.95893	&	624.11295	&	Fe 1	\\
          624.38475	&	624.33945	&	624.41193	&	Si 1	\\
          624.62937	&	624.58407	&	624.69279	&	Fe 1	\\
          624.75621	&	624.69279	&	624.81057	&	Fe 2	\\
          625.25451	&	625.20921	&	625.31793	&	Fe 1	\\
          625.80717	&	625.77093	&	625.83435	&	Ti 1	\\
          625.87059	&	625.83435	&	625.91588	&	Ti 1	\\
          625.96118	&	625.91588	&	626.01554	&	Ni 1	\\
          626.10614	&	626.05178	&	626.16956	&	Ti 1	\\
          627.02120	&	626.95778	&	627.07556	&	Fe 1	\\
          627.12992	&	627.07556	&	627.16616	&	Fe 1	\\
          630.14689	&	630.09253	&	630.21031	&	Fe 1	\\
          631.14348	&	631.08912	&	631.18878	&	Fe 1	\\
          631.57836	&	631.56024	&	631.62366	&	Fe 1	\\
          632.21256	&	632.16726	&	632.23068	&	Ni 1	\\
          632.26692	&	632.23068	&	632.32128	&	Fe 1	\\
          633.00983	&	632.93736	&	633.04607	&	Cr 1	\\
          633.68027	&	633.62591	&	633.77087	&	Fe 1	\\
          633.89771	&	633.81617	&	633.97019	&	Ni 1	\\
          635.86372	&	635.81842	&	635.93620	&	Fe 1	\\
          636.28954	&	636.26236	&	636.38920	&	Fe 1	\\
          636.44356	&	636.38920	&	636.52510	&	Fe 1	\\
          636.64288	&	636.58852	&	636.69724	&	Ni 1	\\
          636.95092	&	636.86032	&	636.98716	&	Fe 2	\\
          637.13212	&	637.07776	&	637.20460	&	Si 2	\\
          637.82973	&	637.77537	&	637.92033	&	Ni 1	\\
          638.07435	&	638.01999	&	638.14683	&	Fe 1	\\
          638.46393	&	638.41863	&	638.50923	&	Ni 1	\\
          638.56359	&	638.50923	&	638.61795	&	Fe 1	\\
          639.26121	&	639.17061	&	639.28839	&	Fe 1	\\
          639.36087	&	639.28839	&	639.47865	&	Fe 1	\\
          640.00412	&	639.90447	&	640.08566	&	Fe 1	\\
          640.72892	&	640.63832	&	640.75610	&	Si 1	\\
          641.16380	&	641.04602	&	641.23628	&	Fe 1	\\
          641.49902	&	641.40842	&	641.59868	&	Si 1	\\
          641.68928	&	641.62586	&	641.73458	&	Fe 2	\\
          641.99732	&	641.90672	&	642.05168	&	Fe 1	\\
          642.13322	&	642.07886	&	642.20570	&	Fe 1	\\
          642.48655	&	642.42313	&	642.53185	&	Ni 1	\\
          643.08451	&	643.00297	&	643.13887	&	Fe 1	\\
          643.90897	&	643.83649	&	644.02675	&	Ca 1	\\
          644.97804	&	644.94180	&	645.09582	&	Ca 1	\\
          645.55788	&	645.52164	&	645.59412	&	Ca 1	\\
          645.63942	&	645.59412	&	645.67566	&	Fe 2	\\
	  \hline
    \end{tabular}
  \end{threeparttable}
\end{table}

\begin{table}
  \centering
  \contcaption{List of lines selected for use with the \textsc{iSpec} script.}
  \begin{threeparttable}
    \begin{tabular}{cccc}
      \hline
      $\lambda_0$ & $\lambda_\text{blue}$ & $\lambda_\text{red}$ & Note \\ %
	  {[nm]} & {[nm]} & {[nm]} & \\ \hline
          646.26456	&	646.20114	&	646.31892	&	Ca 1	\\
          649.15469	&	649.11845	&	649.21811	&	Ti 2	\\
          649.38119	&	649.33589	&	649.41742	&	Ca 1	\\
          649.49896	&	649.41742	&	649.54426	&	Fe 1	\\
          649.65298	&	649.61674	&	649.66204	&	Fe 1	\\
          649.89760	&	649.84324	&	649.92478	&	Fe 1	\\
          649.96102	&	649.92478	&	650.05162	&	Ca 1	\\
          659.75484	&	659.69142	&	659.80920	&	Fe 1	\\
          660.91452	&	660.85110	&	661.01418	&	Fe 1	\\
          662.50001	&	662.44565	&	662.54531	&	Fe 1	\\
          662.75369	&	662.68121	&	662.79899	&	Fe 1	\\
          663.51473	&	663.46037	&	663.55096	&	Ni 1	\\
          663.97678	&	663.92242	&	664.04020	&	Fe 1	\\
          664.69252	&	664.65628	&	664.74688	&	Fe 1	\\
          666.11493	&	666.06057	&	666.16929	&	Cr 1	\\
          666.76725	&	666.70383	&	666.81255	&	Fe 1	\\
          667.80009	&	667.69137	&	667.85445	&	Fe 1	\\
          671.03449	&	670.98919	&	671.09791	&	Fe 1	\\
          671.62339	&	671.57809	&	671.66869	&	Fe 1	\\
          672.53845	&	672.49315	&	672.59281	&	Fe 1	\\
          672.66529	&	672.59281	&	672.75589	&	Fe 1	\\
          674.31420	&	674.23266	&	674.42292	&	Ti 1	\\
          675.01182	&	674.95746	&	675.07524	&	Fe 1	\\
          675.27456	&	675.20208	&	675.31985	&	Fe 1	\\
          677.23151	&	677.16809	&	677.30399	&	Ni 1	\\
          679.35154	&	679.27000	&	679.40590	&	Fe 1	\\
          680.68335	&	680.61993	&	680.74677	&	Fe 1	\\
          681.02763	&	680.97327	&	681.08199	&	Fe 1	\\
          682.85774	&	682.73996	&	682.93928	&	Fe 1	\\
          683.70032	&	683.61878	&	683.79998	&	Fe 1	\\
          683.98118	&	683.93588	&	684.02648	&	Fe 1	\\
          684.13520	&	684.02648	&	684.17144	&	Fe 1	\\
          684.20768	&	684.17144	&	684.23486	&	Ni 1	\\
          684.36170	&	684.31640	&	684.43418	&	Fe 1	\\
          685.51231	&	685.39453	&	685.64821	&	Fe 1	\\
          685.81129	&	685.76599	&	685.89283	&	Fe 1	\\
          686.24617	&	686.21899	&	686.32771	&	Fe 1	\\
          743.07931	&	743.01589	&	743.12461	&	Fe 1	\\
          744.30240	&	744.24804	&	744.38394	&	Fe 1	\\
          744.93660	&	744.85506	&	744.98190	&	Fe 2	\\
          745.39866	&	745.33524	&	745.46208	&	Fe 1	\\
          746.15063	&	746.09627	&	746.18687	&	Fe 1	\\
          746.23217	&	746.18687	&	746.29559	&	Cr 1	\\
          747.35561	&	747.31031	&	747.40997	&	Fe 1	\\
          748.19819	&	748.09853	&	748.27066	&	Fe 1	\\
          749.16760	&	749.10418	&	749.26726	&	Fe 1	\\
          750.60813	&	750.55378	&	750.66249	&	Fe 1	\\
          751.10643	&	750.92523	&	751.20609	&	Fe 1	\\
          752.27517	&	752.20269	&	752.36577	&	Ni 1	\\
          752.51073	&	752.44731	&	752.58321	&	Ni 1	\\
          754.18682	&	754.09622	&	754.24118	&	Fe 1	\\
          757.40311	&	757.35781	&	757.48465	&	Ni 1	\\
          758.38158	&	758.27286	&	758.46312	&	Fe 1	\\
          772.75974	&	772.68726	&	772.83222	&	Ni 1	\\
          774.55361	&	774.47207	&	774.61703	&	Fe 1	\\
          774.65327	&	774.61703	&	774.71669	&	Fe 1	\\
          774.82541	&	774.71669	&	774.86165	&	Fe 1	\\
          774.88883	&	774.86165	&	774.95225	&	Ni 1	\\
          775.11533	&	775.03379	&	775.20593	&	Fe 1	\\
          776.05757	&	775.99415	&	776.19347	&	Si 1	\\
	  \hline
    \end{tabular}
  \end{threeparttable}
\end{table}

\begin{table}
  \centering
  \contcaption{List of lines selected for use with the \textsc{iSpec} script.}
  \begin{threeparttable}
    \begin{tabular}{cccc}
      \hline
      $\lambda_0$ & $\lambda_\text{blue}$ & $\lambda_\text{red}$ & Note \\ %
	  {[nm]} & {[nm]} & {[nm]} & \\ \hline
          780.78687	&	780.70533	&	780.85935	&	Fe 1	\\
          784.99975	&	784.90915	&	785.06317	&	Si 1	\\
          785.53429	&	785.47993	&	785.62489	&	Fe 1	\\
          846.84305	&	846.77057	&	846.96989	&	Fe 1	\\
          847.16920	&	847.07860	&	847.29604	&	Fe 1	\\
          848.19298	&	848.12050	&	848.26546	&	Fe 1	\\
          849.80565	&	849.46137	&	850.09557	&	Ca 2	\\
          851.40927	&	851.34585	&	851.46363	&	Fe 1	\\
          851.50893	&	851.46363	&	851.57235	&	Fe 1	\\
          851.83508	&	851.77167	&	851.89850	&	Ti 1	\\
          855.67651	&	855.59497	&	855.82147	&	Si 1	\\
          858.23142	&	858.13176	&	858.28578	&	Fe 1	\\
          859.59041	&	859.52699	&	859.65383	&	Si 1	\\
          859.70819	&	859.65383	&	859.78067	&	Si 1	\\
          859.88033	&	859.78067	&	859.97093	&	Fe 1	\\
          861.17591	&	861.11249	&	861.23932	&	Fe 1	\\
          862.16344	&	862.10002	&	862.22686	&	Fe 1	\\
          863.69457	&	863.65833	&	863.75799	&	Ni 1	\\
          867.47258	&	867.40916	&	867.65378	&	Fe 1	\\
          868.64131	&	868.54165	&	868.72285	&	Si 1	\\
          868.85875	&	868.72285	&	868.94935	&	Fe 1	\\
          869.94595	&	869.90065	&	869.98219	&	Fe 1	\\		
	  \hline
    \end{tabular}
  \end{threeparttable}
\end{table}

\section[Ca II emission line width measurements of the stellar sample]{\ion{Ca}{ii} emission line width measurements of the stellar sample}
\label{sec:CaII_widths}

\begin{figure*}
  \includegraphics[width=0.425\textwidth]{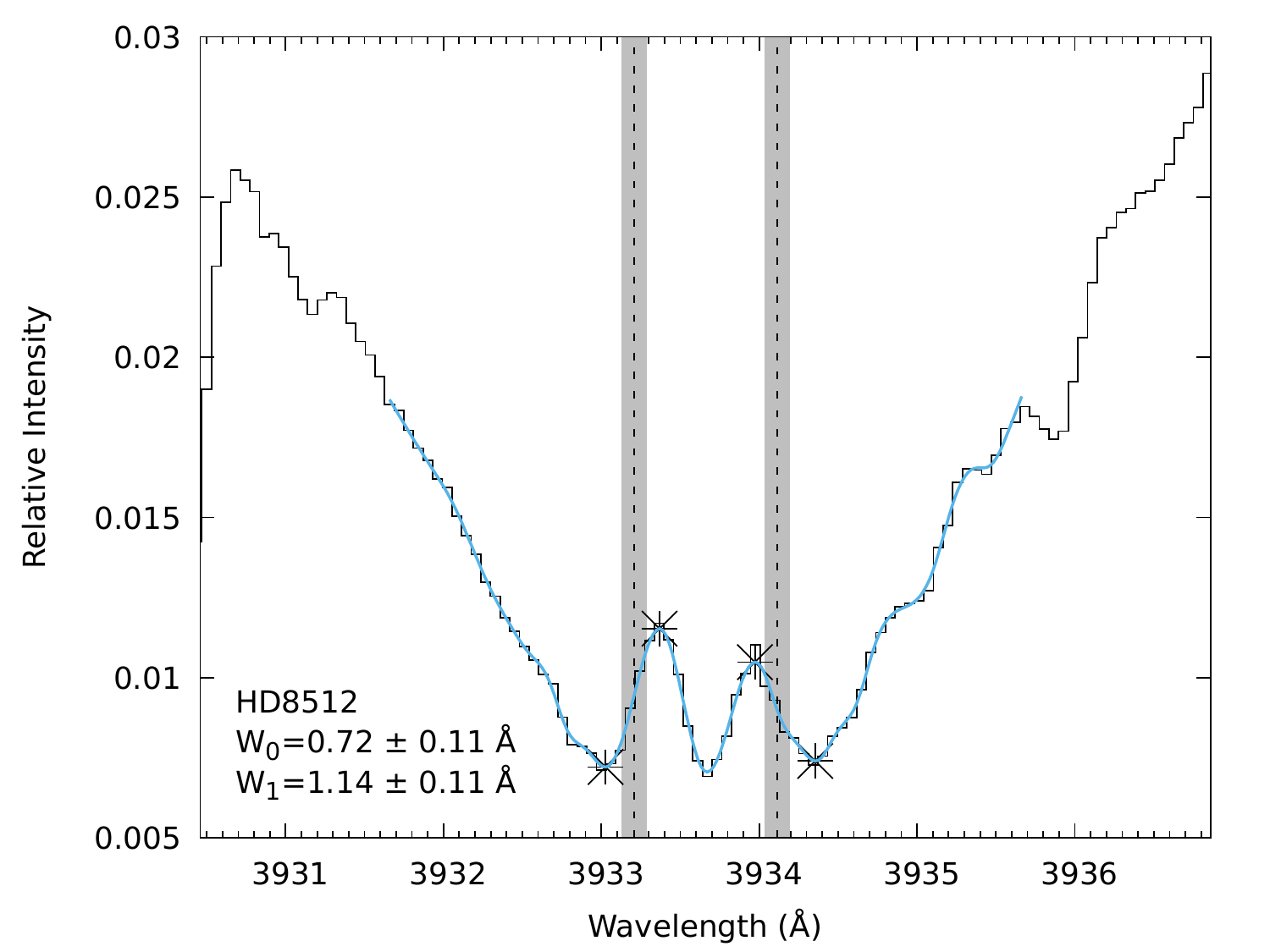}
  \includegraphics[width=0.425\textwidth]{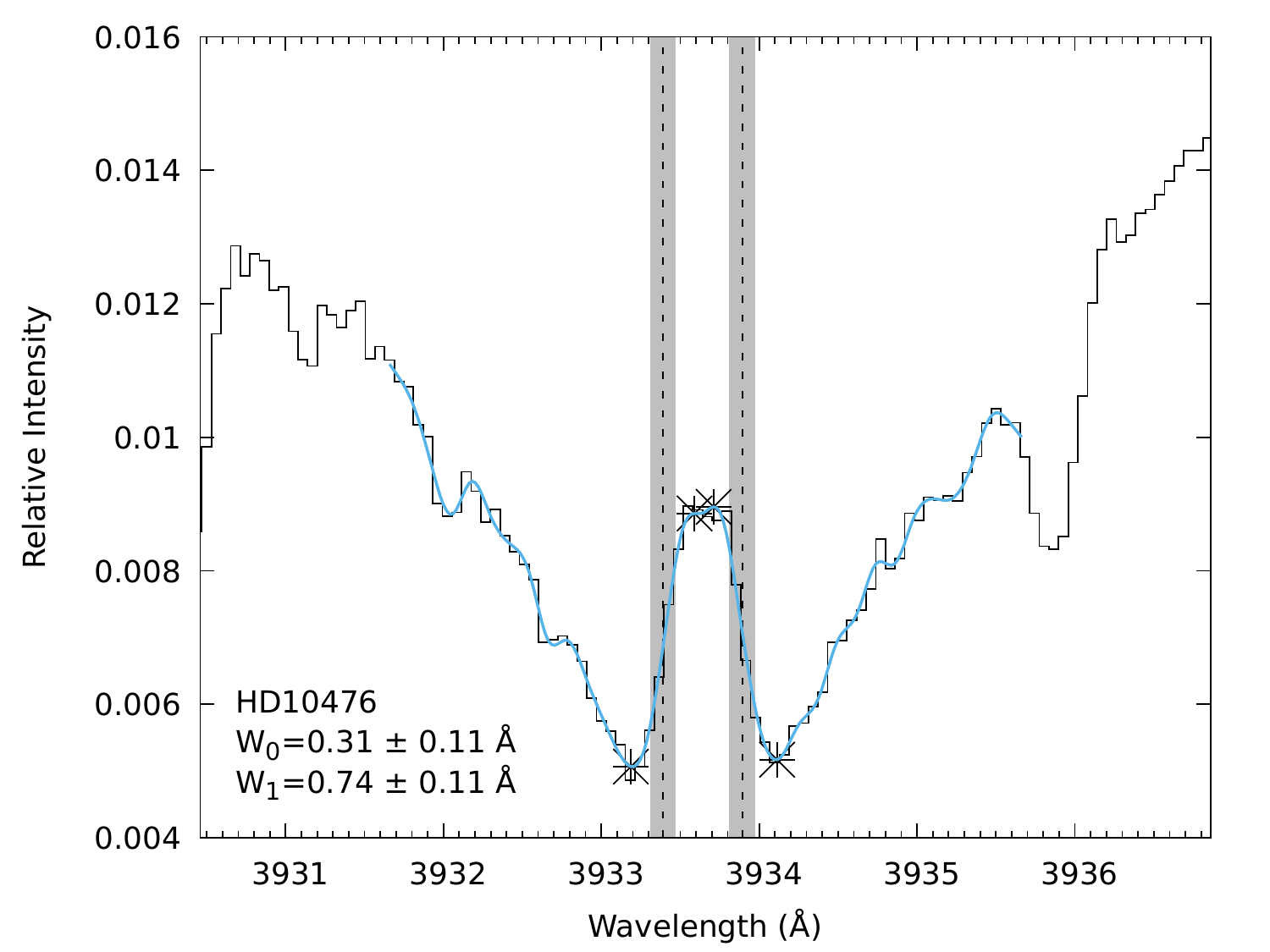}
  \includegraphics[width=0.425\textwidth]{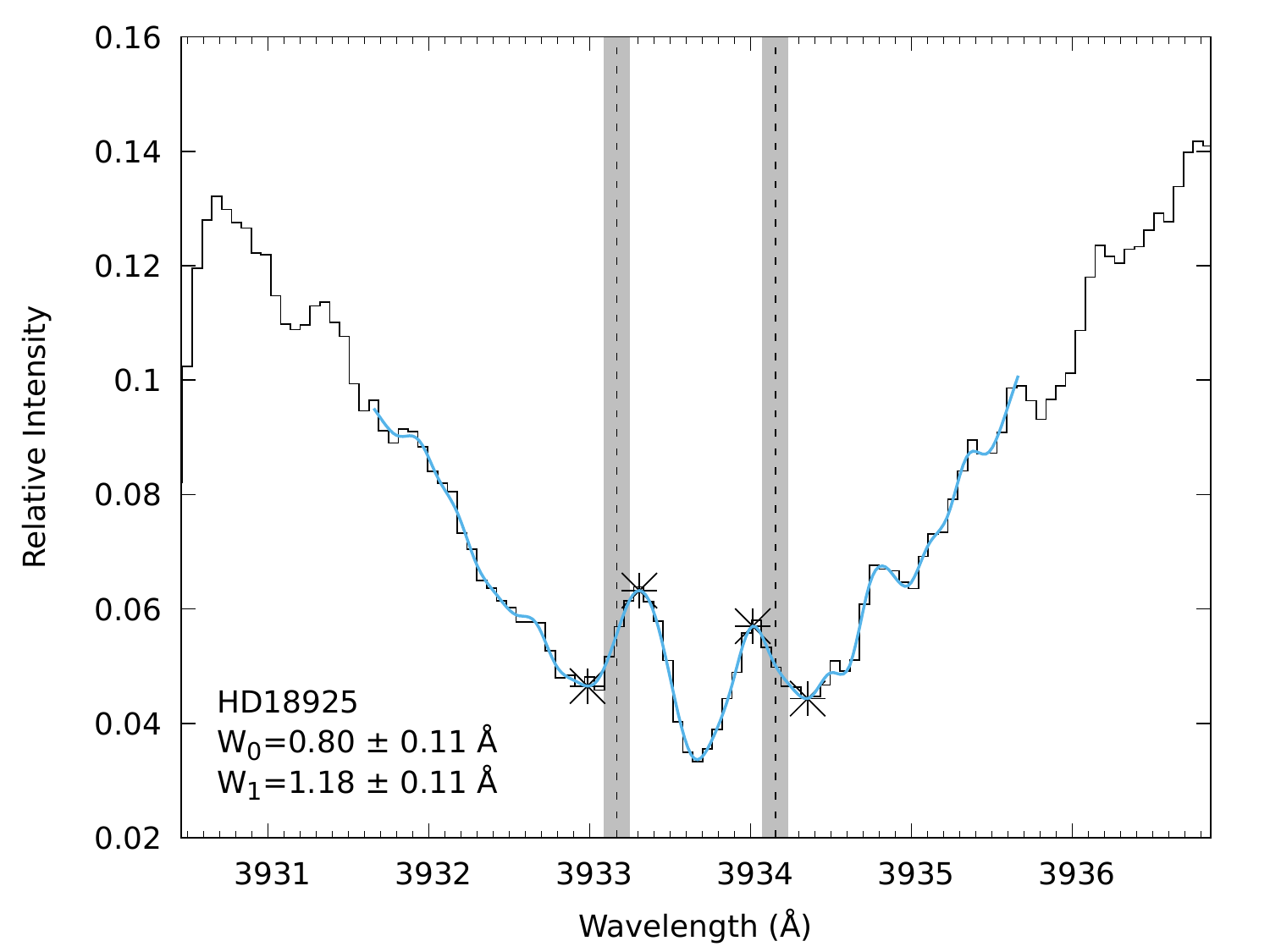}
  \includegraphics[width=0.425\textwidth]{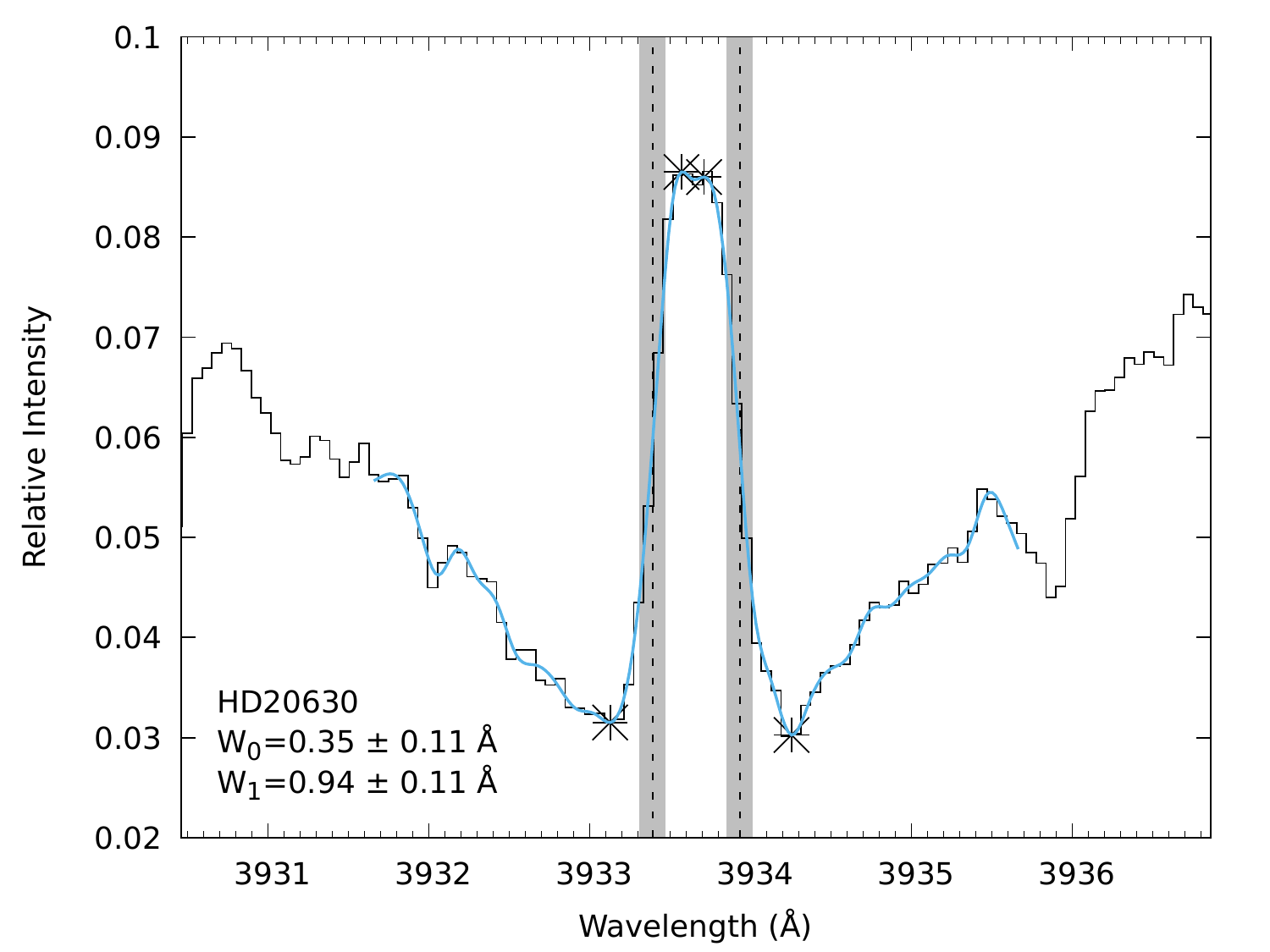}
  \includegraphics[width=0.425\textwidth]{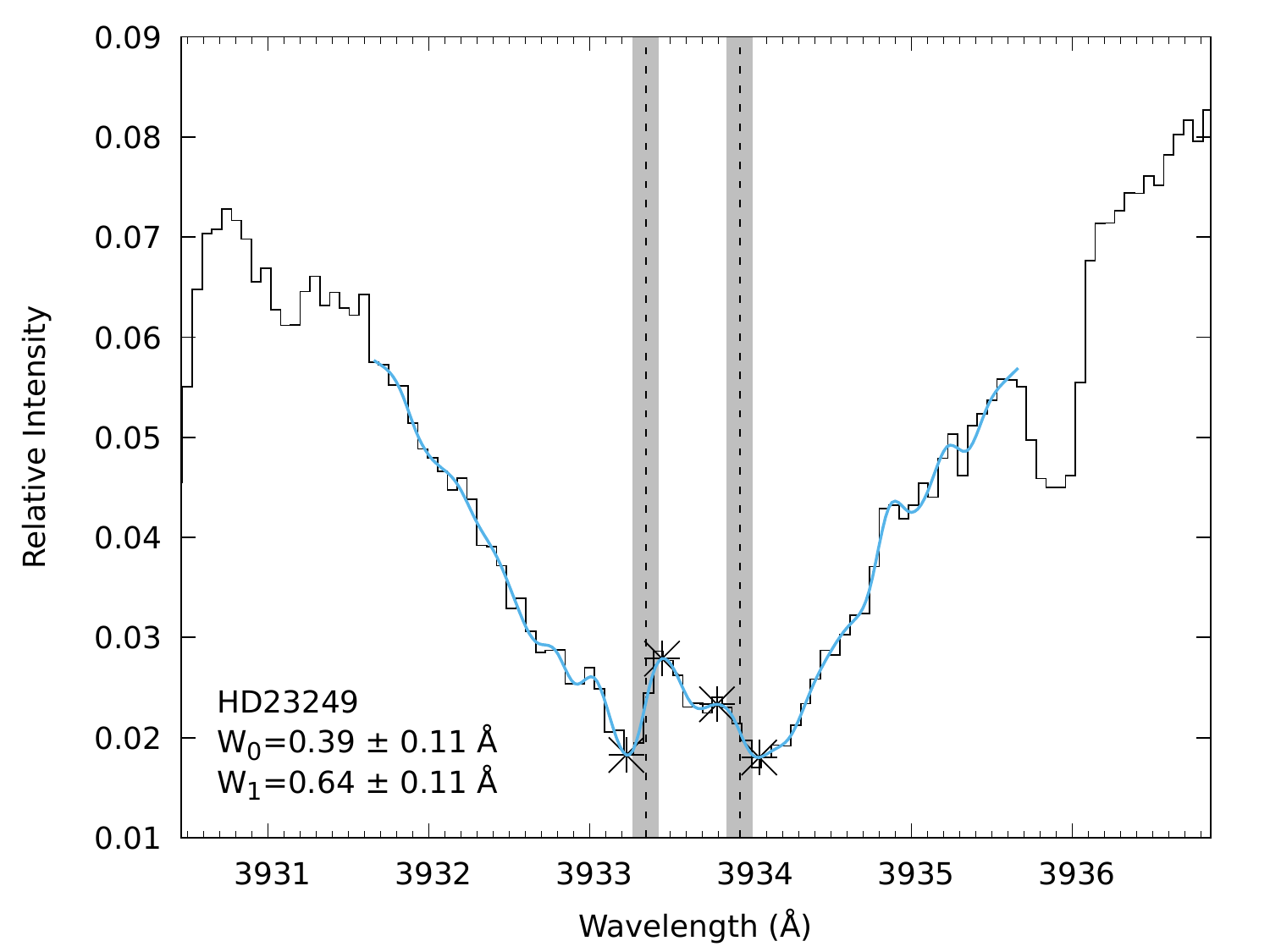}
  \includegraphics[width=0.425\textwidth]{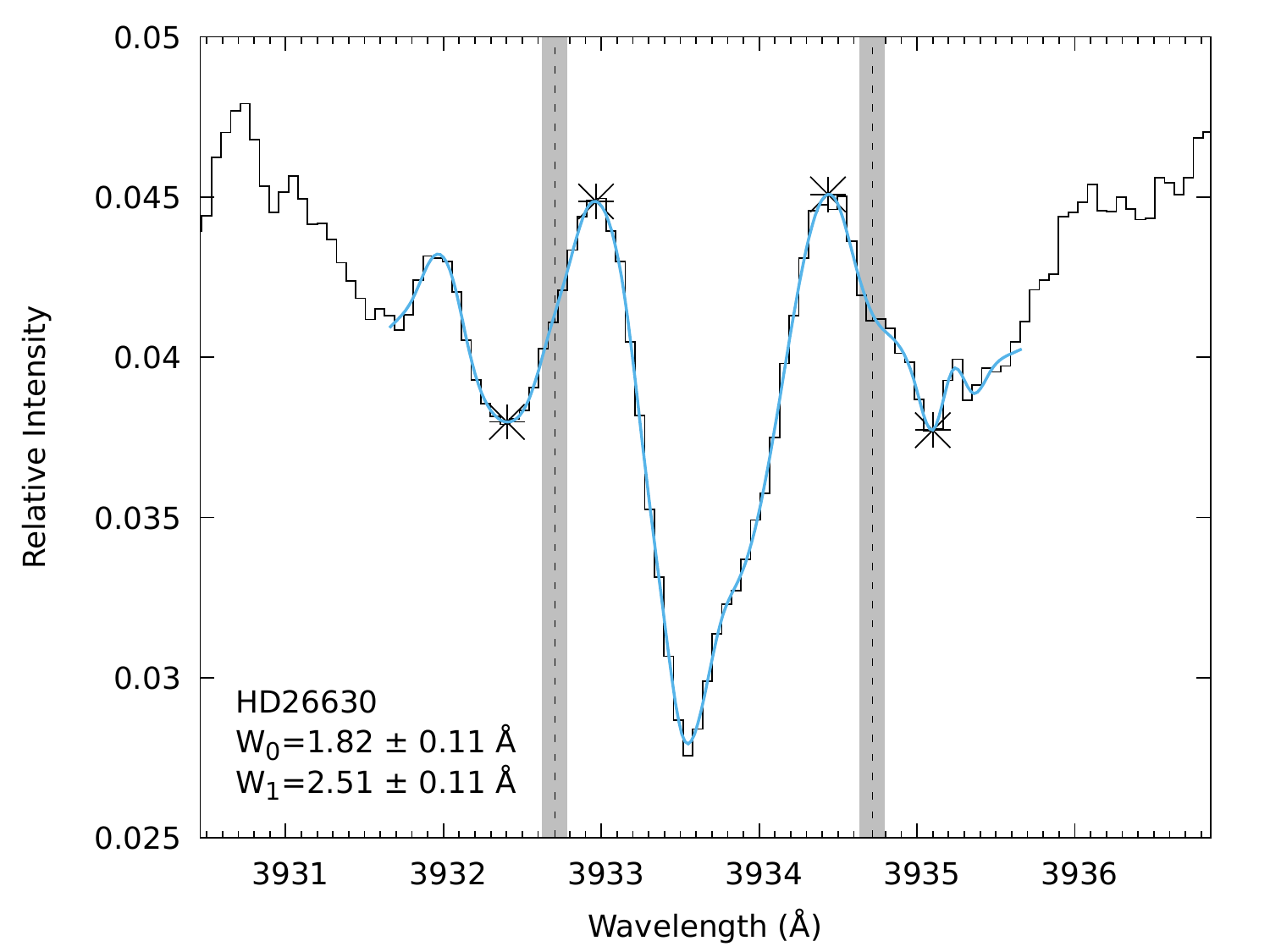}
  \includegraphics[width=0.425\textwidth]{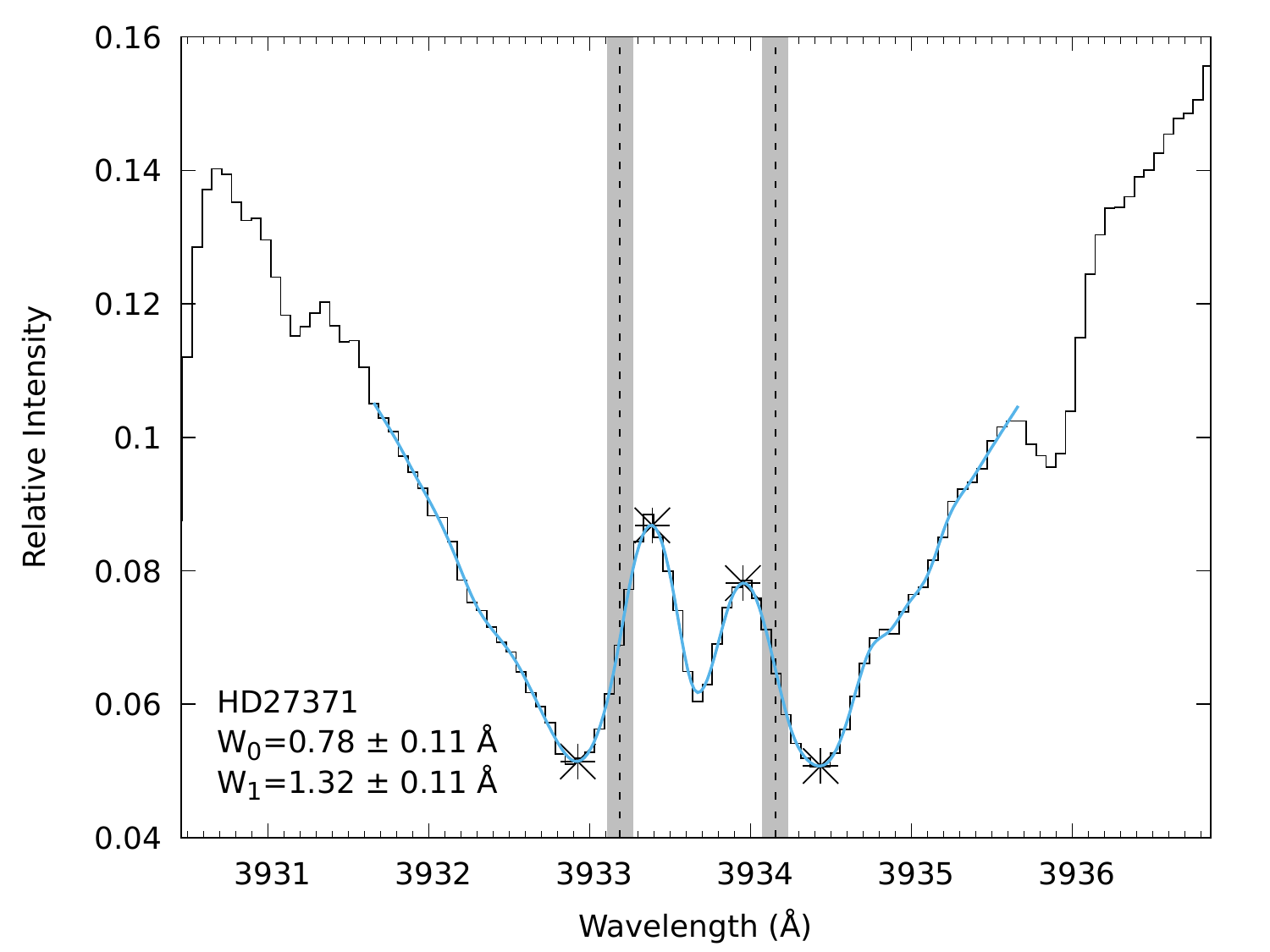}
  \includegraphics[width=0.425\textwidth]{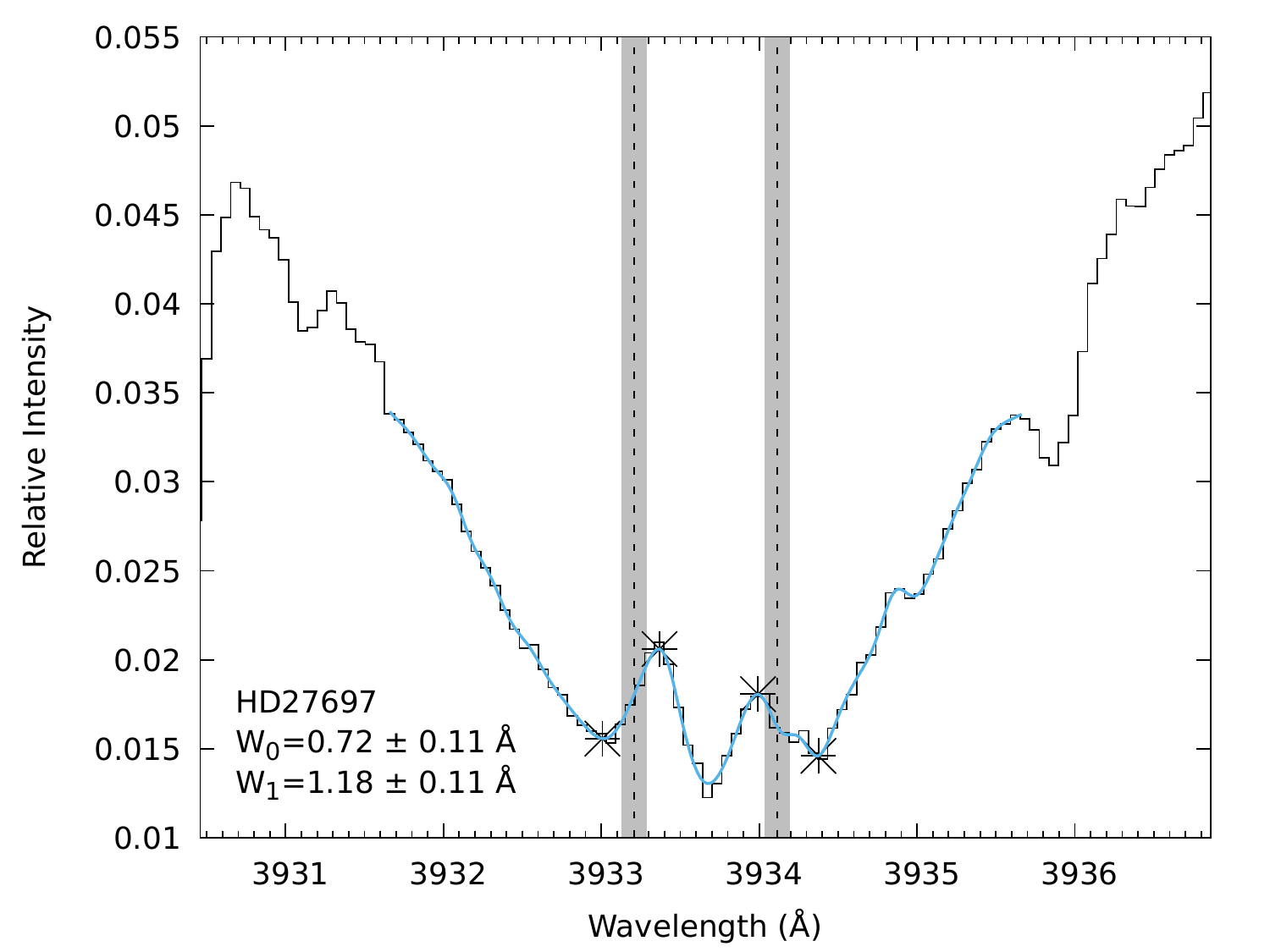}  
  \caption{\ion{Ca}{ii} emission line width measurements of HD 8512, HD 10476, HD 18925, HD 20630, HD 23249, HD 26630, HD 27371 and HD 27697.}
  \label{fig:CaII_Widths_1}
\end{figure*}

\begin{figure*}
  \includegraphics[width=0.425\textwidth]{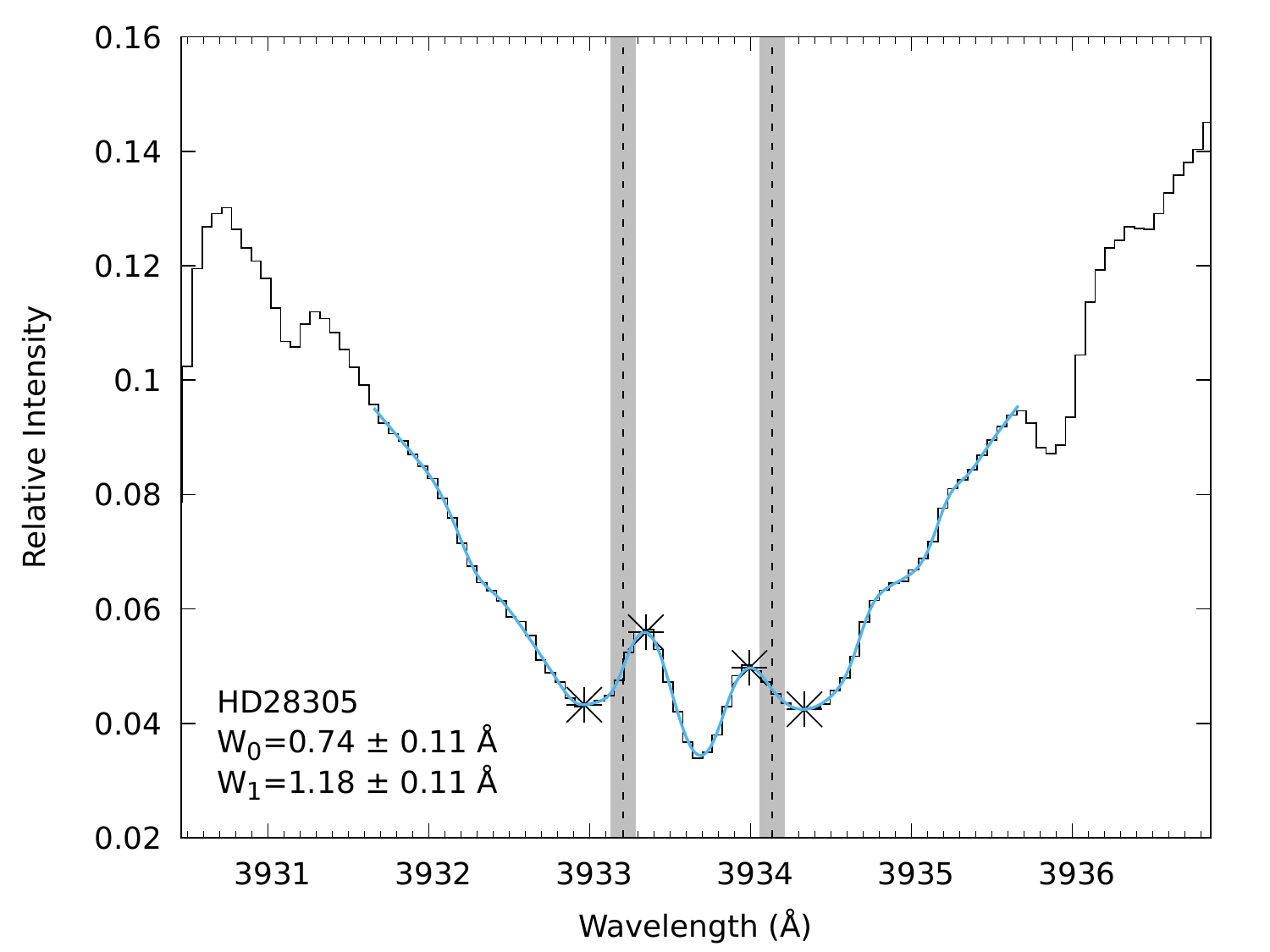}
  \includegraphics[width=0.425\textwidth]{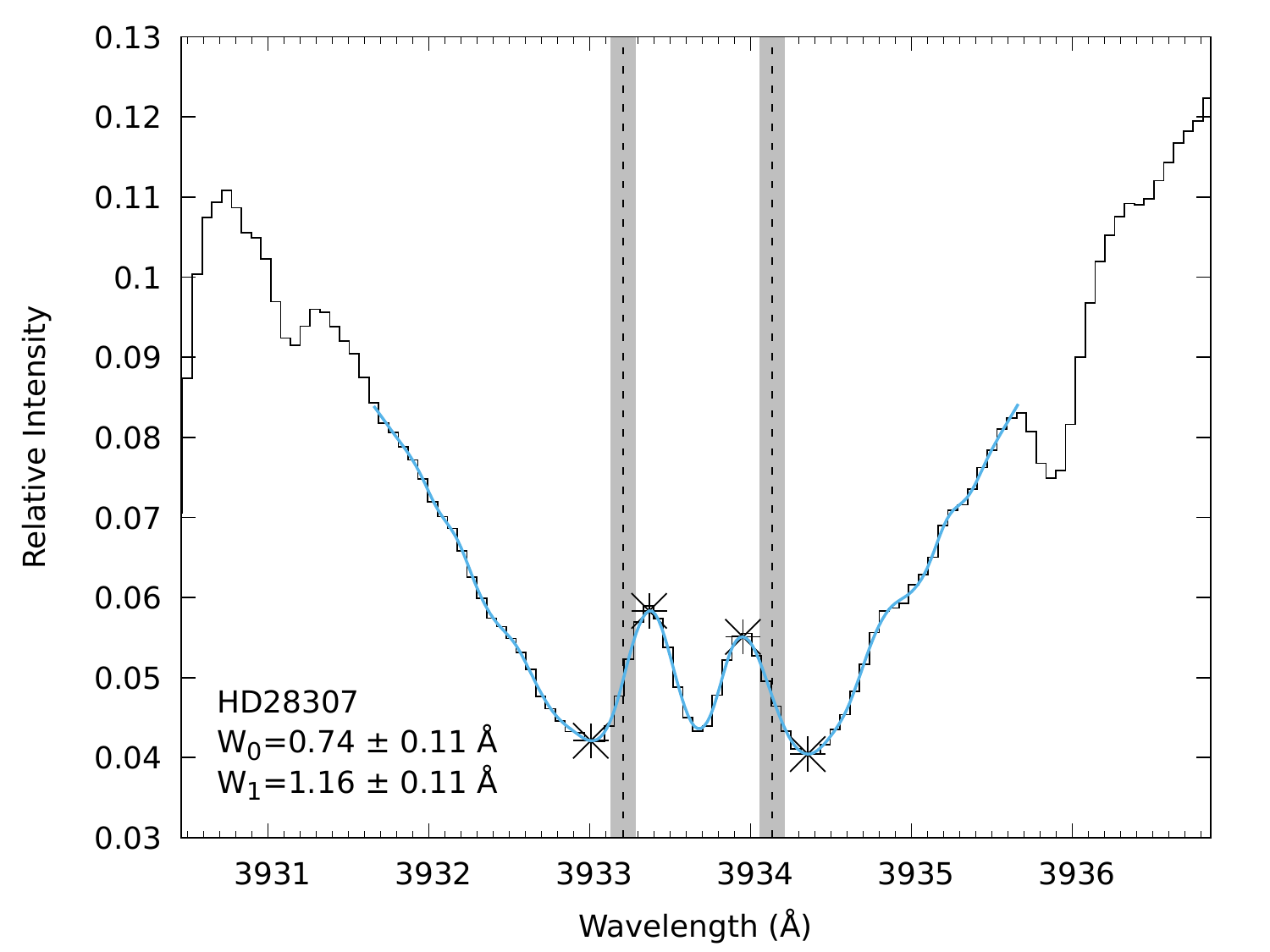}
  \includegraphics[width=0.425\textwidth]{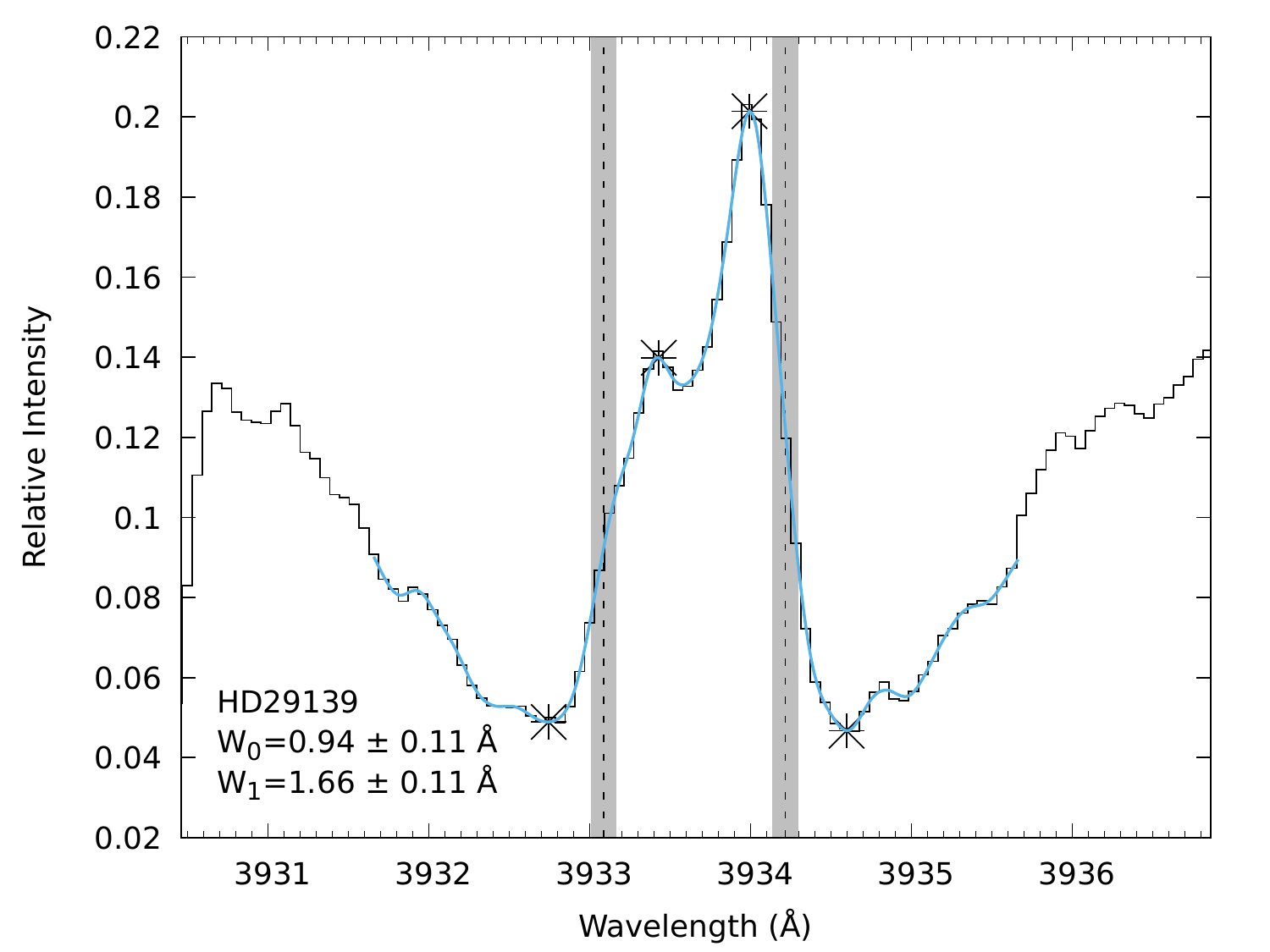}
  \includegraphics[width=0.425\textwidth]{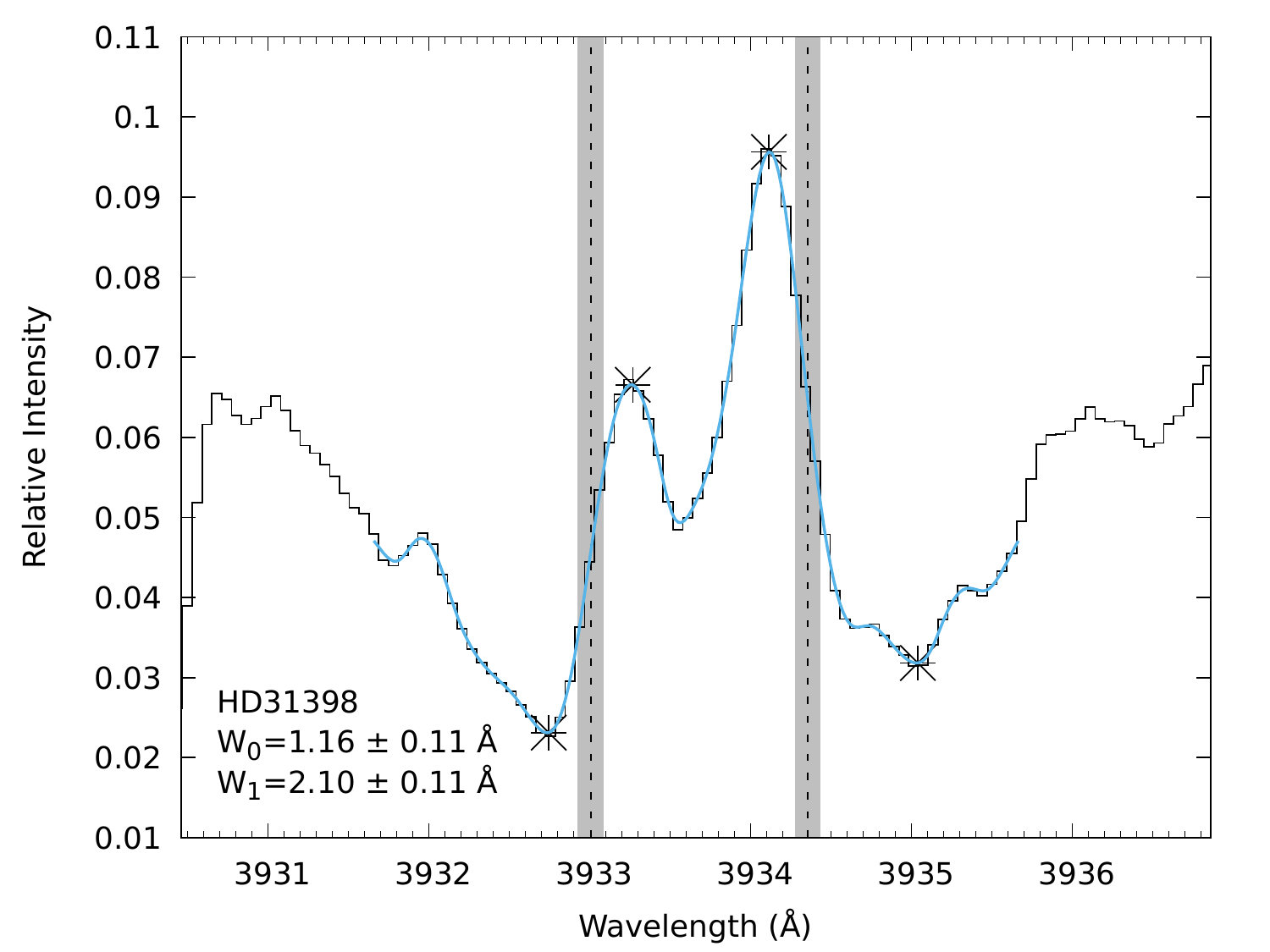}
  \includegraphics[width=0.425\textwidth]{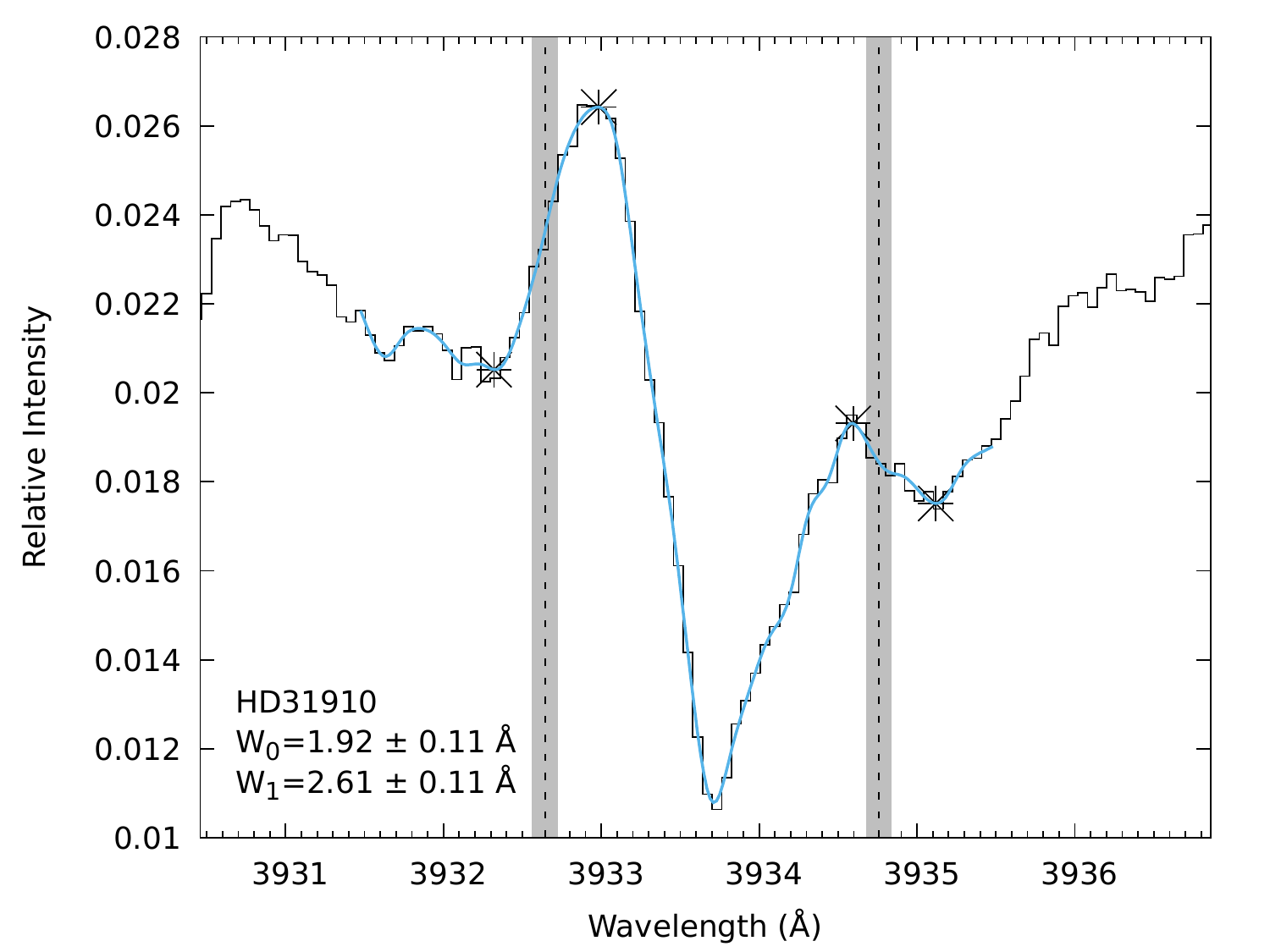}
  \includegraphics[width=0.425\textwidth]{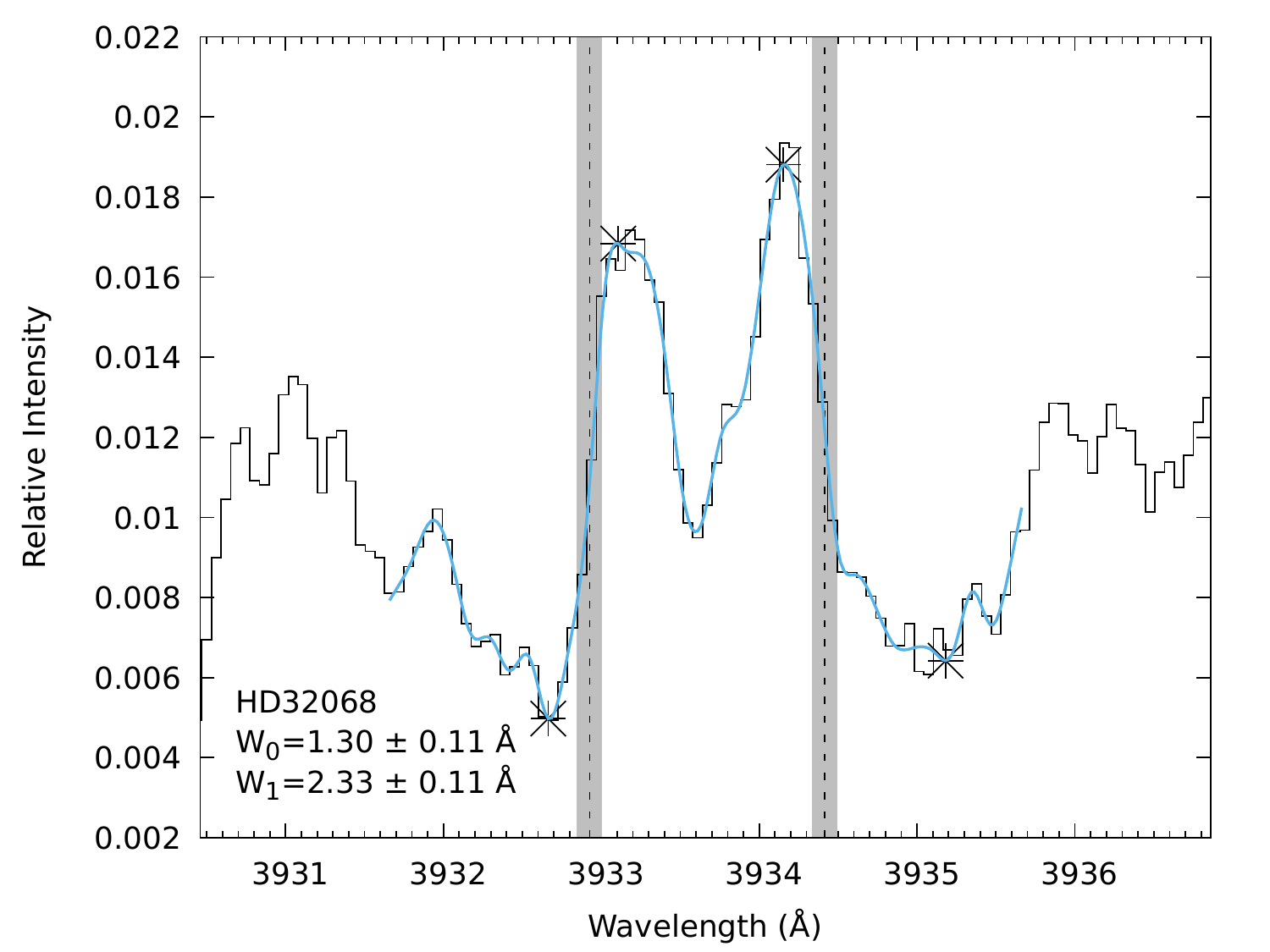}
  \includegraphics[width=0.425\textwidth]{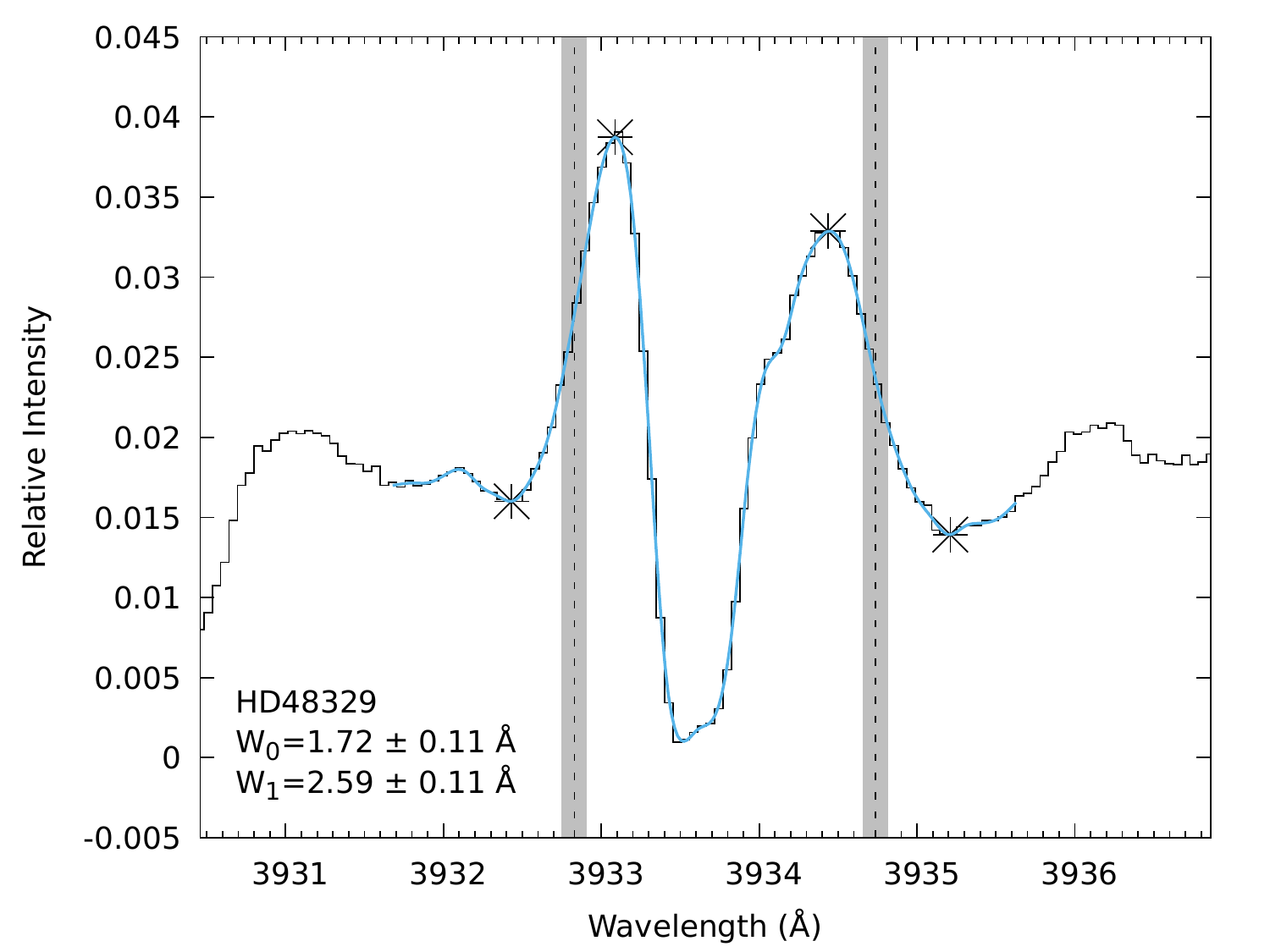}
  \includegraphics[width=0.425\textwidth]{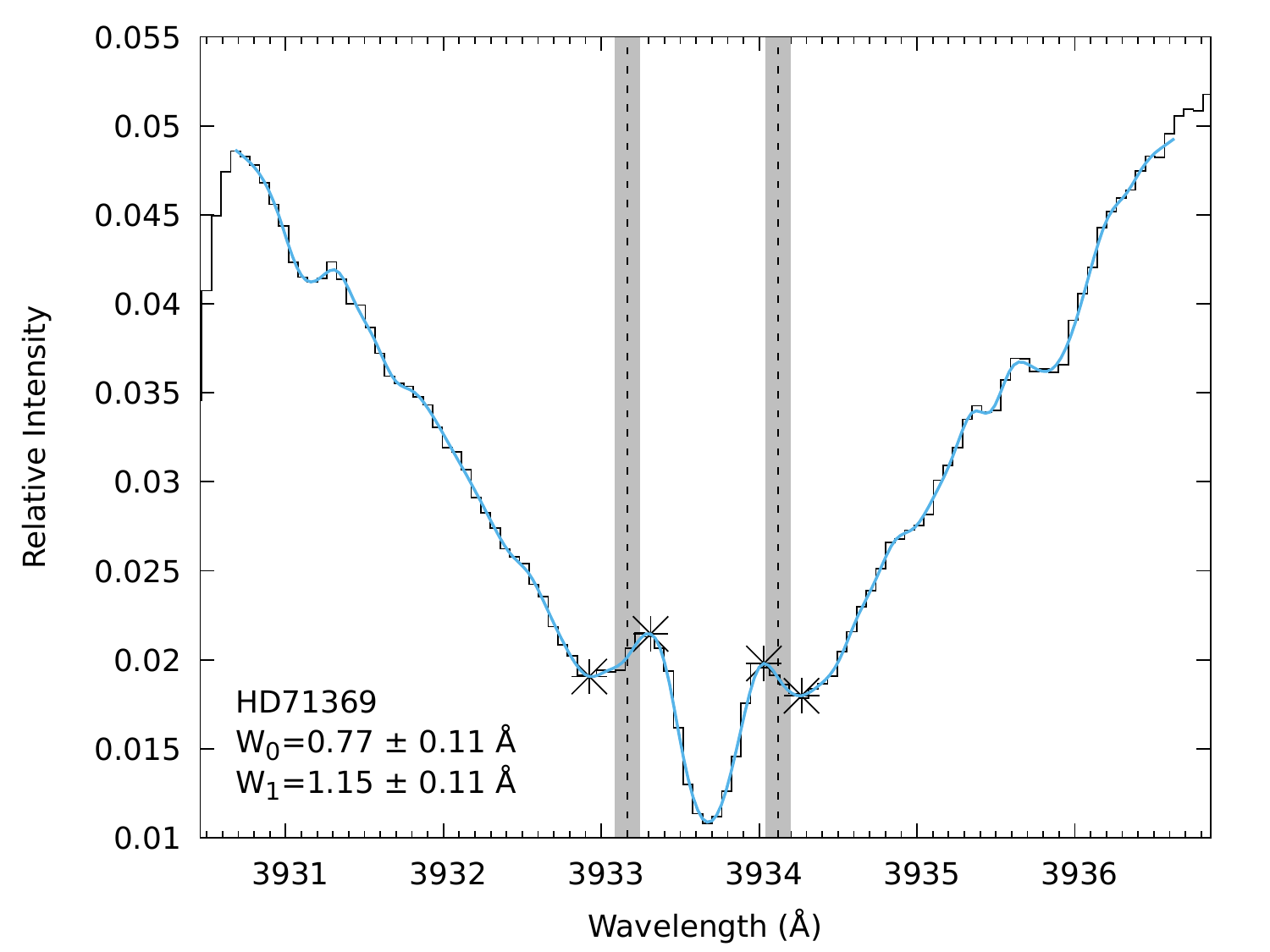}
  \caption{\ion{Ca}{ii} emission line width measurements of HD 28305, HD 28307, HD 29139, HD 31398, HD 31910, HD 32068, HD 48329 and HD 71369.}
  \label{fig:CaII_Widths_2}
\end{figure*}

\begin{figure*}
  \includegraphics[width=0.425\textwidth]{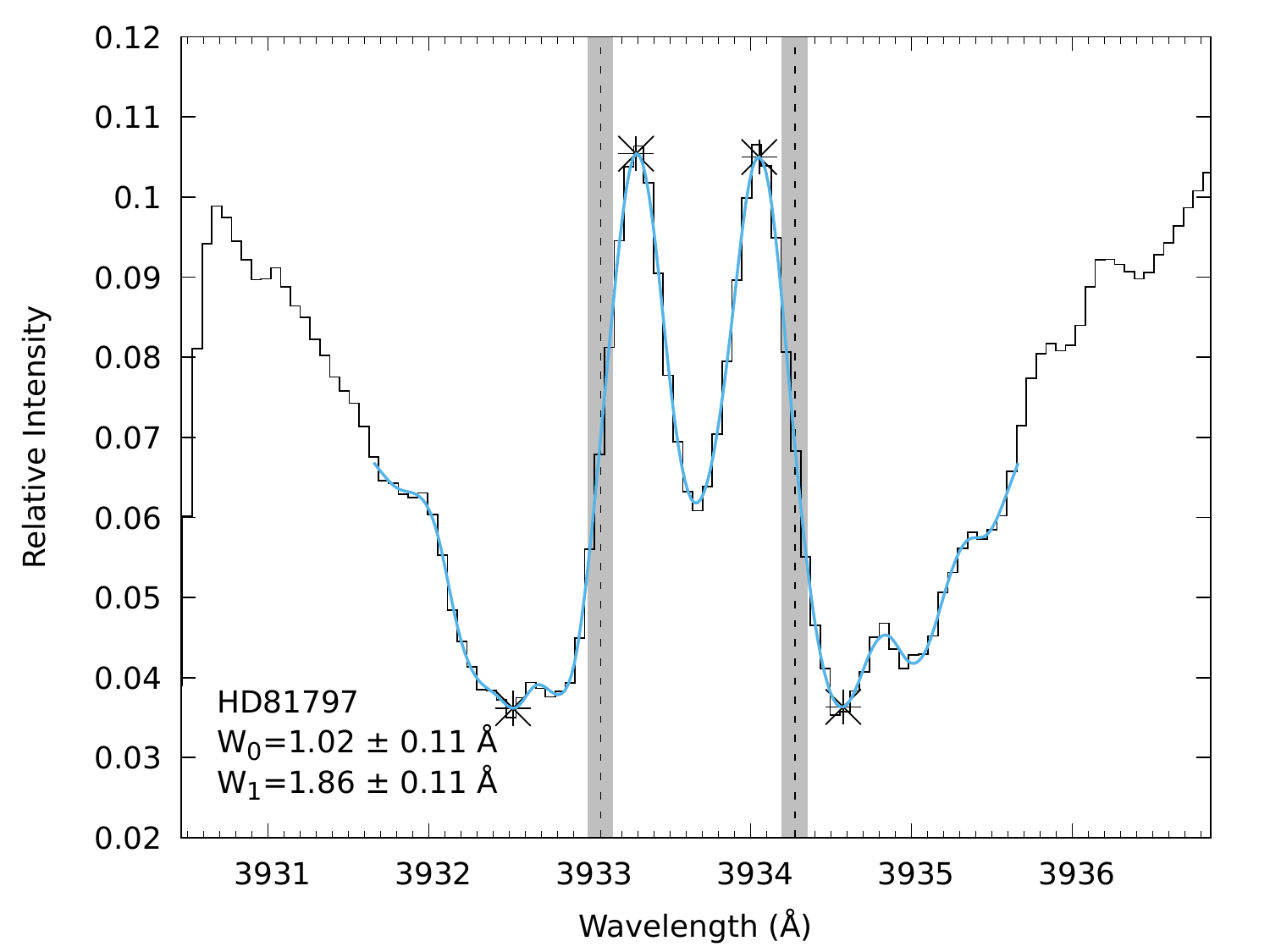}
  \includegraphics[width=0.425\textwidth]{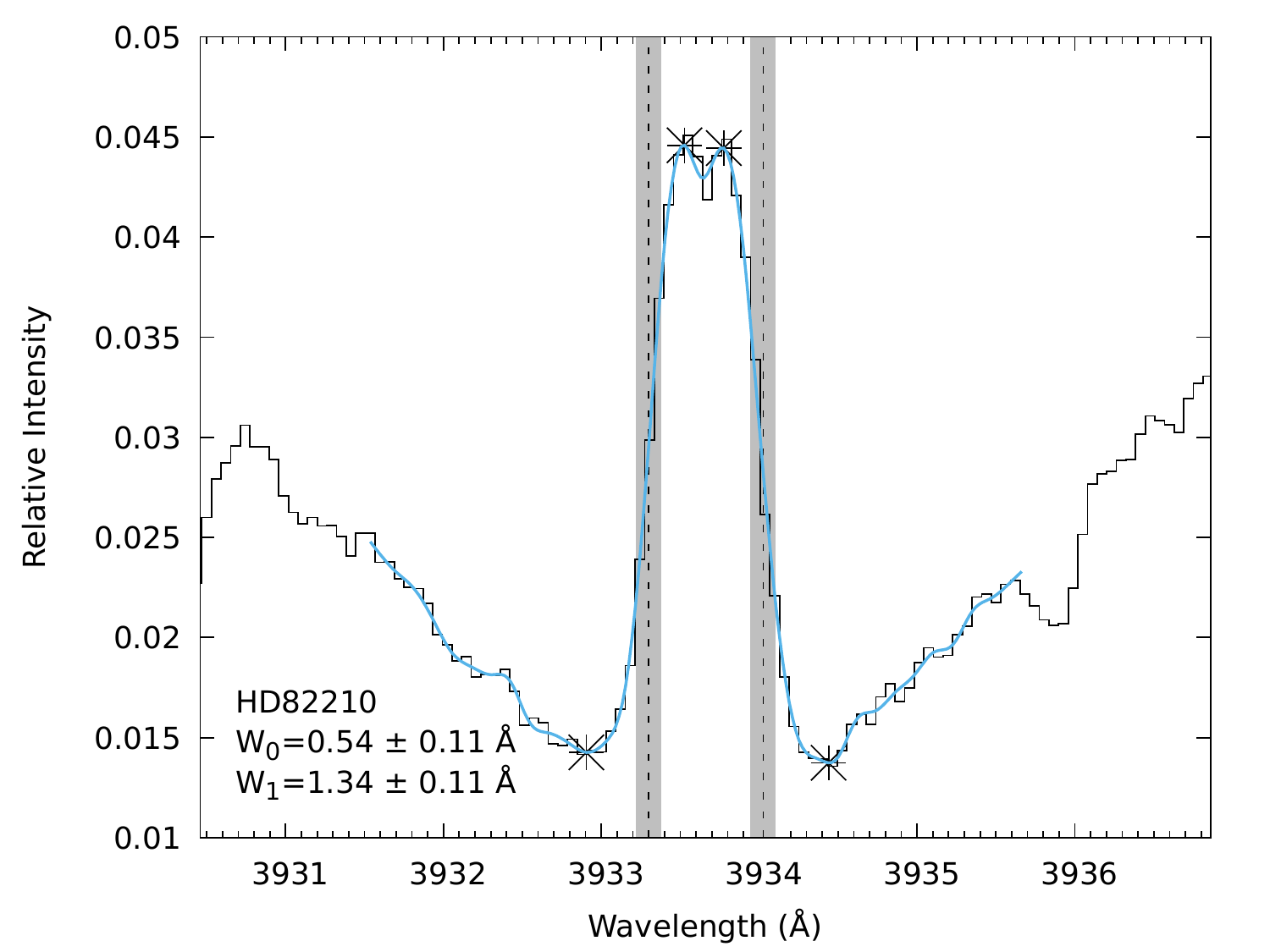}
  \includegraphics[width=0.425\textwidth]{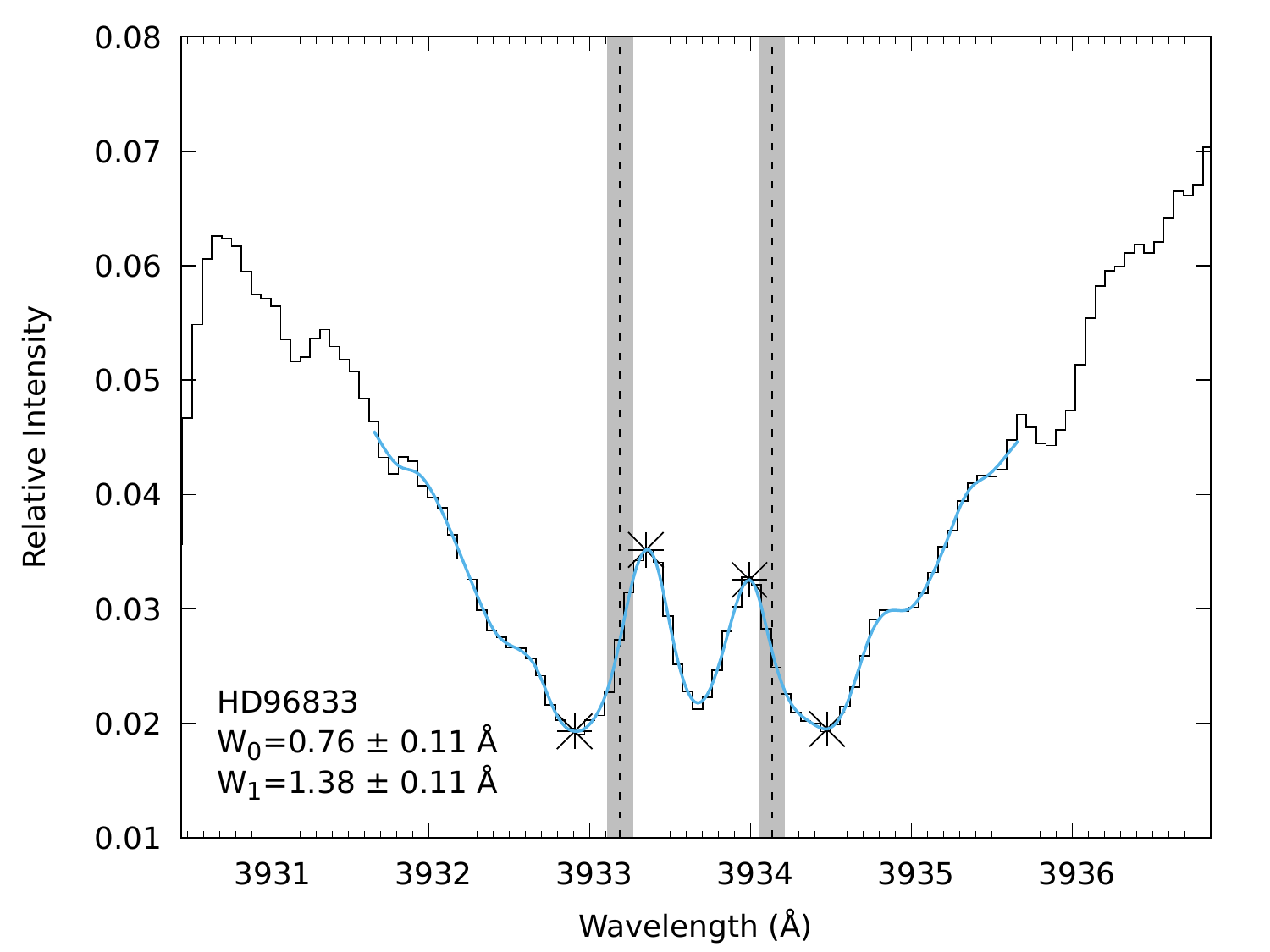}
  \includegraphics[width=0.425\textwidth]{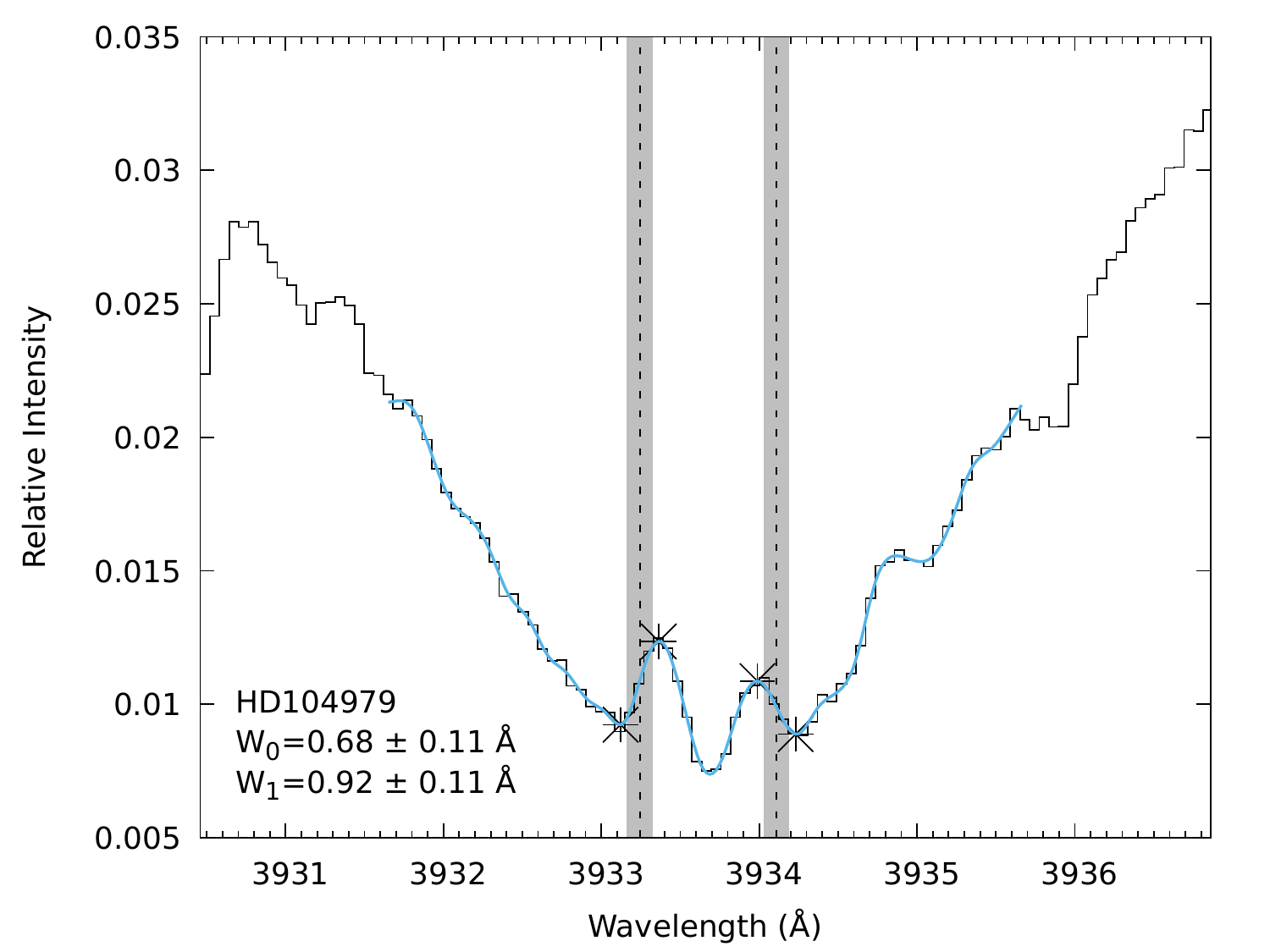}
  \includegraphics[width=0.425\textwidth]{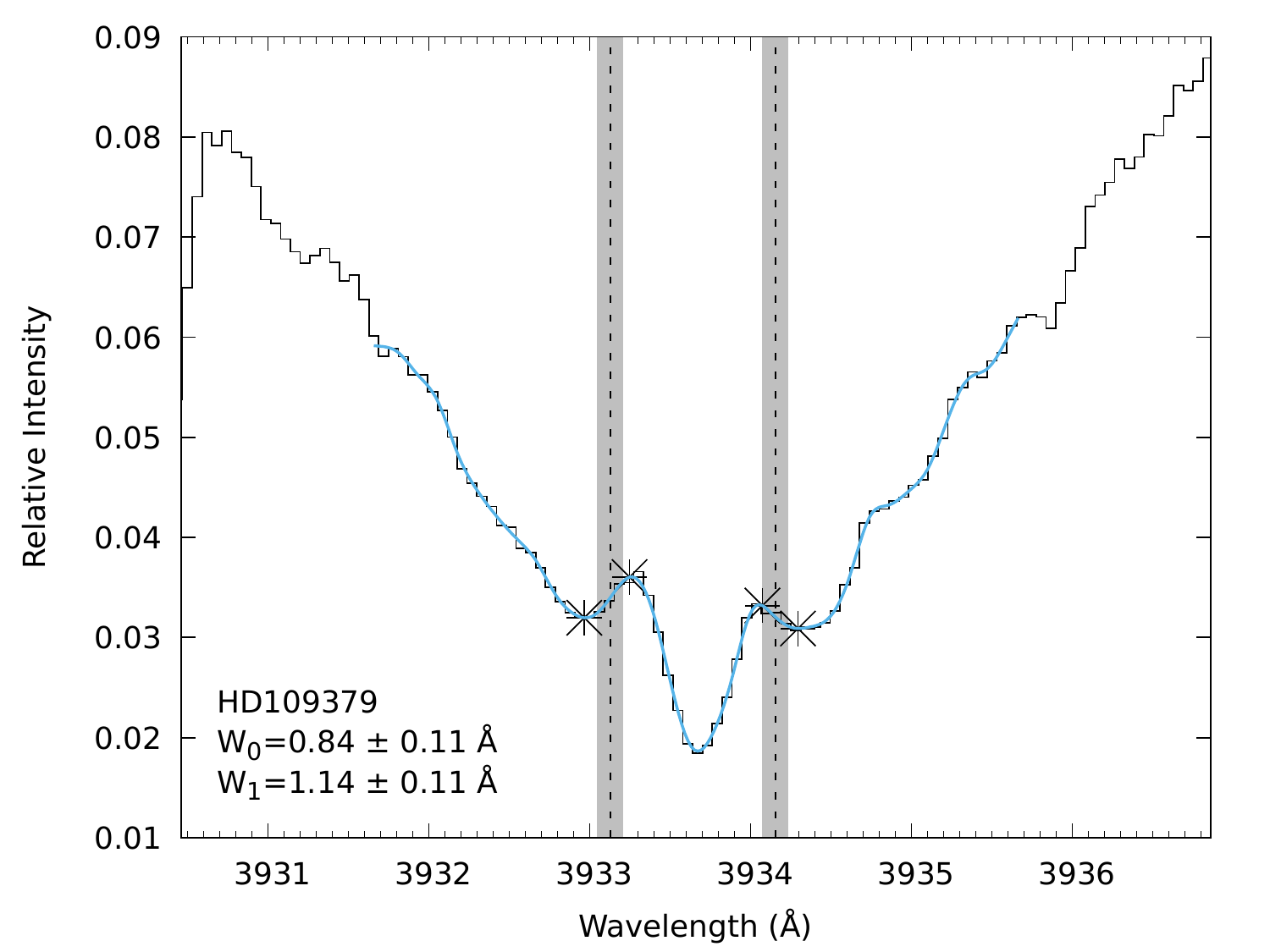}
  \includegraphics[width=0.425\textwidth]{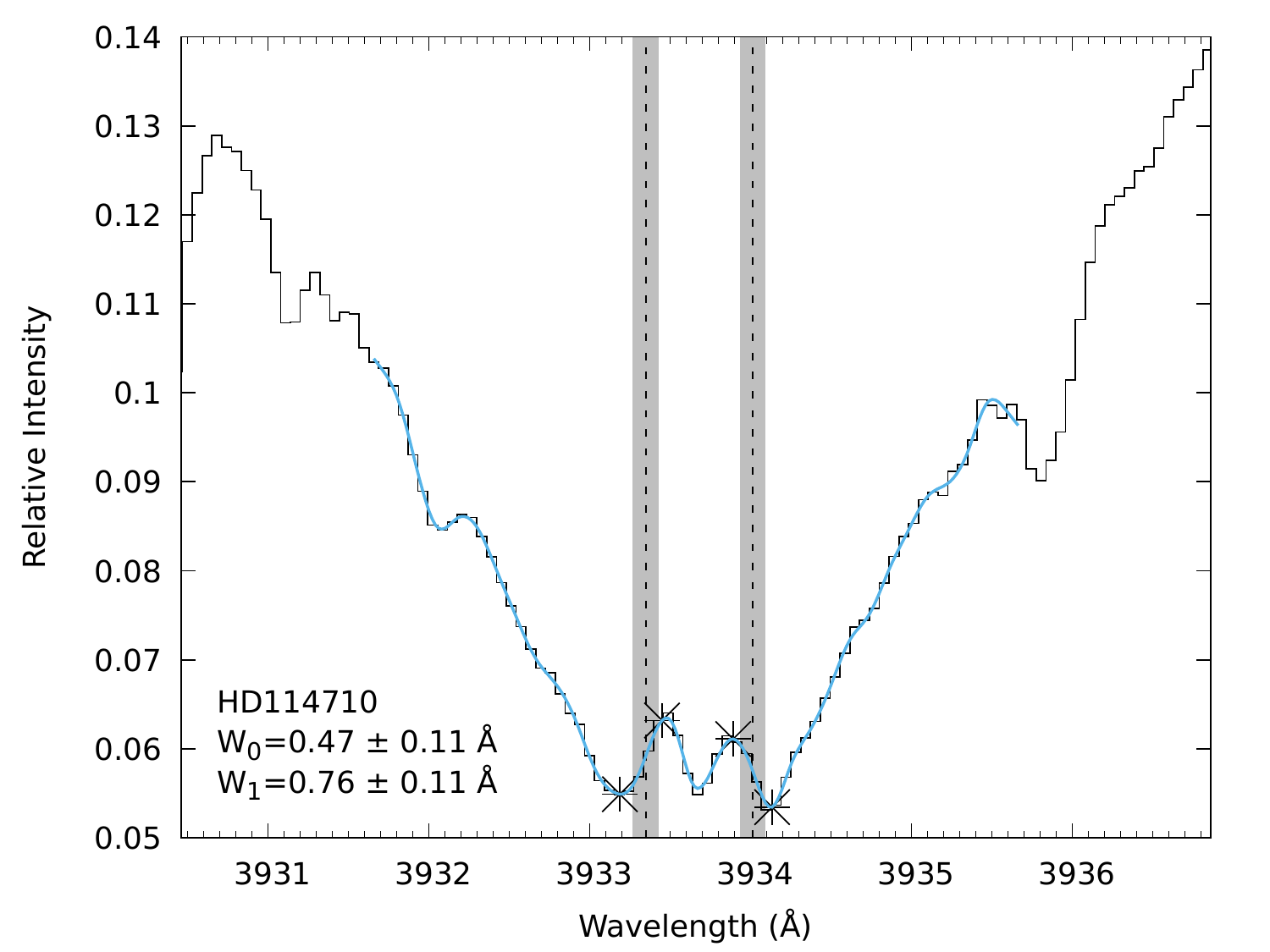}
  \includegraphics[width=0.425\textwidth]{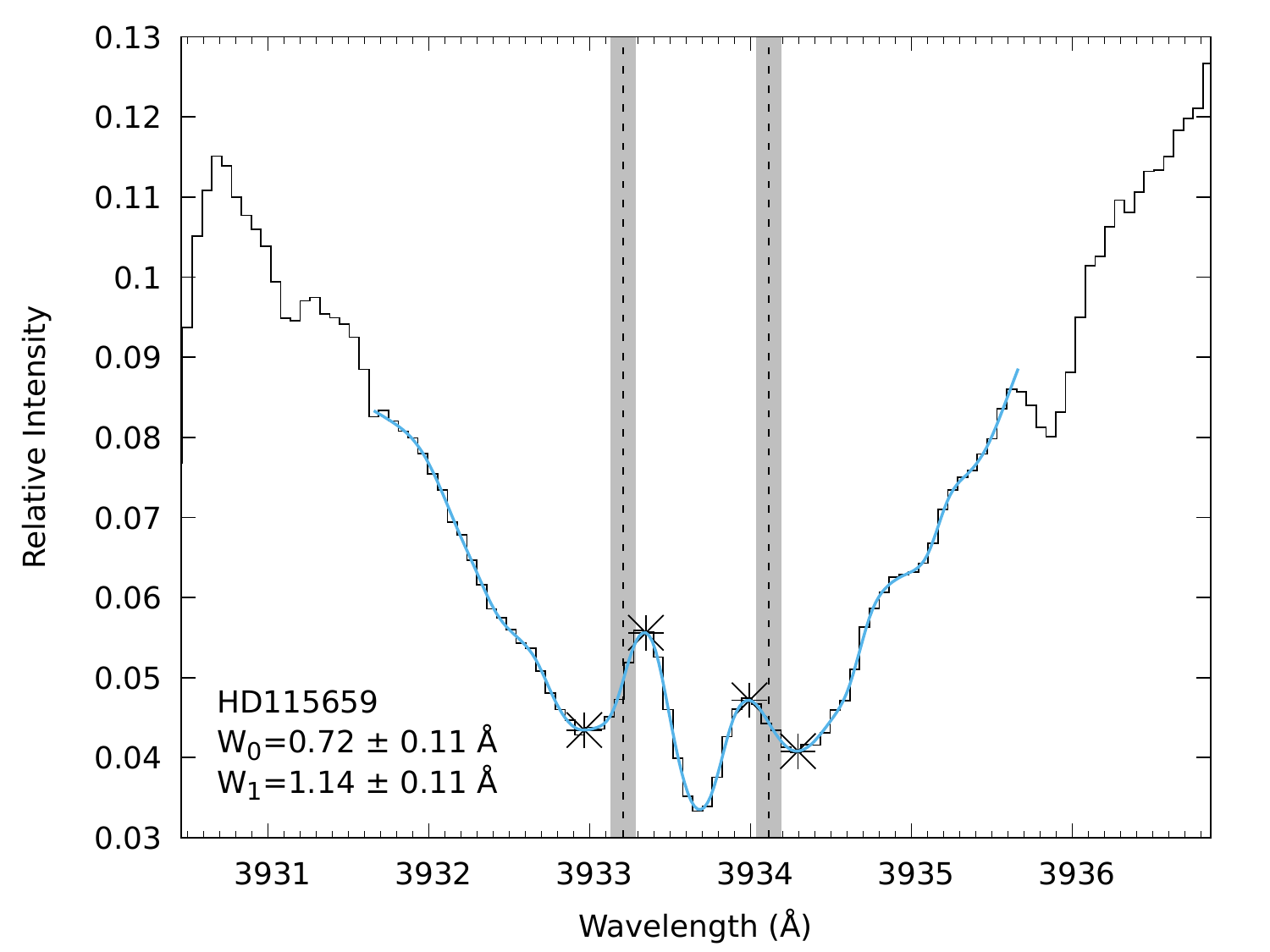}
  \includegraphics[width=0.425\textwidth]{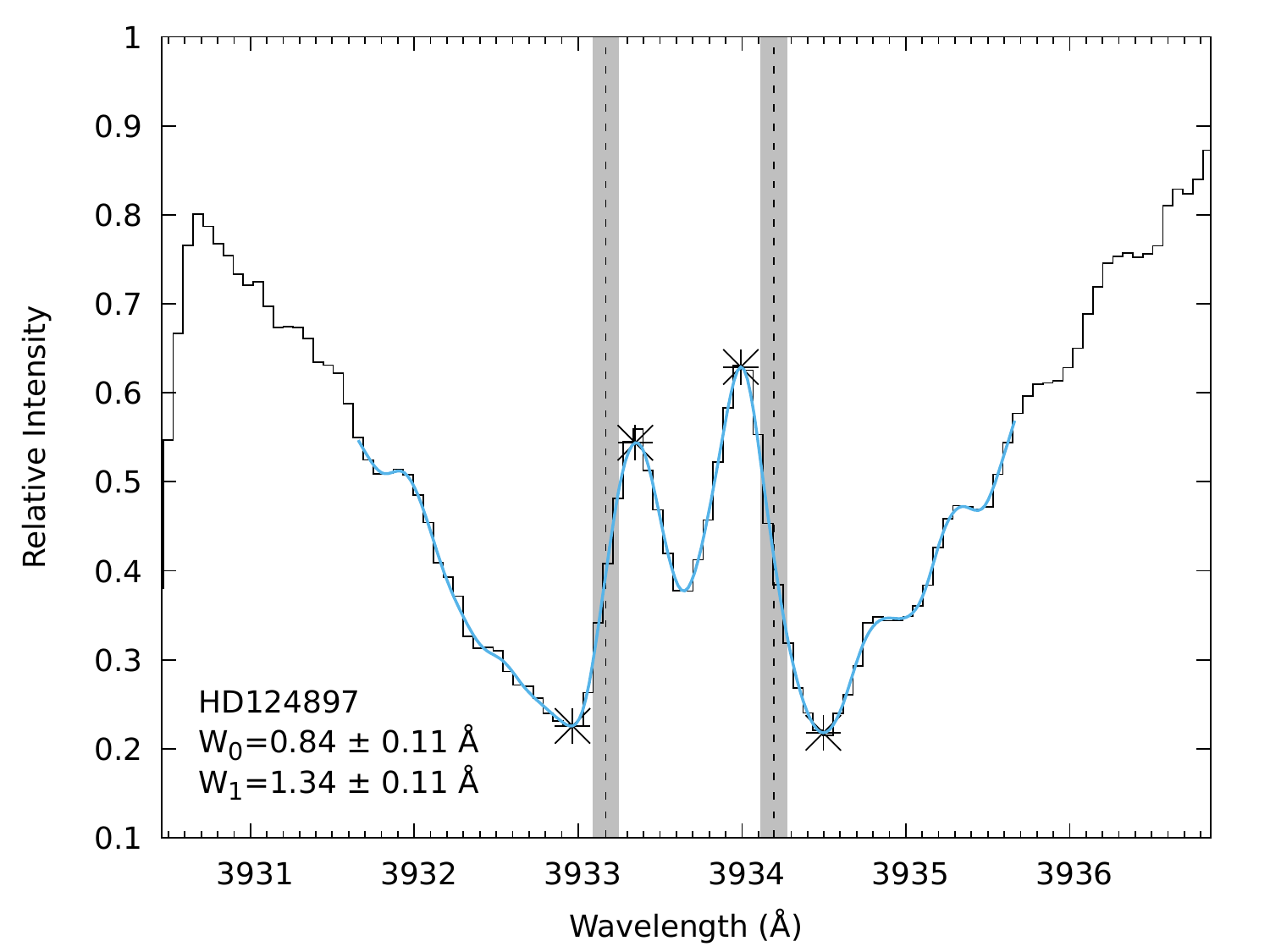}
  \caption{\ion{Ca}{ii} emission line width measurements of HD 81797, HD 82210, HD 96833, HD 104979, HD 109379, HD 114710, HD 115659 and HD 124897.}
  \label{fig:CaII_Widths_3}
\end{figure*}

\begin{figure*}
  \includegraphics[width=0.425\textwidth]{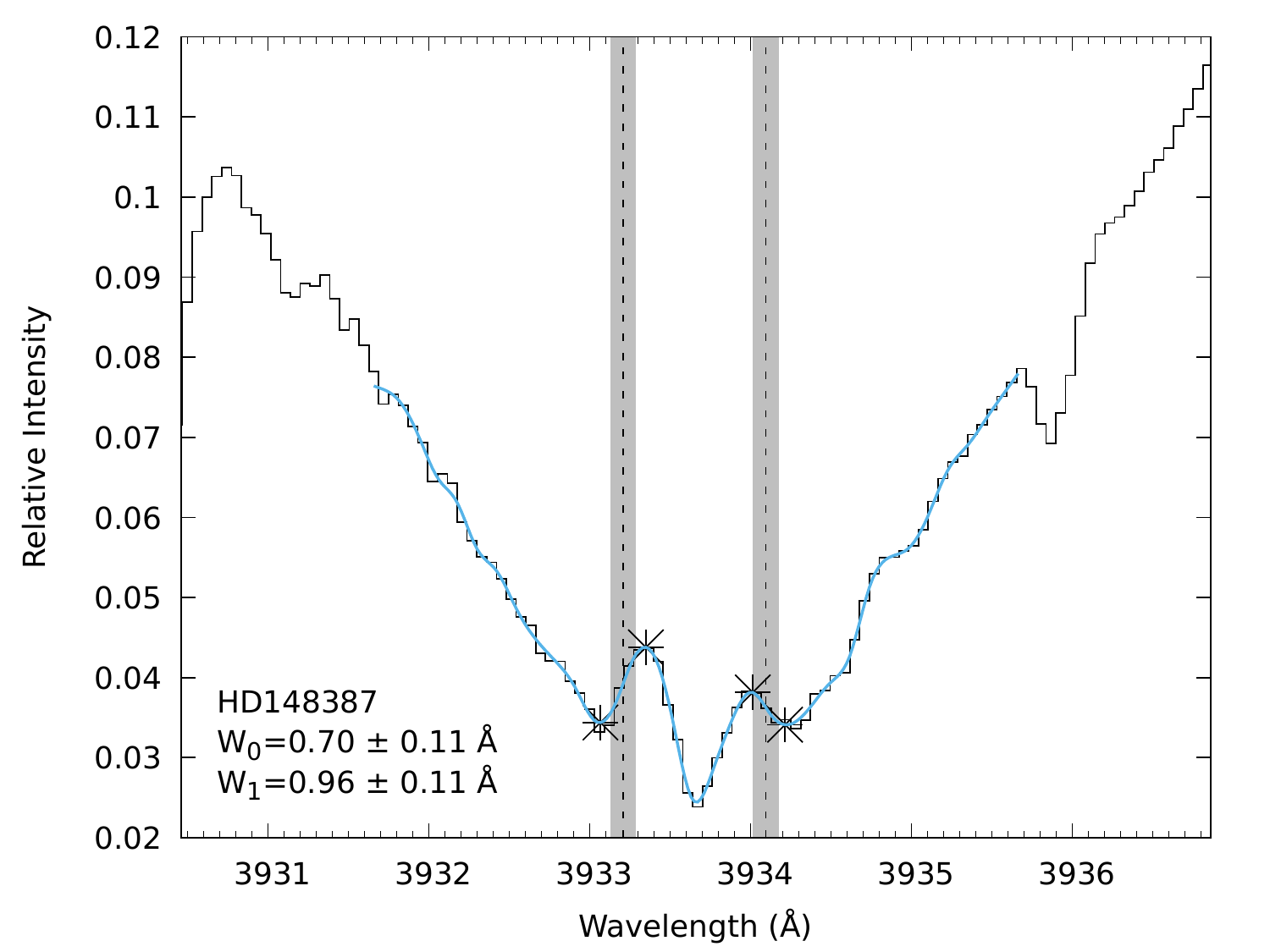}
  \includegraphics[width=0.425\textwidth]{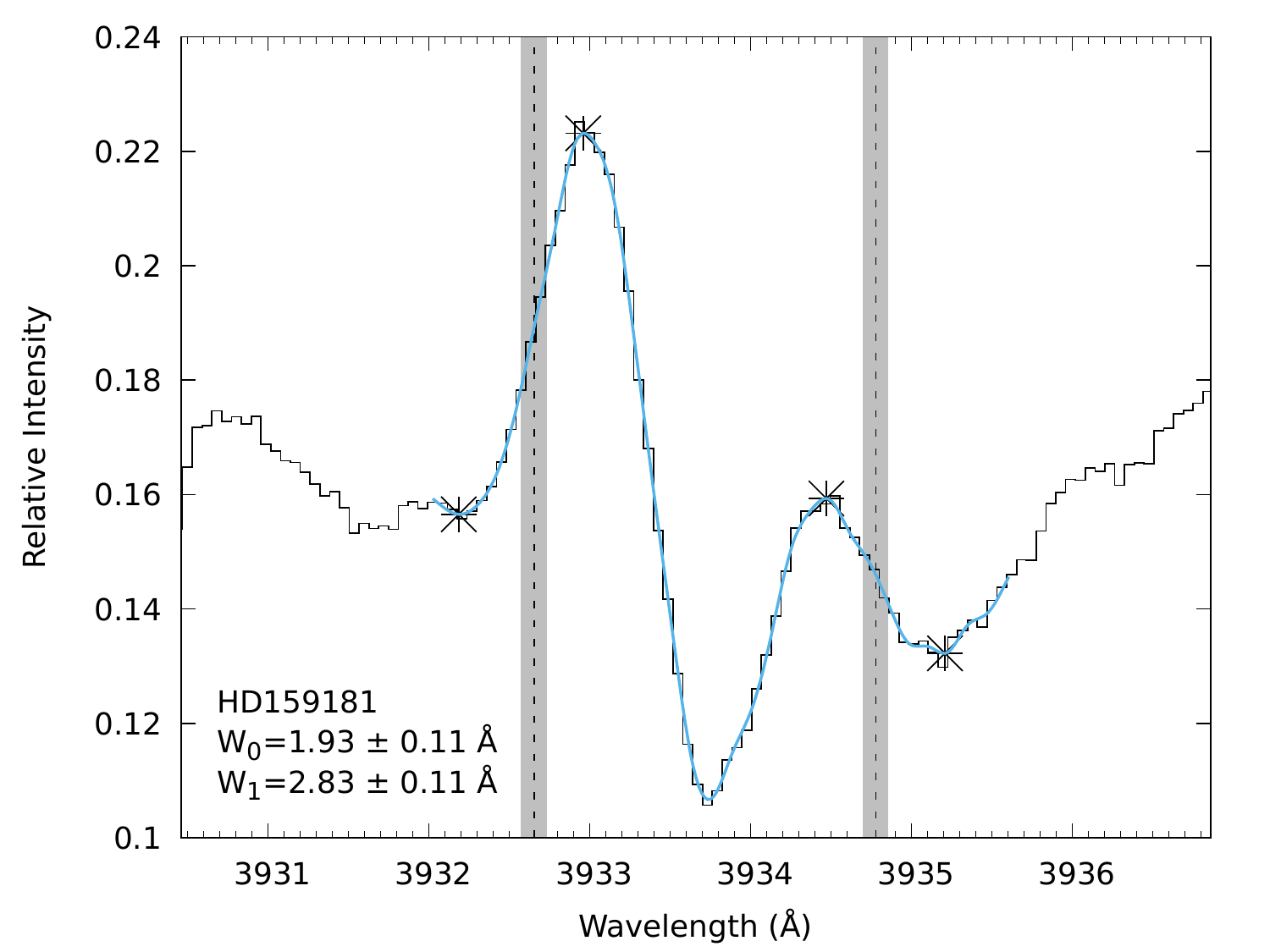}
  \includegraphics[width=0.425\textwidth]{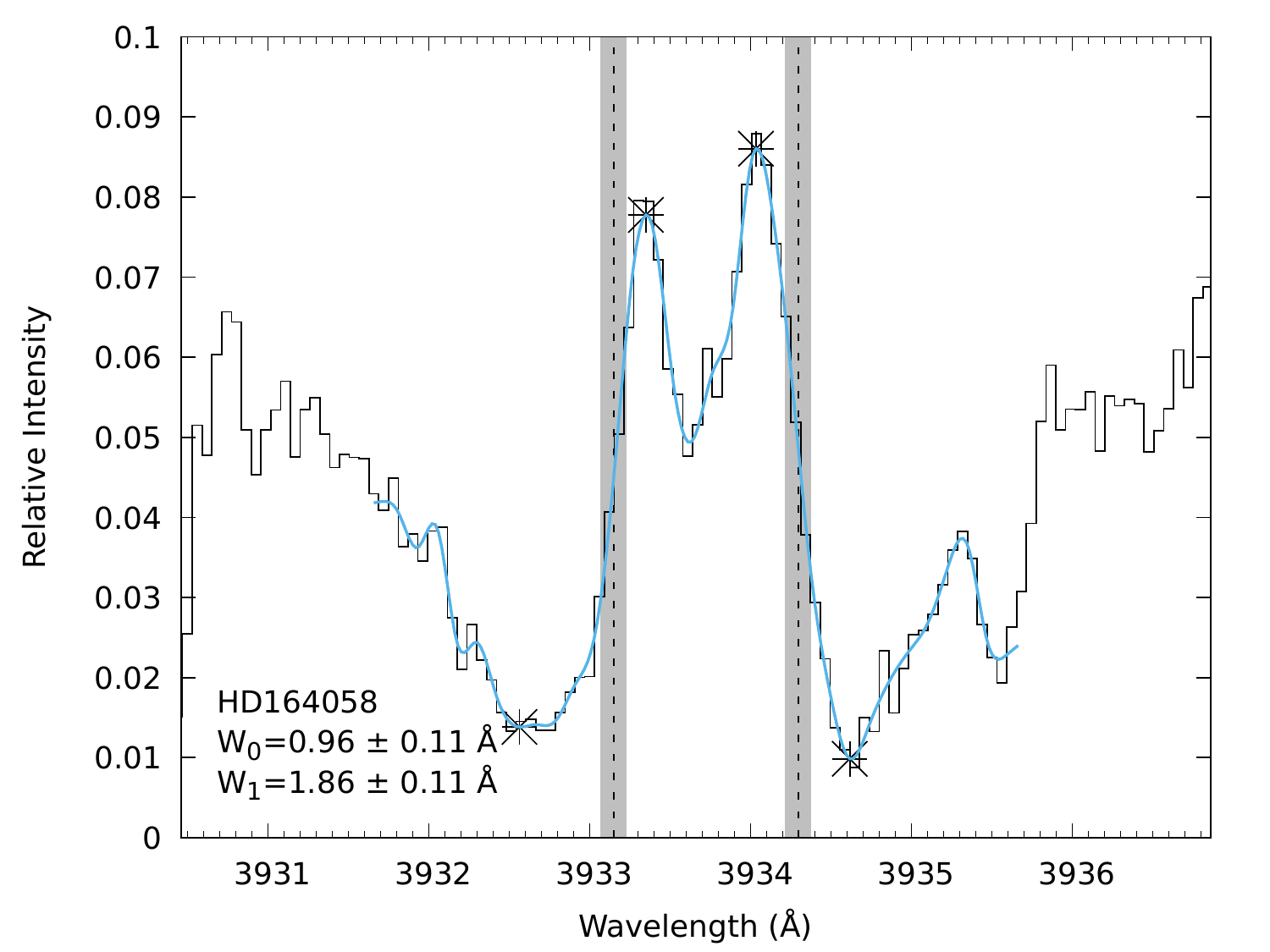}
  \includegraphics[width=0.425\textwidth]{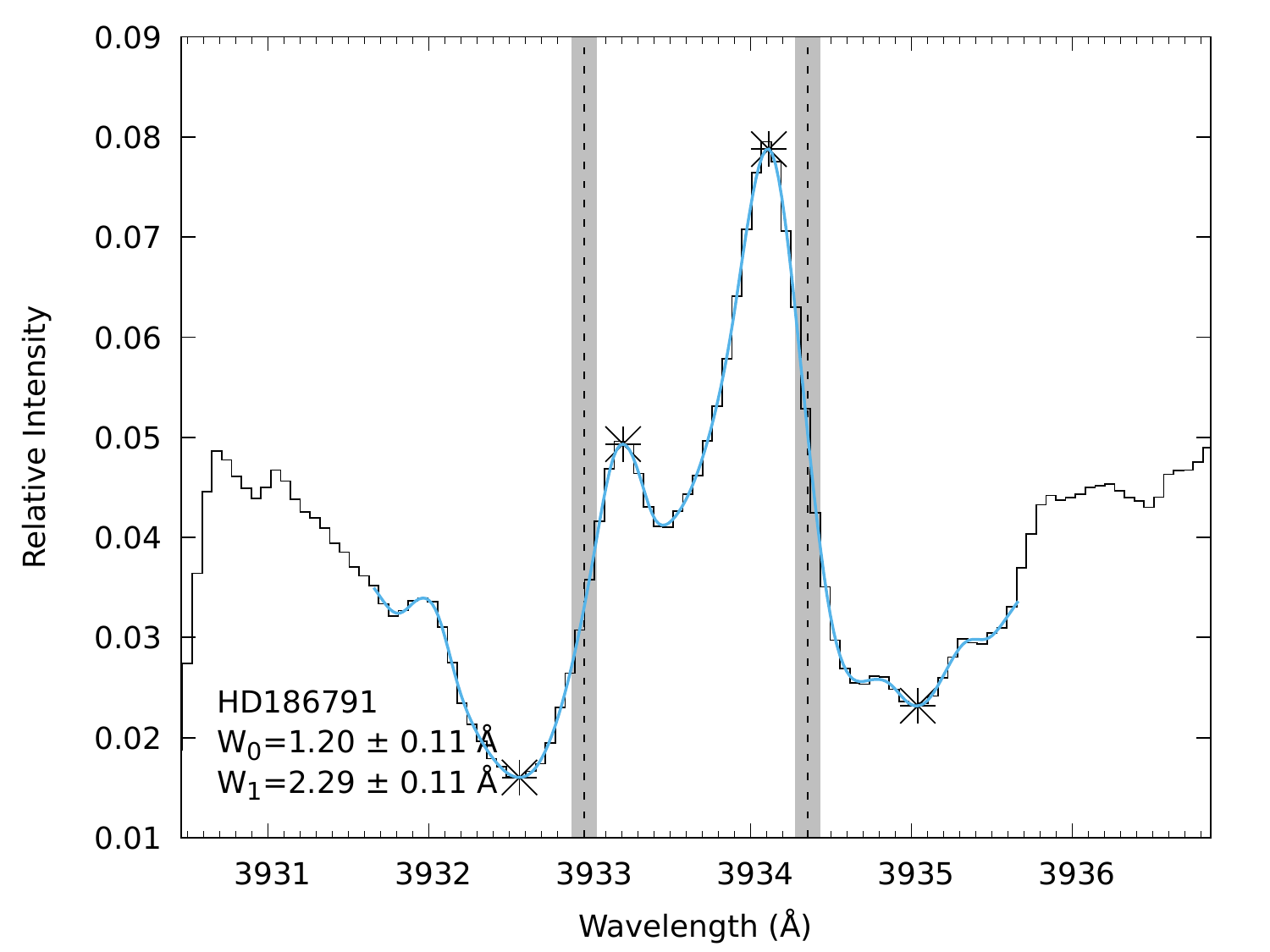}
  \includegraphics[width=0.425\textwidth]{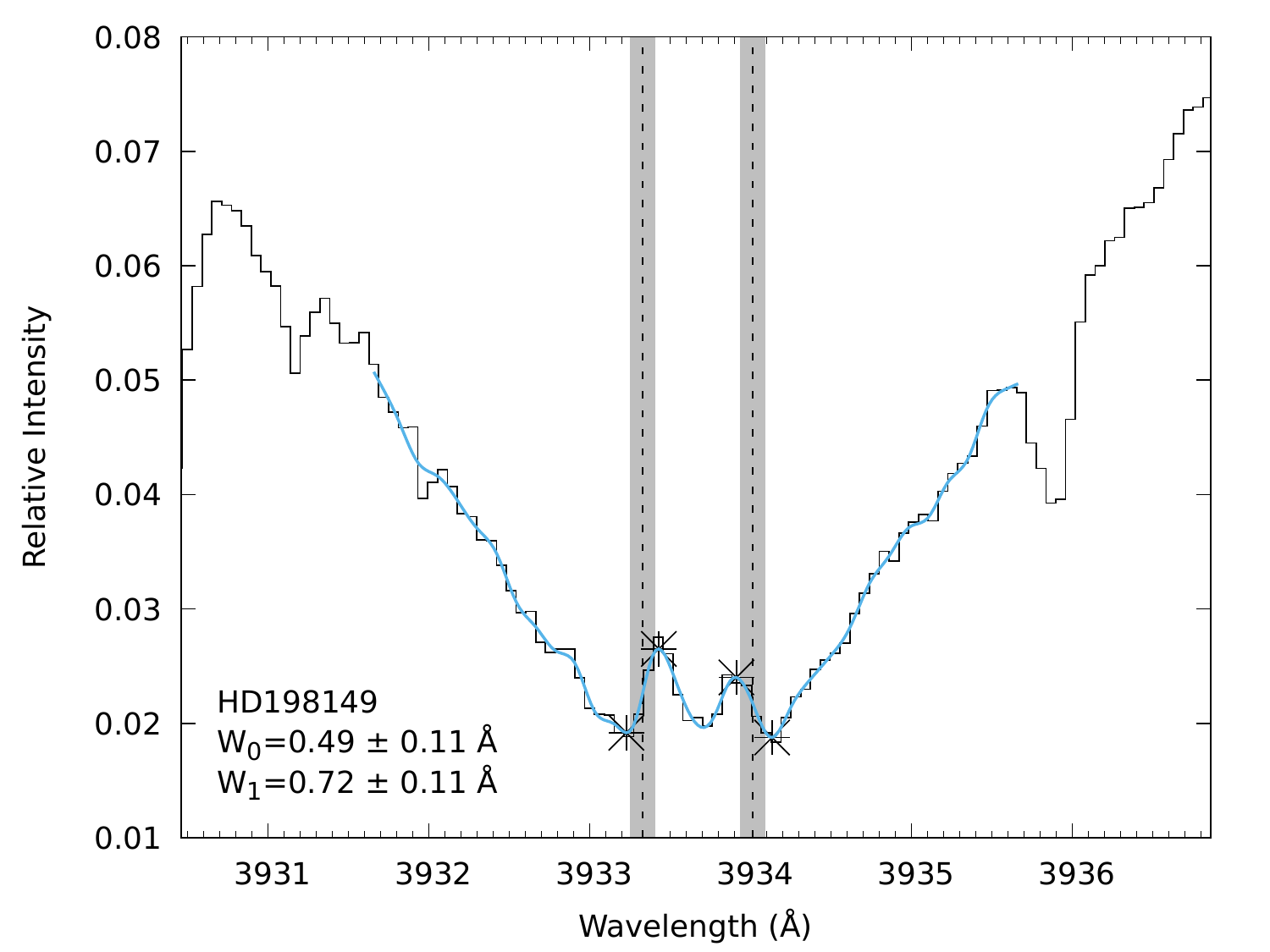}
  \includegraphics[width=0.425\textwidth]{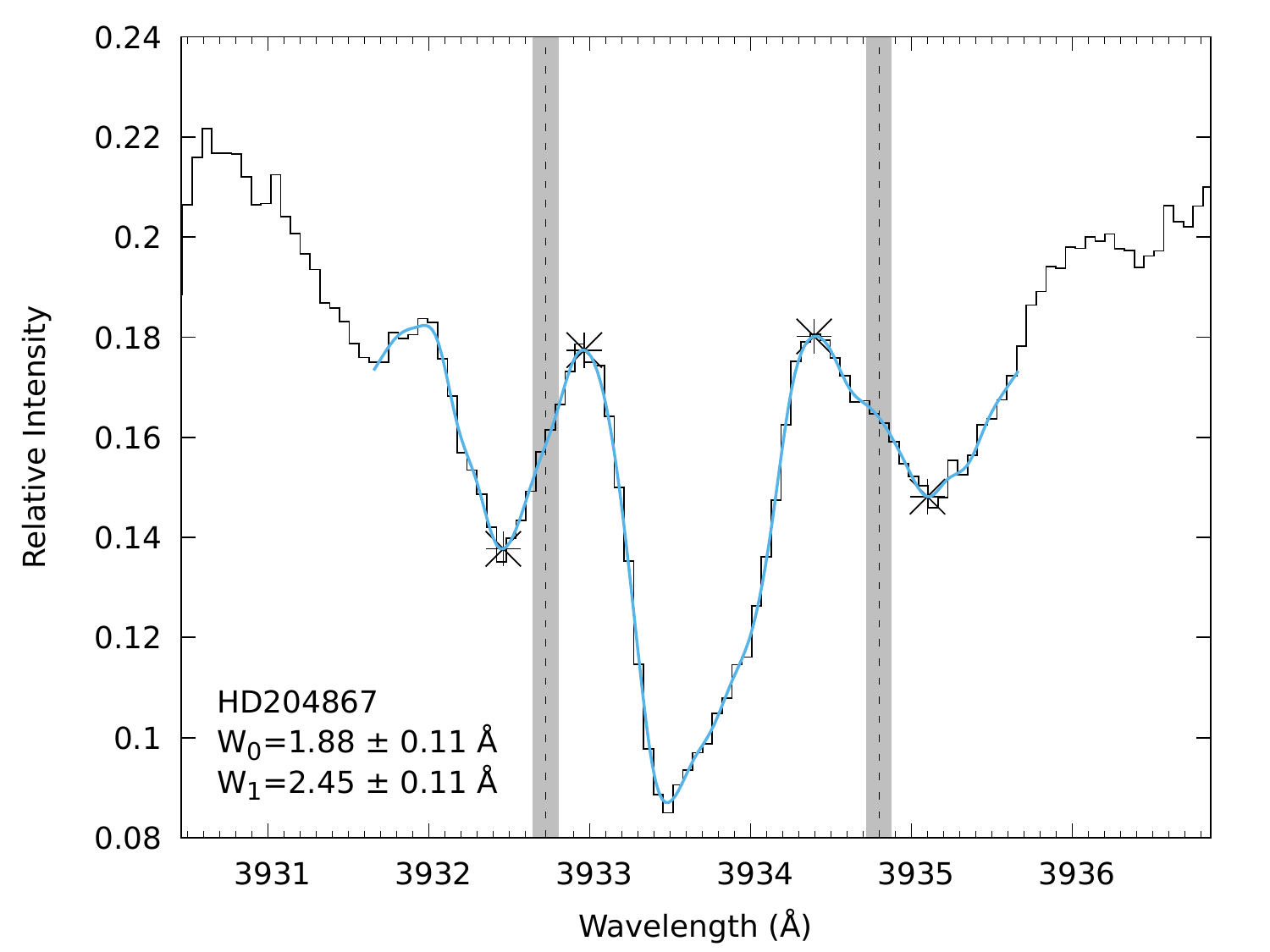}
  \includegraphics[width=0.425\textwidth]{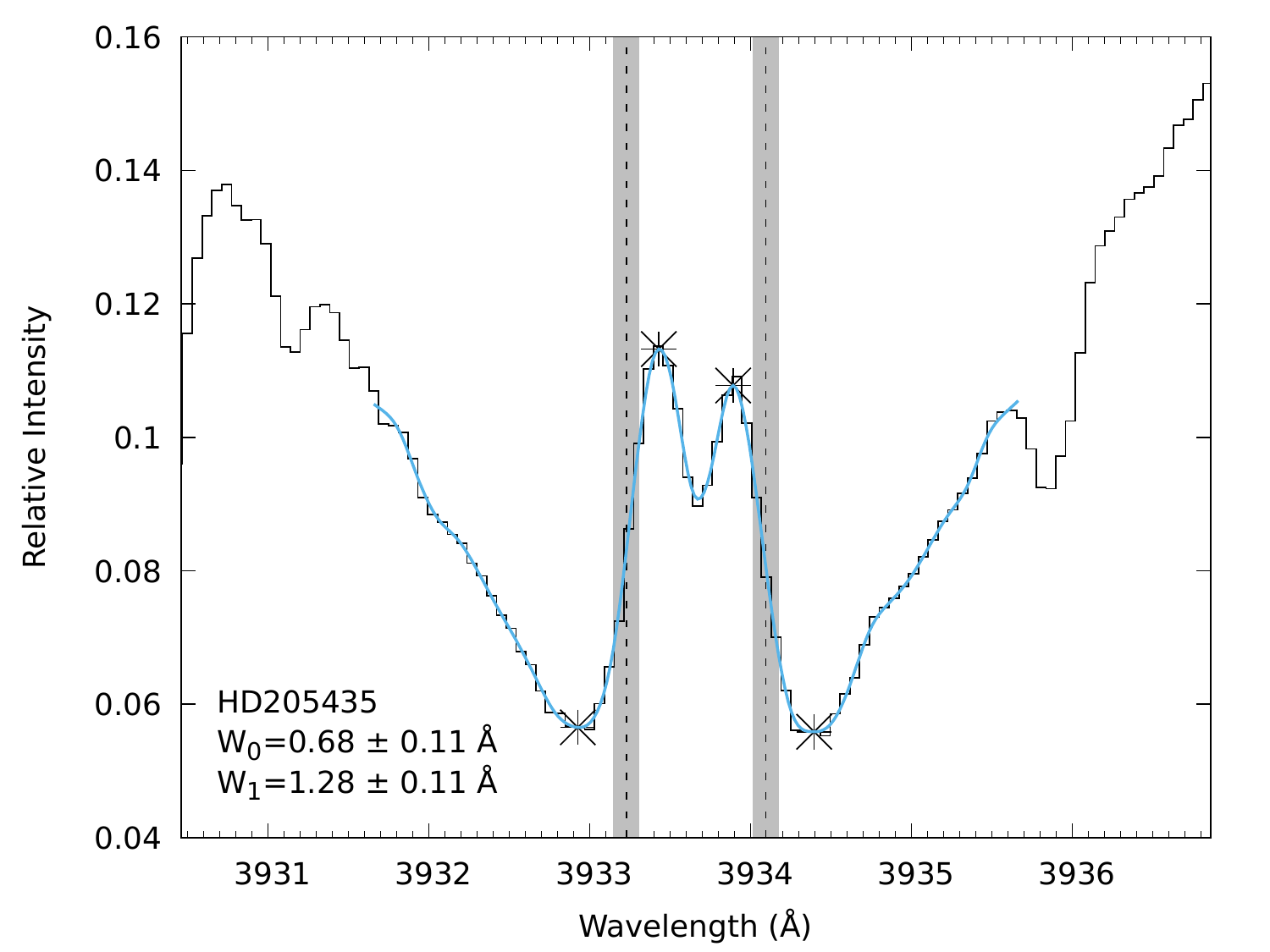}
  \includegraphics[width=0.425\textwidth]{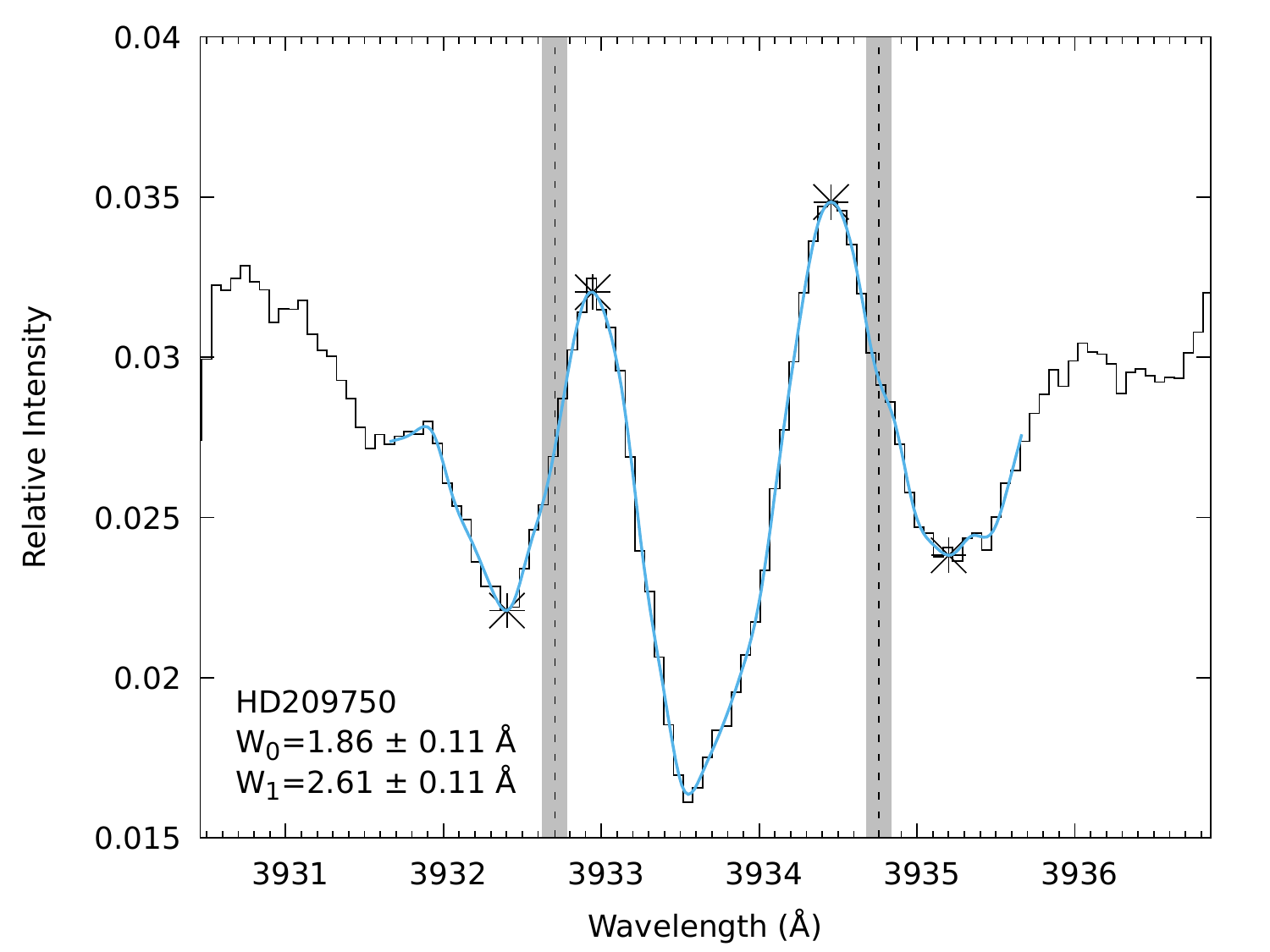}
  \caption{\ion{Ca}{ii} emission line width measurements of HD 148387, HD 159181, HD 164058, HD 186791, HD 198149, HD 204867, HD 205435 and HD 209750.}
  \label{fig:CaII_Widths_4}
\end{figure*}

We present the \ion{Ca}{ii} emission line width measurements of the stellar sample with a fixed wavelength scale to comparison purposes. Fig. \ref{fig:CaII_Widths_1} shows the measurements for HD 8512, HD 10476, HD 18925, HD 20630, HD 23249, HD 26630, HD 27371 and HD 27697. Fig. \ref{fig:CaII_Widths_2} for HD 28305, HD 28307, HD 29139, HD 31398, HD 31910, HD 32068, HD 48329 and HD 71369. Fig. \ref{fig:CaII_Widths_3} for HD 81797, HD 82210, HD 96833, HD 104979, HD 109379, HD 114710, HD 115659 and HD 124897. Fig. \ref{fig:CaII_Widths_4} for HD 148387, HD 159181, HD 164058, HD 186791, HD 198149, HD 204867, HD 205435 and HD 209750. The blue line represents the fit using cubic splines, the dashed line is the half intensity of the peak and gray zone is the measurement uncertainty.

\bsp	
\label{lastpage}
\end{document}